\documentclass[preprints,article,accept,moreauthors,pdftex]{./mdpi}

\firstpage{1} 
\pubvolume{1}
\issuenum{1}
\articlenumber{0}
\pubyear{2022}
\copyrightyear{2022}
\datereceived{} 
\dateaccepted{} 
\datepublished{} 
\hreflink{} 

\usepackage[binary-units,separate-uncertainty=true]{siunitx}
\DeclareSIUnit{\bpm}{bpm}

\Title{Towards Personalized Healthcare in Cardiac Population: The Development of a Wearable ECG Monitoring System, an ECG Lossy Compression Schema, and a ResNet-Based AF Detector}

\TitleCitation{Towards Personalized Healthcare in Cardiac Population: The Development of a Wearable ECG Monitoring System, an ECG Lossy Compression Schema, and a ResNet-Based AF Detector}

\Author{Wei-Ying Yi $^{1,2}$\orcidA{}, Peng-Fei Liu $^{1}$\orcidB{}, Sheung-Lai Lo $^{1,2}$\orcidC{}, Ya-Fen Chan $^{2}$, Yu Zhou $^{2,3}$\orcidE{}, Yee Leung $^{2,3}$, Kam-Sang Woo $^{4}$, Alex Pui-Wai Lee $^{5}$\orcidH{}, Jia-Min Chen $^{1,2}$ and Kwong-Sak Leung $^{1,2,}$*}

\AuthorNames{Wei-Ying Yi, Peng-Fei Liu, Sheung-Lai Lo, Ya-Fen Chan, Yu Zhou, Yee Leung, Kam-Sang Woo, Alex Pui-Wai Lee, Jia-Min Chen and Kwong-Sak Leung}

\AuthorCitation{Yi, W.Y.; Liu, P.F.; Lo, S.L.; Chan, Y.F.; Zhou, Y.; Leung, Y.; Woo, K.S.; Lee, A.P.W.; Chen, J.M.; Leung, K.S.}

\address{%
$^{1}$ \quad Department of Computer Science and Engineering, The Chinese University of Hong Kong, Sha Tin, N.T., Hong Kong, China; wyyi1991@gmail.com (Yi, W.Y.); pfliu@cse.cuhk.edu.hk (Liu, P.F.); \\ \quad~~~lesterlo@link.cuhk.edu.hk (Lo, S.L.); 1155092190@link.cuhk.edu.hk (Chen, J.M.); ksleung@cse.cuhk.edu.hk (Leung, K.S.)\\
$^{2}$ \quad Institute of Future Cities, The Chinese University of Hong Kong, Sha Tin, N.T., Hong Kong, China; yfchan@cuhk.edu.hk (Chan, Y.F.)\\
$^{3}$ \quad Department of Geography and Resource Management, The Chinese University of Hong Kong, Sha Tin, N.T., Hong Kong, China; yuzhou@cuhk.edu.hk (Zhou, Y.); yeeleung@cuhk.edu.hk (Leung, Y.)\\
$^{4}$ \quad Department of Medicine and Therapeutics, The Chinese University of Hong Kong, Sha Tin, N.T., Hong Kong, China; kamsangwoo@cuhk.edu.hk (Woo, K.S.)\\
$^{5}$ \quad Li Ka Shing Institute of Health Science, The Chinese University of Hong Kong, Sha Tin, N.T., Hong Kong, China; alexpwlee@cuhk.edu.hk (Lee, P.W.)
}

\corres{Correspondence: ksleung@cse.cuhk.edu.hk}

\abstract{Cardiovascular diseases (CVDs) are the number one cause of death worldwide. While there is growing evidence that the atrial fibrillation (AF) has strong associations with various CVDs, this heart arrhythmia is usually diagnosed using the electrocardiography (ECG) which is a risk-free, non-intrusive, and cost-efficient tool. Continuously and remotely monitoring the subjects' ECG information unlocks the potentials of prompt pre-diagnosis and timely pre-treatment of AF before the development of any inalterable life-threatening conditions/diseases. Ultimately, the CVDs associated mortality could be reduced. In this manuscript, the design and implementation of a personalized healthcare system embodying a wearable ECG device, a mobile application, and a back-end server are presented. This system continuously monitors the users' ECG information to provide personalized health warnings and feedbacks. The users are also able to communicate with their paired health advisors through this system for remote diagnoses, interventions, etc. The implemented wearable ECG devices have been extensively evaluated and showed excellent intra-consistency ($CV_{RMS}\approx\SI{5.5}{\percent}$), acceptable inter-consistency ($CV_{RMS}\approx\SI{12.1}{\percent}$), and negligible RR-interval errors ($ARE<\SI{1.4}{\percent}$). To boost the battery life of the wearable devices, a lossy compression schema utilizing the quasi-periodic feature of ECG signals to achieve compression was proposed. Compared to the recognized schemata, this approach outperformed the others in terms of compression efficiency and distortion, and achieved at least $2\times$ of CR at a certain PRD or RMSE for the ECG signals from the MIT-BIH database. To enable the capability of automated AF diagnosis/screening in the proposed system, a deep residual network (ResNet)-based AF detector was developed. For the ECG records from the 2017 PhysioNet CinC challenge, this AF detector obtained an average testing $F_{1}=\SI{85.10}{\percent}$ and a best testing $F_{1}=\SI{87.31}{\percent}$, outperforming the state-of-the-art.}


\begin{document}
\section{Introduction}
\label{Sect:Introduction}

In the human sensing context, humans are targets, operators, and data sources of sensing \cite{srivastava2012human}. A particular example is the Internet of things (IoT)-based e-health system. Thanks to technology advances in the micro-electro-mechanical systems (MEMS), bio-related sensors are becoming more power- and cost-efficient while having reduced physical dimensions. Employing these low-cost and tiny sensors in wearable devices for data acquisition and utilizing the ubiquitous smart phones for data visualization and relay is the current trend in the IoT-based e-health systems. These systems enable continuous and remote monitoring of the users' vital information such as respiratory rate, blood pressure, glucose level, and electrocardiography (ECG) \cite{catarinucci2015an,islam2015the,istepanian2011the}.

Among these vital parameters, the ECG information is of primary interest because it is a useful, simple, risk-free, non-intrusive, and cost-efficient tool for the diagnosis of the atrial fibrillation (AF) arrhythmia \cite{dilaveris1998simple}. Despite good progress has been achieved in the management of patients with AF, this arrhythmia is yet prevalently associated with the increased risks of heart failure, sudden death, syncope, dementia, stroke, and cardiovascular morbidity worldwide \cite{fuster2006acc,fuster20112011,kirchhof2016esc,odutayo2016atrial}. Moreover, the number of people suffering from the AF is steeply rising and will reach up to \SIrange{14}{17}{} million in Europe by 2030 \cite{zoni2014epidemiology} and \SIrange{6}{12}{} million in the USA by 2050 \cite{chugh2014worldwide}. A large amount of healthcare expenditure is driven by the AF and the AF-related complications \cite{stewart2004cost,kim2011estimation}. Besides, there is growing evidence that AF has strong associations with numerous cardiovascular diseases (CVDs), including heart failure, coronary artery disease, valvular heart disease, and hypertension \cite{maisel2003atrial,yiu2008hypertension,emdin2016atrial,national2021cvd}. In accordance with the World Health Organization (WHO), the CVDs were still the number one cause of death globally in 2010 \cite{who2010global}. They also accounted for more than two fifths of all deaths in rural and urban areas of China in 2014 \cite{wei2017china}.

Therefore, it is of paramount importance to develop an IoT-based e-health system which monitors the users' ECG information remotely and continuously using the wearable devices, and provides personalized health warnings and feedbacks promptly, such that the users are able to keep track of their health conditions conveniently and take necessary precautions accordingly. Further diagnoses and interventions are available to the users by engaging the health advisors, such as coaches, physicians, and cardiologists, in the e-health system. To improve the labor- and time-efficiency of manual diagnoses and interventions, employing neural networks (NNs)-based automated diagnostic/screening tool in the system is the optimal approach, considering the NNs' great success in a wide range of cardiac classification tasks, including AF detection. Such a system unlocks the potentials of prompt pre-diagnosis and timely pre-treatment of AF before the development of any inalterable life-threatening conditions/diseases, e.g., the CVDs. Hence, it is considered to be an efficient and effective approach for reducing the CVDs associated mortality. The major contributions of this work are:
\begin{itemize}
	\item A personalized healthcare system embodying a wearable device to continuously and remotely monitor the user's ECG information, from which automated/manual health warnings and feedbacks are accessible;
	\item A simple yet effective lossy compression schema for ECG signals aiming at boosting the battery life of the wearable devices for long-term monitoring;
	\item A neural network-based AF detector capable of identifying the arrhythmia types of the short-period single-lead ECG signals with high sensitivity and accuracy.
\end{itemize}

In the remainder of this manuscript, the specific motivations and detailed descriptions of the proposed personalized healthcare system, lossy compression schema, and neural network-based AF detector are respectively presented in Section~\ref{SubSect:E-HealthSystem}, Section~\ref{SubSect:ECGLossy}, and Section~\ref{SubSect:AutomaticAFDetector}. Then, the hardware/software implementation of the comprehensive system, the performance evaluation of the lossy compression schema, as well as the classification performance of the AF detector are detailed in Section~\ref{SubSect:SystemImplementation}, Section~\ref{SubSect:LCSchemaEvaluation}, and Section~\ref{SubSect:PerformanceResNetAFDetector}, respectively; and the limitations and future research are also discussed. Lastly, the conclusion of this work is drawn in Section~\ref{Sect:Conclusion}.

\section{Materials and Methods}
\label{Sect:MaterialsAndMethods}

\subsection{IoT-Based E-Health System with Wearable ECG Device}
\label{SubSect:E-HealthSystem}


Although various vital parameters, namely blood pressure, respiratory rate, glucose level, and electroencephalography (EEG), can be monitored by the IoT-based e-health systems, the proposed system focuses more on the ECG information from which the heart rate variability (HRV) information of the users can be retrieved. The HRV measures the time interval variations of adjacent heartbeats (i.e., the RR-interval as illustrated in Figure~\ref{fig:ECGBeat}), and physiologically, it is influenced by the sympathetic and parasympathetic nervous system activities \cite{stein1994heart,appelhans2006heart,thayer2009claude}. Existing studies have reported that individuals suffering from diabetes, psychiatric disorders, and cardiac disorders manifested decreased HRV \cite{kleiger1987decreased,nolan1998prospective,schroeder2005diabetes,pop2010cardiac,henry2010heart}. Besides, substantial evidence showed that the decreased HRV is a predictor of mortality after myocardial infarction, i.e., heart attack \cite{kleiger1987decreased,bigger1992frequency}. The proposed system was originally designed for a pilot study exploring how HRV is related to the psychosocial stress and risks of stroke, CVDs, and cardiometabolic stress; the ultimate goal of the pilot study is to develop a stochastic model which predicts the risks to the aforementioned diseases using the subject's HRV information acquired by a wearable ECG monitoring device. It has been later extended into a more generalized system that provides personalized health warnings and feedbacks promptly by monitoring the users' ECG information remotely and continuously.

\begin{figure}[H]
	\centering
	\includegraphics[width=0.62\columnwidth]{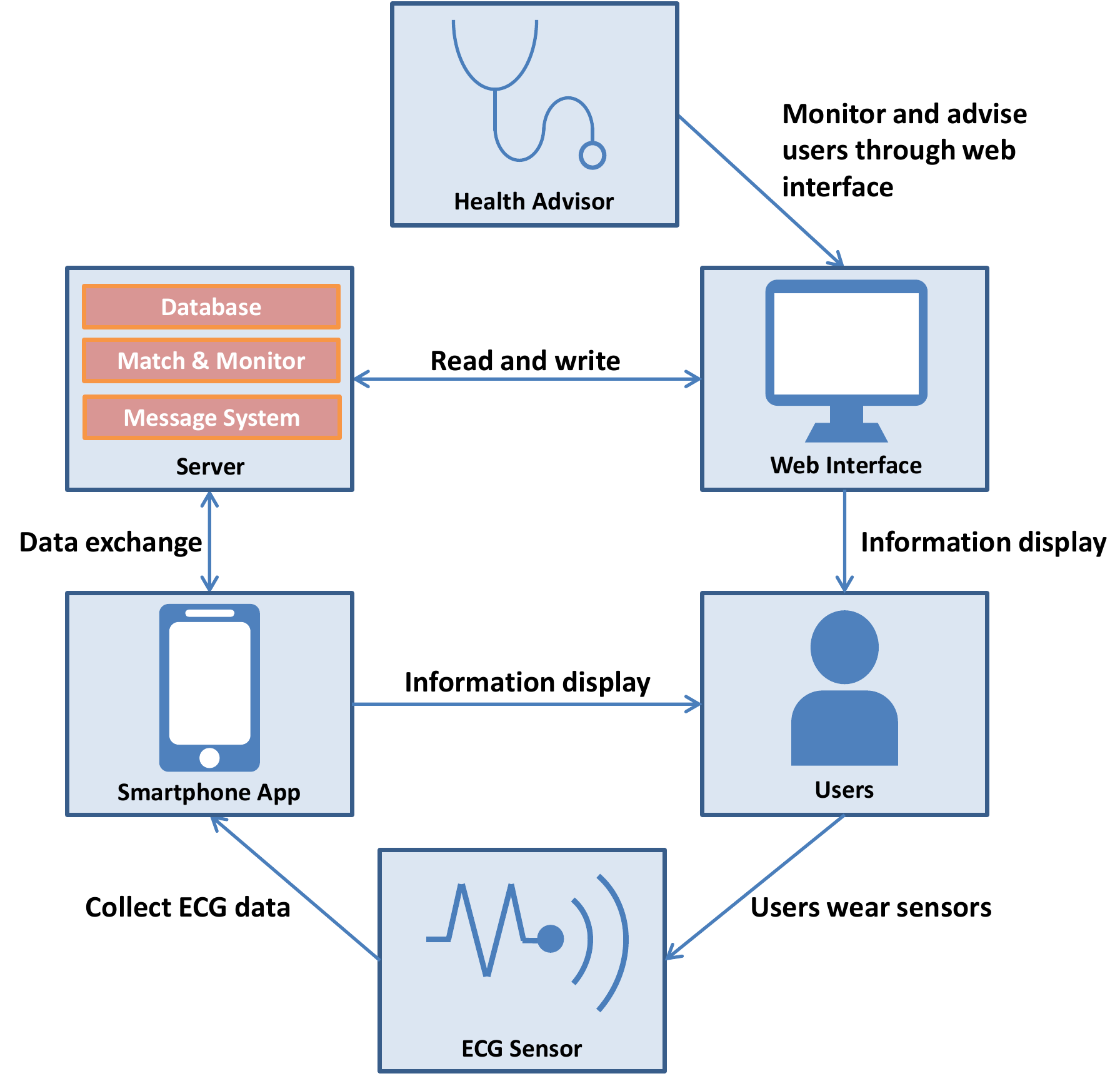}
	\caption{The architecture of the proposed personalized healthcare system.}
	\label{fig:EhealthSystemArchitecture}
\end{figure}

In this manuscript, extending the previous work of Tong et al. \cite{tong2017a}, a personalized healthcare system embodying a wearable ECG monitoring device for data acquisition, a mobile application for data visualization and relay, and a back-end server for data management and analytics was developed. The system architecture is illustrated in Figure~\ref{fig:EhealthSystemArchitecture}. The users wear the monitoring devices to collect their ECG signals. Then, the collected signals will be transmitted to their smart phones through Bluetooth low energy (BLE) for visualization. Therefore, the users are able to access their ECG information in real time. Besides displaying the ECG signals to the users, the smart phones serving as relays will upload the acquired signals to the back-end server through cellular network or Wi-Fi for further analyses. The received ECG signals will be stored in the back-end server's database. A matching service is available on the back-end server to pair the users with the health advisors such that they can communicate with each other using the build-in messenger service. The health advisors could be doctors or coaches, and they are able to monitor their clients' ECG information through the web interface. In addition to the ECG signals, the system also supports other health information such as blood pressure, heart rate, step count, and EEG from other commercial wearable devices.

This system provides a convenient way for the users to keep track of their health conditions via wearable monitoring devices and smart phones in real time. In addition, it enables the health advisors to monitor and analyze their clients' ECG information timely and remotely through a web interface. Promptly and cost-efficiently health caring is available for the users because they are continuously monitored by the health advisors and the back-end server. The implementation of the proposed personalized healthcare system, including the wearable ECG device, the mobile application, and the back-end server and web interface, is presented in Section~\ref{SubSect:SystemImplementation}. Developing such system in-house enables the authors with more controls over it, such as deploying dedicated techniques on the wearable device to enhance its power efficiency, employing advanced algorithms in the mobile application to achieve higher data quality, and applying specified analytic approaches in the back-end server to meet different data consumers' needs.

\subsection{Lossy Compression Schema for ECG Signal}
\label{SubSect:ECGLossy}


In a typical personalized e-health system, the vital parameters are first acquired by the wearable devices and then transmitted to the upper infrastructures (e.g., smart phones and back-end servers) through various wireless technologies (Bluetooth, Wi-Fi, etc.) for continuous monitoring and manual/automated diagnosis. A major issue of these wearable devices, which often have limited storage and computational capacity, is that they are usually battery-driven. Their energy consumption is widely concerned because frequently recharging the batteries poses significant limitations on continuous monitoring of the vital parameters, which is the cornerstone of a personalized e-health system. Although the energy consumption of a wearable device is influenced by various factors, including integrated chip design, embedded software design, and selection of wireless technologies, compressing the acquired signals before transmission provides an opportunity to improve the energy consumption in a high-level way \cite{ma2012assurance,elgendi2017efficient,zhou2017ultra}. Besides, the wireless transmission of vitals is generally one of the most power-demanding elements of a wearable device. In order to provide timely feedbacks to the users based on the acquired vital parameters, the transmission time lags between the wearable devices and the upper infrastructures should be minimized.

In this regard, it is of paramount necessity to design an on-line or even real-time compression schema for the wearable devices such that the compressed vital signals can be effectively cached (i.e., decreased data size) in the limited storage space and energy-efficiently transmitted (i.e., reduced transmission time) to the upper infrastructures. Considering the limited computational capacity and the power-demanding characteristic of the micro-controller employed in a wearable device, the compression schema should be relatively simple and lightweight as well. Compared with the lossless compression schemata, although the lossy ones introduce signal distortion, they are preferred in this work because much higher compression efficiency, which implies shorter transmission time and usually higher energy efficiency, can be achieved. The ultimate goal of the lossy compression schema is to boost the battery life of the wearable devices for personalized long-term monitoring while the reconstructed vitals have sufficient signal quality for the intended applications, e.g., detecting AF from short-period single-lead ECG signals.

The categorization of the existing lossy compression methods specifically for ECG signals is first presented in Section~\ref{SubSubSect:TaxonomyLossy}. Then, the operating principles and the deficiencies of three lossy compression schemata recognized by the authors are described in Section~\ref{SubSubSect:IssuesInExistingLossy}. An intuitive lossy compression schema, which utilizes the ECG signal's quasi-periodic feature to achieve compression, was proposed in Section~\ref{SubSubSect:IntuitiveLossy} to address the deficiencies of the recognized schemata; it eliminates the use of R-peak detector and requires much less memory and computing resources. Finally, a number of evaluation metrics used for quantifying the performance (in terms of compression efficiency and distortion) of a lossy compression schema are introduced in Section~\ref{SubSubSect:EvaluationMetricsLossy}; the performances of the proposed and the recognized schemata were investigated in Section~\ref{SubSect:LCSchemaEvaluation}.

\subsubsection{Taxonomy of Lossy Compression Schemata}
\label{SubSubSect:TaxonomyLossy}

While there exist quite a few lossy compression methods specifically designed for ECG signals in the literature, they can be typically classified into three categories, namely the time domain processing, the transform-based encoding, and the parametric methods \cite{fira2008an,hooshmand2017boosting}. A brief description of these three categories is presented in the following:
\begin{enumerate}
	\item Time Domain Processing: This category is also known as the direct method and it includes the entropy coding \cite{huffman1952a}, the amplitude zone time epoch coding (AZTEC) \cite{cox1968aztec}, the tunning point (TP) \cite{mueller1978arrhythmia}, the coordinate reduction time encoding system (CORTES) \cite{abenstein1982a}, the scan-along polygonal approximation (SAPA) \cite{ishijima1983scan}, and the lightweight temporal compression (LTC) \cite{schoellhammer2004light}. The basic rationale of these lossy compression methods is to represent the original ECG signals with a subset of extracted significant samples.

	\item Transform-based Encoding: The lossy compression methods in this category are based on the linear transformations such as Karhunen-Lo\`eve transform (KLT) \cite{ahmed1975electro}, fast Walsh transform (FWT) \cite{kuklinski1983fast}, fast Fourier transform (FFT) \cite{reddy1986ecg}, discrete wavelet transform (DWT) \cite{djohan1995ecg}, discrete cosine transform (DCT) \cite{lee1999ecg}, and principal component analysis (PCA) \cite{castells2007principal}. The original time series ECG signals are first converted into the transformed domains and then represented by a number of significant transformed coefficients. The coefficients should have been carefully chosen so that the compressed signals can be properly reconstructed with acceptable distortion.

	\item Parametric Methods: This category contains lossy compression methods that are based on pattern recognition \cite{hooshmand2017boosting,qian2020noble}, vector quantization \cite{cohen1990compression,cardenas1999mean,sun2005beat}, neural network \cite{iwata1990data,acharya2017automated}, compressed sensing \cite{polania2011compressed,mamaghanian2011compressed,dixon2012compressed,zhang2013compressed}, denoising autoencoder \cite{del2015light,dasan2021novel}, etc. The fundamental rationale behind these lossy compression methods is to extract additional knowledge from the time series ECG signals and use such knowledge to model the original signals' shape, pattern, or morphology.
\end{enumerate}
Most of the first category methods and those pattern recognition- or vector quantization-based ones compress the original signals by utilizing the ECG signal's quasi-periodic feature, such as the similarity between a subsequent of samples (i.e., ECG beats), the similarity between samples (i.e., adjacent samples), and both.

According to the work of Hooshmand et al. \cite{hooshmand2017boosting}, a pattern recognition-based lossy compression schema named on-line dictionary (OD) was proposed; it demonstrated excellent approximation capability and was able to achieve high compression efficiency at the cost of a reasonably small amount of computational power. For the ECG signals from the PhysioNet MIT-BIH arrhythmia database \cite{moody2001the,goldberger2000physiobank}, the performances of the OD schema were compared with that from most of the aforementioned compression methods, including the LTC, the PCA, different versions of the DCT and the DWT, the vector quantization-based method named gain-shape vector quantization (GSVQ) \cite{sun2005beat}, two compressed sensing-based methods named simultaneous orthogonal matching pursuit (SOMP) \cite{polania2011compressed} and block sparse Bayesian learning (BSBL) \cite{zhang2013compressed}, and a denoising autoencoder (AE)-based approach proposed by Del Testa et al. \cite{del2015light}. Excluding the AE, the OD generally outperformed the others in terms of reconstruction error and energy efficiency at high compression efficiency ($>\SI{25}{}$). In contrast, the other methods except the compressed sensing-based ones demonstrated better reconstruction errors than the OD at low compression efficient ($\leq\SI{25}{}$). Although the AE always surpassed the OD, the major drawback of the former is that it entails an off-line training process which requires a representative dataset; the autoencoders trained from such a dataset do not always guarantee good approximations for the new ECG signals. Besides, the authors believe that deploying the AE on the hardware constrained wearable devices is not feasible.

With comprehensive consideration, the OD \cite{hooshmand2017boosting} and the GSVQ \cite{sun2005beat} are considered as good practices of lossy compression methods specifically designed for the wearable ECG devices utilized in a personalized e-health system. The operating principles and the deficiencies of these schemata are detailed in Section~\ref{SubSubSect:IssuesInExistingLossy}. Serving as the performance baseline of the transform-based encoding category, a PCA-based \cite{rao1964use} lossy compression schema \cite{hooshmand2017boosting} was also included in Section~\ref{SubSubSect:IssuesInExistingLossy}.

\subsubsection{Operating Principles and Deficiencies of Recognized Schemata}
\label{SubSubSect:IssuesInExistingLossy}

The operating principles of the OD, the GSVQ, and the PCA-based lossy compression schemata are detailed in this section. In addition, the deficiencies of each schema, especially when being deployed on the wearable ECG devices with limited memory and computational capacity, are also discussed.

\begin{enumerate}[listparindent=2em]
	\item On-line Dictionary (OD)
		
		The OD lossy compression schema for ECG signals is based on the concepts of motif extraction \cite{danieletto2013} and pattern recognition. The block diagram of the OD's operating principle is illustrated in Figure~\ref{fig:ODBlockDiagram}. In the preprocessing chain, the ECG signal is first filtered by a third-order Butterworth band-pass filter \cite{arzeno2008analysis} with cut-off frequencies from \SIrange{8}{20}{\hertz} in order to eliminate the DC component and the high-frequency noise. With the filtered signal as input, the positions of the R-peaks in the original ECG signal are then identified by the R-peak detector proposed by Elgendi \cite{elgendi2013fast}\footnote{As recommended in the original work \cite{elgendi2013fast}, the sizes of the first and the second moving average windows are $w_1=\SI{97}{\milli\second}$ and $w_2=\SI{611}{\milli\second}$, respectively, whereas the user configurable parameter is $\beta=\SI{0.08}{}$.}, which is fast, lightweight, and suitable for energy constrained wearable devices. Once the positions of the R-peaks are detected, the ECG signal is divided into segments and the segment $\vec{z}_{t}$ at time $t$ contains the consecutive samples between two adjacent R-peaks. Afterwards, the segment $\vec{z}_{t}$ is period normalized using simple linear interpolation with fixed length\footnote{As recommended in the original work \cite{hooshmand2017boosting}, the length $W$ of a codeword and the maximum length of segment $\vec{z}_t$ are \SI{200}{} and \SI{512}{} samples, respectively, for the ECG signals from the MIT-BIH database.} $W$, and amplitude normalized by removing the gain $g_t$ and the offset $o_t$ calculated from the equations (5) to (7) provided in \cite{danieletto2013}. While the original segment length $l_t$, the gain $g_t$, and the offset $o_t$ are pending to be sent in place of the original segment $\vec{z}_{t}$, the period and amplitude normalized segment $\vec{x}_t$ serves as input to the pattern matching block.

		\begin{figure}[H]	
			\centering
			\includegraphics[width=0.63\columnwidth]{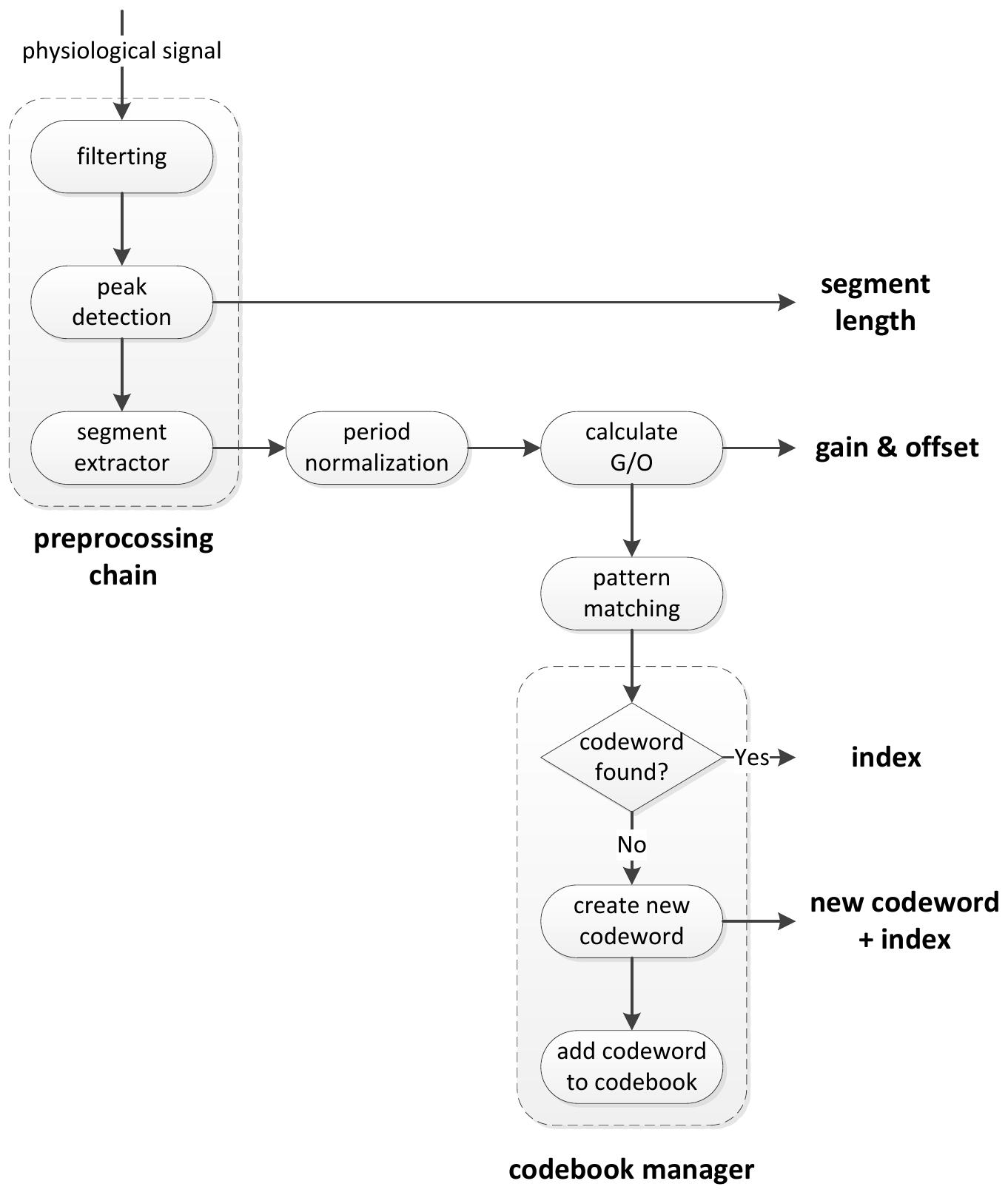}
			\caption{Block diagram illustrating the operating principle of the OD schema.}
			\label{fig:ODBlockDiagram}
		\end{figure}

		In the pattern matching process, the normalized segment $\vec{x}_t$ achieved at time $t$ is compared to the codewords $\vec{c}_i$ within the codebook $\mathbold{C_t}=\{\vec{c}_1,\cdots,\vec{c}_N\}$, which is built and maintained at runtime, to identify the closest codeword $\vec{c}_{i^*}$ at index $i^*$. The distance between the normalized segment $\vec{x}_t$ and the codeword $\vec{c}_i$ is calculated using the distance function\footnote{The $L^\infty$ norm is recommended for the distance function in the original work \cite{hooshmand2017boosting}.} $d(\vec{x}_t, \vec{c}_i)$. If the minimum distance $d(\vec{x}_t, \vec{c}_{i^*})$ is lower than a user configurable threshold $\epsilon$, the corresponding index $i^*$ of the codeword $\vec{c}_{i^*}$ together with the $l_t$, $g_t$, and $o_t$ are transmitted in place of the original ECG segment $\vec{z}_{t}$. Otherwise, the current normalized segment $\vec{x}_t$ is added into the codebook as a new codeword $\vec{c}_{N+1}$ resulting in an updated codebook $\mathbold{C_{t+1}}=\{\vec{c}_1,\cdots,\vec{c}_N, \vec{c}_{N+1}\}$; the new codeword $\vec{c}_{N+1}$ and its index together with the $l_t$, $g_t$, and $o_t$ are transmitted in place of the original ECG segment $\vec{z}_{t}$ such that the codebook on the receiver side remains synchronized all the time. To reconstruct the compressed ECG segment, the decompresser first retrieves the codeword from the synchronized codebook according to the received index and then renormalizes the codeword according to the received gain $g_t$, offset $o_t$, and original segment length $l_t$.

		Unlike the GSVQ lossy compression schema, which will be presented in the following paragraphs, no off-line trained dictionary/codebook is required in this schema because the codebook is generated and maintained at runtime. However, the size of the codebook is therefore dependent upon the user configurable threshold $\epsilon$ and the ECG signals. Different ECG signals with identical $\epsilon$ value may result in different sizes of codebooks, and thus different memory capacity requirements. Also, as time goes on, the size of the codebook shall increase (might become even larger than the allowed memory capacity). Consequently, the number of bits used to encode the codeword index is not guaranteed and a sufficiently large number is preferred, in which case suboptimal compression efficiency is anticipated. In addition to the indeterminable codebook size, the synchronization mechanism of the OD schema suggests that it is not suitable for applications measuring short periods of ECG signals occasionally, in which case the codebook is likely being synchronized frequently. Besides, the R-peak detector requires a sufficiently large set of consecutive samples (large enough for multiple R-peaks) to be temporally stored in the RAM that further limits the OD's applicability in wearable devices. Nevertheless, similar to the GSVQ schema, the key component of the OD schema is the R-peak detector for segmenting the ECG signal; the performances of the OD and the GSVQ are heavily influenced by the accuracy of the R-peak detector.

	\item Gain-Shape Vector Quantization (GSVQ)

		The GSVQ lossy compression schema for ECG signal utilizes the redundant information among adjacent heartbeats. The block diagram of the GSVQ's operating principle is illustrated in Figure~\ref{fig:GSVQBlockDiagram}. In the preprocessing chain, the ECG signal is first divided into segments using the dyadic wavelet transform (DyWT)-based R-peak detector proposed by Kadambe et al. \cite{kadambe1999wavelet}\footnote{As recommended in the original work \cite{kadambe1999wavelet}, the Hamming window length is $L_w=\SI{2.05}{\second}$ with $\SI{75}{\percent}$ overlapping between adjacent windows; the threshold values for identifying local maxima, determining aligned peaks, and distinguishing adjacent peaks are \SI{60}{\percent}, $\pm\SI{0.1}{\second}$, and \SI{0.2}{\second}, respectively. The cubic spline mother wavelet with \SI{120}{\hertz} center frequency and \SI{240}{\hertz} bandwidth is utilized and scales \SI{2}{}, \SI{4}{}, and \SI{8}{} are selected.}, which is robust to noise and time-varying QRS complexes. Then, the segment $\vec{z}_t$ at time $t$ is period normalized using linear interpolation\footnote{The original work used the method proposed by Wei et al. \cite{wei2001ecg} for period normalization. For simplicity's sake, the linear interpolation was adopted for period normalization when reimplementing the GSVQ schema.} and amplitude normalized by diving the gain $g_t$ (i.e., the $L^2$ norm of the period normalized segment). The normalized segment $\vec{x}_t$ has fixed length $K$ and unity amplitude $\lVert\vec{x}_{t}\rVert_{2}=1$. While the original segment length $l_t$ and the gain $g_t$ are pending to be sent in place of the original segment $\vec{z}_t$, the normalized segment $\vec{x}_t$ serves as input to the vector quantization block.

		\begin{figure}[H]	
			\centering
			\includegraphics[width=0.65\columnwidth]{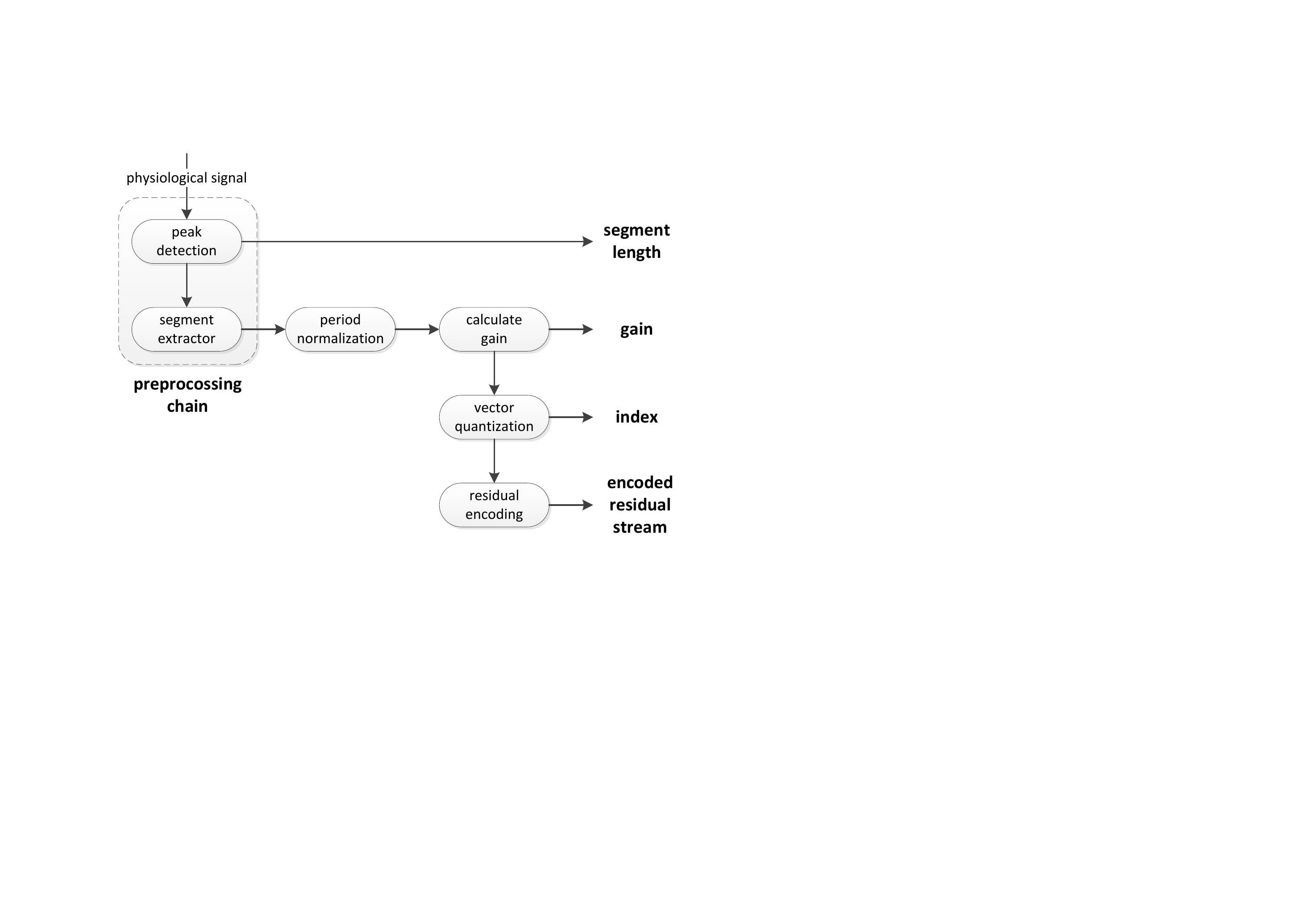}
			\caption{Block diagram illustrating the operating principle of the GSVQ schema.}
			\label{fig:GSVQBlockDiagram}
		\end{figure}

		In the vector quantization process, the normalized segment $\vec{x}_t$ achieved at time $t$ is compared with the codewords $\vec{c}_i$ inside a pre-built codebook $\mathbold{C}=\{\vec{c}_1,\cdots,\vec{c}_N\}$ to identify the closest\footnote{\label{Fn:GSVQDisFunc}The codebook is initiated and updated according to the work of Ramakrishnan et al. \cite{ramakrishnan1996ecg}, in which the distance between two codewords is the $L^2$ norm of their difference. Therefore, the $L^2$ norm of the difference between the normalized segment $\vec{x}_t$ and a codeword $\vec{c}_i$ is considered as distance.} codeword $\vec{c}_{i^*}$ at index $i^*$. The original segment length $l_t$, the gain $g_t$, and the index $i^*$ are transmitted in place of the original ECG segment $\vec{z}_{t}$. Besides, in the residual encoding process, the residuals between the original segment $\vec{z}_t$ and the renormalized closest codeword are encoded using the patented AREA algorithm proposed by Tai \cite{tai1997adaptive}, which is an adaptive sampling schema for one dimensional digital signals. The AREA algorithm requires a user configurable distortion threshold $A_{th}$ to identify a significant point\footnote{As recommended in the original work \cite{sun2005beat}, the buffer size of the AREA algorithm is set to \SI{16}{}.}, and therefore, it controls the compression efficiency and distortion of the GSVQ. In order to achieve better compression efficiency, the processed residuals are further encoded by employing the arithmetic encoding\footnote{This step was omitted in the GSVQ schema reimplemented by the authors because employing the lossless arithmetic encoding might conceal the true performance of the GSVQ lossy compression schema. Besides, other lossy compression schemata could as well apply such technique to achieve better compression efficiency.}, which is a lossless compression method. The encoded residuals are transmitted to the receiver side to improve the quality of the reconstructed ECG segment. To reconstruct the compressed ECG segment, the decompresser first retrieves the codeword from the pre-build codebook according to the received index $i^*$ and then renormalizes the codeword according to the received gain $g_t$ and original segment length $l_t$. The decoded residuals are finally added to the renormalized codeword to generate the reconstructed ECG segment.

		Note that a pre-built codebook $\mathbold{C}=\{\vec{c}_1,\cdots,\vec{c}_N\}$ with $N$ codewords of fixed length $K$ is needed in this schema\footnote{For the ECG signals from the MIT-BIH database, the codebook size $N=\SI{64}{}$ and the codeword size $K=\SI{256}{}$ are recommended in the original work \cite{sun2005beat}, whereas the the maximum length of segment $\vec{z}_t$ is \SI{512}{} samples.}. This codebook is constructed using the Linde-Buzo-Gray (LBG) algorithm \cite{linde1980an} through an off-line training process\textsuperscript{\ref{Fn:GSVQDisFunc}}. In order to ensure good approximation for the ECG signals, a representative training dataset is essential in the training process. Otherwise, the performances of the GSVQ schema are not guaranteed for new ECG signals. While a sufficiently large codebook might result in better compression efficiency and lower distortion, it is not appreciated by the wearable devices with limited memory capacity. Moreover, the $L^2$ norm distance function involving expensive operators is also not preferred by the wearable devices due to limited computational capacity. Similar to the OD, the GSVQ's performances are heavily influenced by the accuracy of the R-peak detector.

		\item Principal Component Analysis (PCA)

		While there exist quite a few PCA-based lossy compression schemata for ECG signals \cite{castells2007principal}, the one introduced in \cite{hooshmand2017boosting} was discussed in this section. Before applying PCA, the ECG signal first goes through the preprocessing chain as illustrated in Figure~\ref{fig:ODBlockDiagram}; the retrieved segments $\mathbold{Z}=\{\vec{z}_1,\cdots,\vec{z}_N\}$ are then period and amplitude normalized using the normalization method introduced in the OD schema\footnote{Preliminary results suggested that the PCA-based lossy compression schema utilizing the OD's preprocessing chain and normalization method generally demonstrated lower distortion under the same compression efficiency compared to that employing the GSVQ's.}. While the original segment lengths $\vec{l}=\{l_1,\cdots,l_N\}$, the gains $\vec{g}=\{g_1,\cdots,g_N\}$, and the offsets $\vec{o}=\{o_1,\cdots,o_N\}$ of the segments $\mathbold{Z}$ are pending to be transmitted in place of the original ECG signal, the normalized segments $\mathbold{X}=\{\vec{x}_1,\cdots,\vec{x}_N\}$, where $\vec{x}_i\in\mathbb{R}^{W}$, are fed into the PCA encoder\footnote{As recommended in the work of Hooshmand et al. \cite{hooshmand2017boosting}, the length of a normalized segment $\vec{x}_i$ is $W=200$ samples for the ECG signals from the MIT-BIH database.}.
		
		The PCA encoder generates a vector $\vec{\mu}\in\mathbb{R}^W$ containing the column-wise averages of the matrix $\mathbold{X}\in\mathbb{R}^{N \times W}$, a matrix $\mathbold{\Psi}\in\mathbb{R}^{W \times K}$ containing the eigenvectors corresponding to the $K$ largest eigenvalues, and a matrix $\mathbold{Y}\in\mathbb{R}^{N \times K}$ achieved by transforming $\mathbold{X}$ into the subspace $\mathbold{\Psi}$. The user configurable parameter $K$ ($K<W$) is the desired number of principle components that control the compression schema's efficiency and distortion. Due to page limitation, the mathematics behind PCA are omitted. Together with the vectors $\vec{l}$, $\vec{g}$, and $\vec{o}$, the vector $\vec{\mu}$ and matrices $\mathbold{\Psi}$ and $\mathbold{Y}$ are transmitted to the receiver side in place of the original ECG signal. Because it is difficult to perform PCA on a large matrix, a long ECG signal should be divided into small time series signals such that the number $N$ is within a reasonable range\footnote{Preliminary results suggested that the number of normalized segments $N=50$ is a good choice for the ECG signals from the MIT-BIH database.}. To reconstruct the compressed ECG signal, the decompresser first reconstructs the matrix $\hat{\mathbold{X}}\in\mathbb{R}^{N \times W}$, where $\hat{\mathbold{X}} = \mathbold{Y} \cdot \mathbold{\Psi}^T + \vec{\mu}$. Each row of matrix $\hat{\mathbold{X}}$, denoted by $\hat{\vec{x}}_i$ where $i=1,\cdots,N$, represents the normalized ECG segment with distortion. Finally, the reconstructed ECG signal is achieved by renormalizing the $\hat{\vec{x}}_i$ according to the received gain $g_i$, offset $o_i$, and original segment length $l_i$, and combining all the renormalized segments.

		As this lossy compression schema utilizes a R-peak detector to construct the matrix $\mathbold{X}\in\mathbb{R}^{N \times W}$ for the PCA encoder, its performances are heavily influenced by the accuracy of the R-peak detector. This schema requires significant memory space to store a fair number of ECG samples and multiple matrices in order to perform PCA, in which case the memory requirement might exceed the maximum capacity of the wearable devices and the real-time capability of this schema might be diminished. For example, with parameters $N=50$ and $W=200$, a matrix $\mathbold{X}$ already occupied around $\SI{20}{\kilo\byte}$ of RAM (assuming 16-bit per ECG sample), and at least \SI{50}{} heartbeats of time lags between the transceiver and the receiver were introduced. While a long-period ECG signal should be divided into short-period ones before performing compression by this schema, the preliminary results indicated that the ends (i.e., tail to head) of adjacent reconstructed short-period signals were generally misaligned. Therefore, further distortion was introduced.
\end{enumerate}

To sum up, all of the lossy compression schemata presented in this section utilize the R-peak detectors to segment the ECG signal. Their compression efficiency and distortion are heavily influenced by the accuracy of the R-peak detectors. While the OD schema's codebook has an indeterminable size which increases as time goes on, the codebook in the GSVQ schema is generated through an off-line training process which entails a representative dataset for optimal performances. Both of their codebooks and the matrices in the PCA schema require considerable memory space for storage, and might occupy space larger than the allowed memory capacity of a wearable device\footnote{\label{Fn:LCsMemory}The experiment in Section~\ref{SubSect:LCSchemaEvaluation} showed that the sizes of codebooks generated in the GSVQ and the OD schemata were around \SI{25}{\kilo\byte} and \SI{12.1}{\kilo\byte} (typical value), respectively. In the PCA-based schema, the matrix $\mathbold{X}$ only with $N=50$ and $W=200$ already occupied around \SI{20}{\kilo\byte} of memory.}. Besides, the use of expensive operators, such as the $L^2$ norm distance function and the calculation of covariance matrix, in these schemata may impair their real-time capability due to the limited computational capacity of a wearable device. Nonetheless, the synchronization mechanism of the OD's codebook suggests that it is not suitable for applications that will experience frequent synchronization; further distortion is introduced by the misaligned ends of the adjacent reconstructed short-period ECG signals from the PCA schema.

\subsubsection{Intuitive Lossy Compression (InLC) Schema}
\label{SubSubSect:IntuitiveLossy}

To address the deficiencies of the existing lossy compression schemata presented in Section~\ref{SubSubSect:IssuesInExistingLossy}, an intuitive lossy compression (InLC) schema, which is simple but effective, was proposed in this section. The InLC eliminates the use of R-peak detector to segment the ECG signal and only requires three \SI{2}{\kilo\byte} buffers to perform compression\footnote{The buffer size is empirically determined for the ECG signals from the MIT-BIH database while representing each ECG sample with 16 bits.}. It utilizes the quasi-periodic feature of the ECG signals to achieve compression and could be categorized as a hybrid of the Time Domain Processing and the Parametric Methods. The core of the InLC schema is a relative inexpensive distance function based on the $L^\infty$ norm. The results in Section~\ref{SubSect:LCSchemaEvaluation} indicated that the InLC outperformed all the lossy compression schemata detailed in Section~\ref{SubSubSect:IssuesInExistingLossy}, in terms of compression efficiency and distortion, for the long-period (around \SI{30}{\minute}) and short-period (first \SI{1}{\minute} samples of a record) ECG signals from the MIT-BIH database. The operating principle of the InLC schema is presented in the following paragraphs.

Given a bank $\vec{B}_b$ of size $s_b$ and a buffer $\vec{B}_f$ of size $s_f$ containing the consecutive ECG samples that have been previously transmitted to the receiver side and those that are currently acquired but have not yet been transmitted, respectively, the fundamental rationale behind the InLC is to identify the longest fragment $\vec{f}^*$ from the buffer $\vec{B}_f$ such that a similar fragment $\vec{b}^*$ from the bank $\vec{B}_b$ can be found to approximate the $\vec{f}^*$. The start index $i^*$ and length $l^*$ of fragment $\vec{b}^*$ together with the gain $g^*$ and offset $o^*$ between fragments $\vec{b}^*$ and $\vec{f}^*$ are then transmitted in place of the original fragment $\vec{f}^*$ to achieve lossy compression. Note that the fragments $\vec{f}^*$ and $\vec{b}^*$ have identical length, and the start index of any fragment $\vec{f}_{l}$ with length $l$ is always zero, i.e., at the beginning of buffer $\vec{B}_f$. The details of the InLC schema are described in the following:
\begin{enumerate}
	\item\label{step1}Define the bank $\vec{B}_{b}$ of size $s_{b}$ and the buffer $\vec{B}_{f}$ of size $s_{f}$ ($s_{f} \leq s_{b}$), the minimum length $l_{min}$ and maximum length $l_{max}$ of fragment $\vec{f}_{l}$ with length $l$ from $\vec{B}_{f}$ ($l_{min} \leq l \leq l_{max} \leq s_{f}$), and the threshold $\epsilon$ determining whether the fragment $\vec{f}_{l}$ from $\vec{B}_{f}$ and the fragment $\vec{b}_{l}^i$ with start index $i$ and length $l$ from $\vec{B}_{b}$ are similar or not;

	\item\label{step2}Push the currently acquired ECG samples into $\vec{B}_{f}$ until $|\vec{B}_{f}|=s_{f}$\footnote{The symbol $|\vec{B}_{f}|$ and $|\vec{B}_{b}|$ represent the number of samples inside the buffer $\vec{B}_{f}$ and the bank $\vec{B}_{b}$, respectively.};

	\item\label{step3}Consider the first $l_{min}$ samples of $\vec{B}_{f}$ as $f_{l_{cur}}$, where $l_{cur}=l_{min}$; if $|\vec{B}_{b}| \neq 0$, jump to Step~\ref{step5};

	\item\label{step4}Push the samples from $\vec{f}_{l_{cur}}$ into $\vec{B}_{b}$ if $|\vec{B}_{b}|<s_{b}$ and transmit these $l_{cur}$ samples to the receiver side directly; remove $\vec{f}_{l_{cur}}$ from $\vec{B}_{f}$ and jump back to Step~\ref{step2};

	\item\label{step5}Calculate the distance $d^{i}_{l_{cur}}$ between $\vec{f}_{l_{cur}}$ and every fragment $\vec{b}^{i}_{l_{cur}}$ in the $\vec{B}_{b}$, where $i=0,\cdots,(s_b-l_{cur})$;

	\item\label{step6}Identify the local minima from the calculated distances $\vec{d}_{l_{cur}}$; whenever $d^i_{l_{cur}}$ is a local minimum and $d^i_{l_{cur}} \leq \epsilon$, push the start index $i$ and the distance $d^i_{l_{cur}}$ into $\vec{B}_{id}$\footnote{For the calculated distances $\vec{d}_l$, the number of local minima will not exceed $(s_b-l)/2$. Therefore, a buffer $\vec{B}_{id}$ of size $s_b$ is sufficient for storing both the indexes and distance values.}; if $|\vec{B}_{id}|=0$, jump back to Step~\ref{step4} (no similar fragment $\vec{b}^{i}_{l_{cur}}$ can be found in $\vec{B}_b$);

	\item\label{step7}Calculate the length of next fragment $\vec{f}_{l_{nxt}}$ containing the first $l_{nxt}$ samples of $\vec{B}_{f}$; if $l_{nxt}>|\vec{B}_{b}|$ or no more $l_{nxt}$ can be found, jump to Step~\ref{step10} (longest fragment $\vec{f}^*$ with length $l_{cur}$ is found);

	\item\label{step8}Calculate the distance $d^{i}_{l_{nxt}}$ between $\vec{f}_{l_{nxt}}$ and every fragment $\vec{b}^{i}_{l_{nxt}}$ in the $\vec{B}_{b}$, where the starting index $i$ is from $\vec{B}_{id}$;

	\item\label{step9}Identify the local minima with value $d^{i}_{l_{nxt}}\leq\epsilon$ from the calculated distances $\vec{d}_{l_{nxt}}$; if no constrained local minimum is found, jump back to Step~\ref{step7} directly; otherwise, empty $\vec{B}_{id}$ and push the start index $i$ and the distance $d^{i}_{l_{nxt}}$ of every constrained local minimum into $\vec{B}_{id}$, assign $l_{nxt}$ to $l_{cur}$ and jump back to Step~\ref{step7} (longer fragment $\vec{f}_{l_{nxt}}$ is found);

	\item\label{step10}Push the samples from $\vec{f}_{l_{cur}}$ into $\vec{B}_{b}$ if $|\vec{B}_{b}|<s_b$; identify the minimum distance value $d^{i^*}_{l_{cur}}$ in $\vec{B}_{id}$; transmit the corresponding start index $i^*$, the length $l_{cur}$, and the ordinary least-squares (OLS) regression parameters (i.e., gain $g^*$ and offset $o^*$) between $\vec{f}_{l_{cur}}$ and $\vec{b}^{i^{*}}_{l_{cur}}$ in place of the original fragment $\vec{f}_{l_{cur}}$; remove $\vec{f}_{l_{cur}}$ from $\vec{B}_{f}$ and then jump back to Step~\ref{step2}.
\end{enumerate}
It is apparent that the user configurable threshold $\epsilon$ controls the InLC schema's compression efficiency and distortion when the $s_{b}$, $s_{f}$, $l_{min}$ and $l_{max}$ are settled. To reconstruct the compressed ECG fragment $\vec{f}^{*}$, the decompresser on the receiver side first determines whether or not the received data package is the ECG fragment $\vec{f}_{l_{min}}$ directly transmitted in Step~\ref{step4}. If it is so\footnote{A data package of ECG fragment $\vec{f}_{l_{min}}$ can be easily identified as its size is generally different from that of a data package containing the compressed information.}, the direct fragment $\vec{f}_{l_{min}}$ is considered as the reconstructed one with zero distortion, and it is pushed into the bank $\vec{B}'_{b}$ on the receiver side when $|\vec{B}'_{b}| < s_{b}$ to synchronize with the bank $\vec{B}_{b}$. Otherwise, the decompresser retrieves the fragment $\vec{b}^{*}$ from $\vec{B}'_{b}$ according to the received start index $i^{*}$ and length $l^{*}$, and then reconstructs the fragment $\vec{f}^{*}$ according to the received gain $g^{*}$ and offset $o^{*}$. The reconstructed fragment $\hat{\vec{f}}^{*}$ is then appended to the bank $\vec{B}'_{b}$ on the receiver side when $|\vec{B}'_{b}| < s_{b}$.

\begin{figure}[H]
	\centering
	\includegraphics[width=0.95\columnwidth]{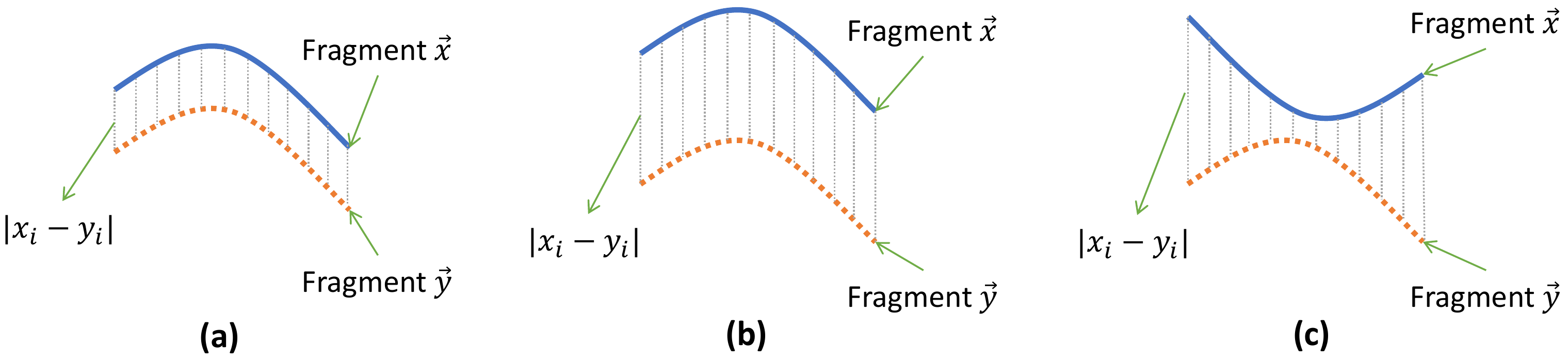}
	\caption{Illustration of the absolution differences between: (\textbf{a}) Fragments with identical shapes and a small offset. (\textbf{b}) Fragments with identical shapes and a substantial offset. (\textbf{c}) Horizontally mirrored fragments with moderate offset.}
	\label{fig:IllustrationDistFunct}
\end{figure}

Considering the memory and computational limitations of a wearable device, a relatively inexpensive distance function was proposed to quantify the similarity between fragments $\vec{x}$ and $\vec{y}$ ($|\vec{x}|=|\vec{y}|=n$):
\begin{equation}
\label{eq:OldDist}
d\left(\vec{x}, \vec{y}\right)=\rVert\vec{x}-\vec{y}\rVert_{\infty} - \overline{|\vec{x}-\vec{y}|}
\end{equation}
where $\rVert\vec{x}-\vec{y}\rVert_{\infty}$ is the $L^{\infty}$ norm, i.e., the maximum absolute difference between fragments $\vec{x}$ and $\vec{y}$. The rationale of Equation~\ref{eq:OldDist} is illustrated in Figure~\ref{fig:IllustrationDistFunct}. When the fragments $\vec{x}$ and $\vec{y}$ have similar shapes, the variance of the absolute differences $\{|x_i-y_i| \mid i=0,\cdots,n\}$ should be insignificant or negligible regardless of the offset between them, and vise versa. Instead of calculating the expensive variance value, the proposed distance function calculates the difference between the maximum and the average values of set $\{|x_i-y_i| \mid i=0,\cdots,n\}$ using multiple simple operators (subtraction and comparison) and a single division operator. Compared to the similarity measures that are frequently utilized in the literature, such as cross correlation (CC), root-mean-square error (RMSE), and $L^2$ norm, the proposed one is much less computationally costly. Besides, in contrast to the latter two, the Equation~\ref{eq:OldDist} is not dependent on the offset between fragments $\vec{x}$ and $\vec{y}$ (see Figure~\ref{fig:IllustrationDistFunct}a and Figure~\ref{fig:IllustrationDistFunct}b); such characteristic is preferred for similarity quantification. Note that a near-zero distance value calculated from Equation~\ref{eq:OldDist} does not guarantee the negligible variance of set $\{|x_i-y_i| \mid i=0,\cdots,n\}$, i.e., the former is the necessary but insufficient condition of the latter. Consequently, the short-period curvilinear waveforms in the original ECG signal are likely being replaced by a few straight lines (samples within the TP-intervals) in the reconstructed one as discussed in Section~\ref{SubSubSect:AveragePerformanceLCs}. To alleviate such issue, the Equation~\ref{eq:OldDist} is further constrained as follows:
\begin{equation}
\label{eq:NewDist}
	d\left(\vec{x}, \vec{y}\right)=
	\begin{cases}
		\infty,\quad\text{if}~\sum_{i=1}^{n} \left|x_i-y_i\right| \neq \left|\sum_{i=1}^{n}(x_i-y_i)\right|\\
		\max \left\{ \rVert\vec{x}-\vec{y}\rVert_{\infty}-\overline{|\vec{x}-\vec{y}|}, \overline{|\vec{x}-\vec{y}|}-\rVert\vec{x}-\vec{y}\rVert_{-\infty} \right\},\quad\text{otherwise}
	\end{cases}
\end{equation}
where $\rVert\vec{x}-\vec{y}\rVert_{-\infty}$ is the $L^{-\infty}$ norm, that is, the minimum absolute difference between fragments $\vec{x}$ and $\vec{y}$. The rationale of Equation~\ref{eq:NewDist} is also discussed in Section~\ref{SubSubSect:AveragePerformanceLCs}.

\begin{figure}[t]
	\widefigure
	\centering
	\includegraphics[width=0.98\columnwidth]{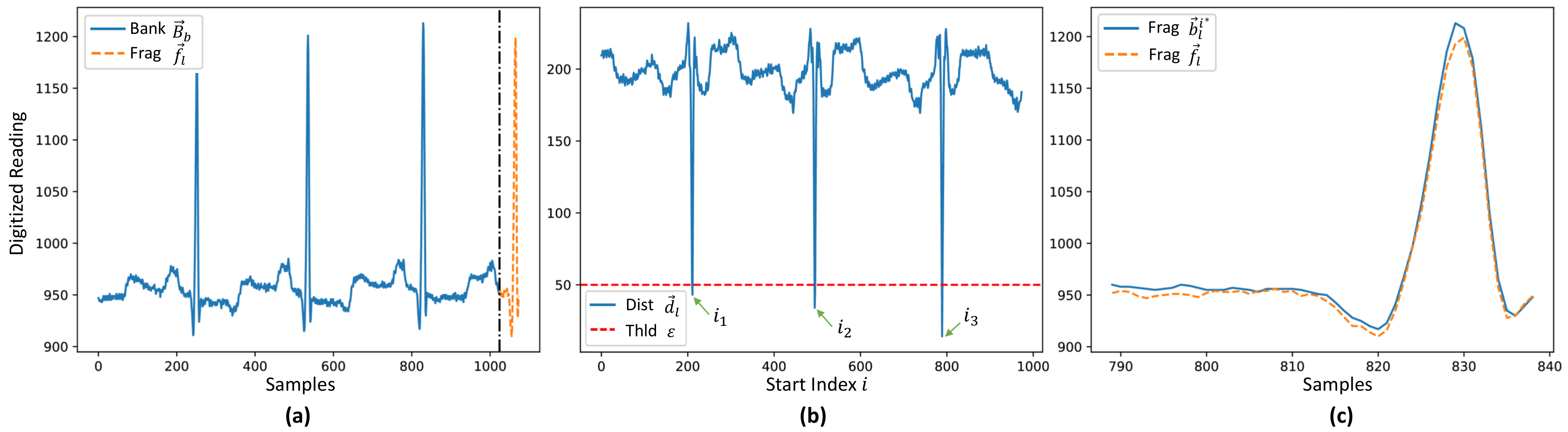}
	\caption{Illustration of the InLC schema's key steps: (\textbf{a}) ECG samples in the bank $\vec{B}_{b}$ (blue solid line) and those in the current fragment $\vec{f}_{l_{cur}}$ (orange dashed line) from $\vec{B}_{f}$ where $|\vec{B}_{b}|=s_{b}=s_{f}=1024$ and $l_{cur}=l_{min}=50$. (\textbf{b}) Distances between the fragment $\vec{f}_{l_{cur}}$ and every fragment $\vec{b}^{i}_{l_{cur}}$ from $\vec{B}_{b}$, where threshold $\epsilon=50.0$ (red dashed line). (\textbf{c}) The longest fragment $\vec{f}_{l_{cur}}$ (orange dashed line) in $\vec{B}_{f}$ having a similar fragment $\vec{b}^{i^{*}}_{l_{cur}}$ (blue solid line) been found in $\vec{B}_{b}$ with minimum distance $d^{i^{*}}_{l_{cur}} < \epsilon$.}
	\label{fig:IllustrationInLC}
\end{figure}

In order to further boost the computational efficiency of the InLC schema, the binary search approach\footnote{Initially, the current minimum and maximum lengths of fragment $\vec{f}_{l}$, where $l=l_{min}$, are $l_{cur\_min}=l_{min}$ and $l_{cur\_max}=l_{max}$, respectively. The first $l_{nxt}$ always equals to the integer part of $(l_{min}+l_{max})/2$. If no local minima with distance $\leq\epsilon$ could be identified in Step~\ref{step9}, the $l_{nxt}$ is assigned to the $l_{cur\_max}$; otherwise, the $l_{nxt}$ is assigned to the $l_{cur\_min}$. Then the $l_{nxt}$ in the next iteration equals to the integer part of $(l_{cur\_min}+l_{cur\_max})/2$; no more $l_{nxt}$ can be found when $|l_{cur\_max}-l_{cur\_min}| \leq 1$.}\textsuperscript{,}\footnote{Although the Equation~\ref{eq:OldDist} and Equation~\ref{eq:NewDist} are mathematically non-monotonic, the preliminary results in Section~\ref{SubApped:MonotonicityInLCDistFunct} suggested that they can be empirically approximated as functions monotonically increasing along with the number of samples $n$ in fragment $\vec{x}$ or $\vec{y}$.} is employed in Step~\ref{step7} to determine the length $l_{nxt}$. In addition, as described in Step~\ref{step8}, only the distances between the fragment $\vec{f}_{l_{nxt}}$ and the fragment(s) $\vec{b}^{i}_{l_{nxt}}$ with start index $i$ from $\vec{B}_{id}$ will be calculated. The rationale behind is that the calculated distances between these pairs of fragments are intuitively and empirically sub-minimal, if not minimal. Besides, reducing the sizes of the bank $\vec{B}_{b}$ and buffer $\vec{B}_{f}$ generally results in fewer iterations (higher computational efficiency) at the cost of lower compression efficiency (shorter $\vec{f_{l}}$ or lower chance of finding the $\vec{b}^{i}_{l}$ in $\vec{B}_{b}$); preliminary experiments suggested that a good trade-off can be achieved by setting $s_{b}=s_{f}=l_{max}=1024$ and $l_{min}=10$ for the ECG signals from the MIT-BIH database. The key steps of the InLC schema are illustrated in Figure~\ref{fig:IllustrationInLC}. Notice that, as depicted in Figure~\ref{fig:IllustrationInLC}b, only the distances between the next fragment $\vec{f}_{l_{nxt}}$ and the fragments $\vec{b}^{i}_{l_{nxt}}$ with start indexes in $\{i_1, i_2, i_3\}$, instead of $\{0,\cdots,s_{b}-l_{nxt}\}$, will be calculated in Step~\ref{step8}, in which case the computational efficiency is significantly improved.

\end{paracol}
\begin{paracol}{2}
\switchcolumn

In summary, the InLC schema eliminates the use of R-peak detector and employs a relatively inexpensive distance function to improve its applicability in those energy, memory, and computational power constrained wearable devices dedicated for personalized long-term ECG monitoring, while maintaining good compression efficiency and low distortion. Compared to the OD, the GSVQ, and the PCA lossy compression schemata recognized by the authors, the proposed InLC requires much smaller memory capacity\textsuperscript{\ref{Fn:LCsMemory},}\footnote{The experiment in Section~\ref{SubSect:LCSchemaEvaluation} showed that the proposed InLC with $s_{b}=s_{f}=l_{max}=1024$ and $l_{min}=10$ required only three \SI{2}{\kilo\byte} buffers (the resolution of an ECG sample is 16-bit) to perform compression.}. Multiple techniques have been adopted to further improve its computational efficiency and real-time capability. Utilizing the metrics provided in Section~\ref{SubSubSect:EvaluationMetricsLossy}, the performance of the InLC was compared with that of the OD, the GSVQ, and the PCA lossy compression schemata, and the results are presented in Section~\ref{SubSubSect:AveragePerformanceLCs}. One may notice that the bank $\vec{B}_{b}$ remains constant once it is full (refer to Step~\ref{step4} and Step~\ref{step10}), and therefore, it always contains the first $s_{b}$ samples of an ECG record. Intuitively, it would become more and more difficult to find a similar fragment $\vec{b}^{i}_{l}$ in $\vec{B}_{b}$ for the fragment $\vec{f}_{l}$ with large length $l$ as time goes on (and vise versa), in which case the compression efficiency is anticipated to be significantly reduced. The performances of the InLC with different setups, such as updating the bank $\vec{B}_{b}$ constantly and eliminating the gain $g^{*}$ in the compressed information, were investigated in Section~\ref{SubSubSect:AveragePerformanceInLC}.

\subsubsection{Evaluation Metrics}
\label{SubSubSect:EvaluationMetricsLossy}

The efficiency of a compression schema is generally quantified using the compression ratio (CR) defined as follows:
\begin{equation}
\label{eq:CR}
CR = \frac{\#bits~of~original~signal}{\#bits~of~compressed~signal}
\end{equation}
where the numerator and the denominator represent the number of bits of the original signal and the number of bits of the compressed signal, respectively. By definition, the lossy compression methods use inexact approximations and partial data to represent the original signal with smaller data size; the alteration of the reconstructed signal with respect to the original one is inevitable. Such alteration is qualified by the term distortion, and various metrics have been widely employed in the literature to quantify it \cite{lee1999ecg,lu2000wavelet,wei2001ecg,al2006quality,chou2006effective,rodrigues2008ecg,fira2008an,lee2011real}. The definitions of the distortion measures recognized by the authors are listed in the following:

\begin{itemize}
	\item Percentage Root-Mean-Square Difference (PRD): PRD is frequently adopted in the existing studies developing lossy compression methods specifically for ECG signals. The definition of the PRD is expressed as:
	\begin{equation}
		\label{eq:PRD}
		PRD = 100 \cdot \sqrt{\frac{\sum_{n=1}^{N}{(x(n)-\hat{x}(n))^2}}{\sum_{n=1}^{N}{x^2(n)}}}
	\end{equation}
	where $x(n)$ is the original signal, $\hat{x}(n)$ is the reconstructed signal, and $N$ is the number of samples from which the PRD is calculated. Because the value of the calculated PRD is dependent on the original signal's mean magnitude, a normalized version, named PRDN, has been introduced in the literature and defined as follows:
	\begin{equation}
		\label{eq:PRDN}
		PRDN = 100 \cdot \sqrt{\frac{\sum_{n=1}^{N}{(x(n)-\hat{x}(n))^2}}{\sum_{n=1}^{N}{(x(n)-\bar{x})^2}}}
	\end{equation}
	where $\bar{x}$ is the mean magnitude of the original signal $x(n)$.

	\item Root-Mean-Square Error (RMSE): In addition to the PRD and PRDN, the evaluation metric RMSE is as well widely employed in the existing studies to quantify the distortion introduced by the lossy compression methods. It is formulated as:
	\begin{equation}
		\label{eq:RMSE}
		RMSE = \sqrt{\frac{\sum_{n=1}^{N}{(x(n)-\hat{x}(n))^2}}{N}}.
	\end{equation}
	In order to immediately estimate the error against the original signal's amplitude, a variant of the RMSE was introduced by Hooshmand et al. \cite{hooshmand2017boosting}. For the sake of convenience, this variant is named RMSEP in this manuscript and its formula is expressed as:
	\begin{equation}
		\label{eq:RMSEP}
		RMSEP = \frac{100}{p2p} \cdot \sqrt{\frac{\sum_{n=1}^{N}{(x(n)-\hat{x}(n))^2}}{N}}
	\end{equation}
	where $p2p$ represents the average peak-to-peak magnitude of the original signal.

	\item Signal-to-Noise Ratio (SNR): According to Mueller, the SNR is yet another metric that could be used to evaluate the distortion between the original signal and the reconstructed one \cite{mueller1978arrhythmia}. It is defined as the ratio of the sum of the squared original signal that has been centered (i.e., signal) to the sum of squared differences between the original signal and the reconstructed one (i.e., noise). Because the ratios may have a very wide dynamic range, especially when the denominator is extremely small, the SNR is often expressed using the logarithmic decibel scale as follows:
	\begin{equation}
		\label{eq:SNR}
		SNR = 10 \cdot \log_{10}{\left(\frac{\sum_{n=1}^{N}{(x(n)-\bar{x})^2}}{\sum_{n=1}^{N}{(x(n)-\hat{x}(n))^2}}\right)}.
	\end{equation}
	It is apparent that the SNR can be expressed in terms of PRDN:
	\begin{equation}
		\label{eq:SNRtoPRDN}
		\begin{aligned}
			PRDN & = 10^{\frac{-SNR}{20}} \cdot 100,~\mbox{or} \\
			SNR & = 40 - 20 \cdot \log_{10}{(PRDN)}.
		\end{aligned}
	\end{equation}

	\item Maximum Amplitude Error (MAE): The MAE, which represents the maximum absolute difference between the original signal and the reconstructed one, is often used as a distortion measure in the literature for lossy compression methods. Its formula is expressed as:
	\begin{equation}
		\label{eq:MAE}
		MAE = \max_{n}\left \{ \left | x(n)-\hat{x}(n) \right | \right \}.
	\end{equation}

	\item Cross Correlation (CC): The CC quantifies the distortion introduced by the lossy compression methods by measuring the similarity between the original signal and the reconstructed one. The magnitude of CC ranges from \SIrange{-1}{1}{}. The closer the magnitude is to \SI{1}{}, the more similar the two signals are, and vise versa. The CC is expressed as below:
	\begin{equation}
		\label{eq:CC}
		CC = \frac{\sum_{n=1}^{N}(x(n)-\bar{x})(\hat{x}(n)-\bar{\hat{x}})}{\sqrt{\sum_{n=1}^{N}(x(n)-\bar{x})^{2}}\sqrt{\sum_{n=1}^{N}(\hat{x}(n)-\bar{\hat{x}})^{2}}}
	\end{equation}
	where $\bar{\hat{x}}$ is the mean magnitude of the reconstructed signal.
\end{itemize}

In addition to the distortion measures introduced above, there exist other feature-based evaluation metrics in the literature, such as the clinical distortion index (CDI) based on the clinical features retrieved from the original and the reconstructed ECG signals \cite{olmos1999a,alesanco2009clinical}, the weighted diagnosis distortion (WDD) determined by comparing the PQRST complexes between the original and the reconstructed ECG signals \cite{zigel2000the}, and the diagnostic distortion measure (DDM) coupling the CDI and the WDD with improved performances for self-guided unsupervised compression methods in real-time applications \cite{santos2017feature}. Besides, a metric named quality score (QS), which has taken both the compression efficiency and the reconstruction distortion into consideration, was proposed by Fira et al. \cite{fira2008an}.

In this manuscript, the efficiency of a compression method is quantified utilizing the CR defined in Equation~\ref{eq:CR}. As for the distortion measures, the PRD defined in Equation~\ref{eq:PRD} and the RMSE defined in Equation~\ref{eq:RMSE} are empirically selected to evaluate the alteration of the reconstructed signal with respect to the original one. For the ECG signals from the MIT-BIH database, the evaluation results of the InLC, the OD, the GSVQ, and the PCA-based lossy compression schemata are presented in Section~\ref{SubSect:LCSchemaEvaluation}. Note that, for vital parameters such as the ECG signals, the CR should be maximized under the premise that the distortion of the reconstructed signals will not alter the diagnostic results.

\subsection{Automatic AF Detection using Deep Residual Network (ResNet)}
\label{SubSect:AutomaticAFDetector}


The authors anticipate that an automated and effective AF detector, which is able to classify different arrhythmias from the short-period single-lead ECG signals with high sensitivity and accuracy, could relax the current healthcare burden because it unlocks the potentials of prompt pre-diagnosis and timely pre-treatment of AF before the development of any inalterable life-threatening conditions/diseases. Deploying such AF detector to the personalized healthcare system proposed in this work (detailed in Section~\ref{SubSect:SystemImplementation}) will enable continuous monitoring of users' ECG rhythms and provide timely warnings to the users and health advisors. Serving as a screening tool, the automated AF detector allows the informed health advisors (e.g., cardiologists) to establish further diagnosis and intervention in a more labor- and time-efficient way. 

However, existing AF detectors tend to demonstrate substantial rates of misclassification \cite{poon2005diagnostic,shah2007errors,schlapfer2017computer}, and possess deficiencies such as binarily classifying AF and normal rhythms only and employing cherry-picked dataset or small dataset with limited number of patients for evaluation \cite{lee2012atrial,petrenas2015low,xia2018detecting,andersen2019deep}. The applicability and the generalizability of the AF detector in a real-life personalized e-health system may be limited. Although there exist quite a few studies in the literature that classified the heart arrhythmias using dedicated algorithms \cite{larburu2011comparative,petrenas2015low,datta2017identifying} or machine learning techniques \cite{guler2005ecg,lei2007afc,zabihi2017detection,mahajan2017cardiac}, those utilizing the deep neural network (DNNs) are more recognized by the authors with the great success of the DNN in a wide range of classification tasks \cite{liu2017survey,ball2017comprehensive,nweke2018deep,labati2019deep}. To the best of the authors' knowledge, the convolutional neural networks (CNNs), including the native or deep CNNs \cite{xia2018detecting,hannun2019cardiologist}, the convolutional recurrent neural networks (CRNNs) \cite{zihlmann2017convolutional,andersen2019deep}, and the CNNs with long-short term memory (LSTM) structures \cite{teijeiro2017arrhythmia,hong2017encase}, were frequently adopted in the existing studies for classifying various heart arrhythmias, especially for AF detection. It is well established that the classification performance of a neural network is heavily dependent on the network depth, and numerous recognition tasks have substantially benefited from a very deep network \cite{simonyan2014very,szegedy2015going,long2015fully,he2016deep}. However, the CNNs generally suffer from a degradation problem, i.e., the saturated accuracy of the network degrades greatly when network depth increases; this problem is not ascribed to overfitting but the training errors introduced by adding more layers to a sufficiently deep network \cite{he2015convolutional,srivastava2015highway,he2016deep}.

In order to address the aforementioned issues, a deep residual network (ResNet)-based \cite{he2016deep} heart arrhythmia classifier specifically designed for AF detection was proposed in this work. A network structure with \SI{52}{} layers was adopted in this AF detector, and it was trained and tested utilizing the dataset from the 2017 PhysioNet CinC challenge (hereinafter referred to as the CinC challenge or the challenge) \cite{goldberger2000physiobank,clifford2017af,physionet2021the}, which contains \SI{8528}{} records of short-period (from \SIrange{9}{61}{\second}) single-lead ECG signals labeled into four classes. In the remainder of this section, the dataset and the evaluation metrics are first introduced in Section~\ref{SubSubSect:AFDatasetAndMetrics}. Afterwards, the preprocessing techniques applied to convert a 1D time series ECG signal into a 2D image/spectrogram for the ResNet are presented in Section~\ref{SubSubSect:AFPreprocessing}. Finally, the architecture of the ResNet-based AF detector is detailed in Section~\ref{SubSubSect:AFResNet}; its classification performance was investigated in Section~\ref{SubSect:PerformanceResNetAFDetector}.

\subsubsection{Dataset and Evaluation Metrics}
\label{SubSubSect:AFDatasetAndMetrics}

The CinC challenge dataset is consisted of \SI{12186}{} records of short-period single-lead ECG signals. These ECG records has been labeled as normal rhythm, AF rhythm, others rhythm, or noise\footnote{The dataset with labels in version 2 \cite{clifford2017af} was utilized when developing the ResNet-based AF detector. This set of labels were as well used by the participants in the challenge.}, and were divided into two groups: a public dataset containing \SI{8528}{} records and a hidden dataset containing the remaining ones. Note that the hidden dataset is not accessible. Therefore, the training and testing of the proposed AF detector were performed using the public dataset only. Each record in the public dataset contains \SIrange{9}{61}{\second} of 16-bit ECG measurements collected by the AliveCor Kardia\textsuperscript{TM} Band \cite{alivecor2021peace} sampling at \SI{300}{\hertz}. The frequency band and the amplitude of the ECG signals are from \SIrange{0.5}{40}{\hertz} and $\pm\SI{5}{\milli\volt}$, respectively. The classification profile of the records in the public dataset is listed in Table~\ref{Tab:AFDatasetProfile}. It is apparent that \SI{70}{\percent} of the records are \SI{30}{\second} long and the classification profile of them is similar to that of the entire public dataset. To simplify the preprocessing chain, only the \SI{30}{\second} records were utilized in this work, and it is assumed that the representative of the results will not be influenced.

\begin{specialtable}[H]
\caption{Classification profile of the public dataset from the CinC challenge.}
\label{Tab:AFDatasetProfile}
\small
\begin{tabular}{ccc}
\toprule 
\textbf{Classification Type} & \textbf{Number of Records} & \textbf{Number of \SI{30}{\second} Records} \\
\midrule
Normal 	& \SI{5050}{} (\SI{59.2}{\percent}) & \SI{3678}{} (\SI{61.5}{\percent}) \\
\midrule
AF 		& \SI{738}{} (\SI{8.7}{\percent}) 	& \SI{499}{} (\SI{8.4}{\percent}) 	\\
\midrule
Others 	& \SI{2456}{} (\SI{28.8}{\percent}) & \SI{1675}{} (\SI{28.0}{\percent}) \\
\midrule
Noise 	& \SI{284}{} (\SI{3.3}{\percent}) 	& \SI{125}{} (\SI{2.1}{\percent}) 	\\
\bottomrule
\end{tabular}
\end{specialtable}

The $F_{1}$ measure introduced in the CinC challenge \cite{clifford2017af} was adopted for evaluating the AF detector's classification performance. It is the average $F_{1}$ value of the classification types. For a particular classification type $c$, the $F_{1}$ value is defined as:
\begin{equation}
\label{eq:F1Value}
F_{1}(c) = \frac{2 \cdot TP(c)}{2 \cdot TP(c) + FP(c) + FN (c)}
\end{equation}
where $TP(c)$ is the number of inputs with label $c$ classified as $c$, $FP(c)$ is the number of inputs with label not $c$ classified as $c$, and $FN(c)$ is the number of inputs with label $c$ classified as not $c$\footnote{In addition to the $TP(c)$, $FP(c)$, and $FN(c)$ achieved from a confusion matrix, the $TN(c)$ represents the number of inputs with label not $c$ classified as not $c$.}. Then, the $F_{1}$ measure calculated from the four classification types is represented as:
\begin{equation}
\label{eq:F1Measure}
F_{1} = \frac{F_{1}(N) + F_{1}(A) + F_{1}(O) + F_{1}(\sim)}{4}
\end{equation}
where $F_{1}(N)$, $F_{1}(A)$, $F_{1}(O)$, and $F_{1}(\sim)$ are the $F_{1}$ values of normal class, AF class, others class, and noise class, respectively. In accordance with the challenge, the $F_{1}(\sim)$ value was omitted considering that there are only a small portion ($\approx\SI{3.3}{\percent}$) of noise records in the public dataset. The equation for the final $F_{1}$ measure is expressed as:
\begin{equation}
\label{eq:FinalF1Measure}
F_{1} = \frac{F_{1}(N) + F_{1}(A) + F_{1}(O)}{3}.
\end{equation}
Besides, the top-1 accuracy, denoted by $A_{1}$, was also adopted for evaluating the classification performance. It is defined as:
\begin{equation}
\label{eq:Top1Accuracy}
A_{1} = \frac{No.~of~correct~predictions}{Total~No.~of~predictions}.
\end{equation}
Note that a prediction from the AF detector is considered as correct if the predicted label with the highest probability matches the exact label.

\subsubsection{Data Preprocessing}
\label{SubSubSect:AFPreprocessing}

As mentioned in the previous section, only the \SI{30}{\second} records, i.e., records with \SI{9000}{} samples, were retained in this work. Since the ResNet was originally designed for 2D image recognition \cite{he2016deep}, the time series ECG signals are needed to be converted into 2D images at first. Inspired by the work of Zihlmann et al. \cite{zihlmann2017convolutional}, the time series ECG signals were transformed into one-sided spectrograms utilizing the short-time Fourier transform (STFT). Given that the spectrogram generated from the STFT is heavily influenced by the window function employed, a Tukey window with width $W$ and overlapping $O$ was empirically selected. In accordance with the experimental results detailed in Section~\ref{SubApped:WidthOverlappingOfTukey}, employing the Tukey window with $W=60$ and $O=55$ samples (i.e., corresponding to \SI{200}{\milli\second} width at \SI{300}{\hertz} sampling rate and \SI{92}{\percent} overlapping percentage) in the AF detector's preprocessing phase produced satisfactory saturated top-1 accuracy and $F_{1}$ measure with a fast convergence rate. The experimental results also suggested that the saturated top-1 accuracy and $F_{1}$ measure were primarily determined by the overlapping parameter $O$, and eliminating the components with frequencies higher than \SI{75}{\hertz} of the spectrograms had negligible impact on the proposed AF detector's classification performance but significantly reduced its memory requirement\footnote{As presented in Section~\ref{SubSect:PerformanceResNetAFDetector}, each model was trained on a single graphics card; therefore, lower memory requirement generally results in a faster training time and unlocks the potential of increasing the network depth.}. Hence, the final output of a preprocessed ECG record is a 2D spectrogram with dimensions of $1789\times15$ as illustrated in Figure~\ref{fig:ECGtoSpectrogram}; each column contains the STFT result of a particular window of ECG samples and the colors represent the frequency components' magnitudes.

\begin{figure}[H]
	\centering
	\includegraphics[width=0.95\columnwidth]{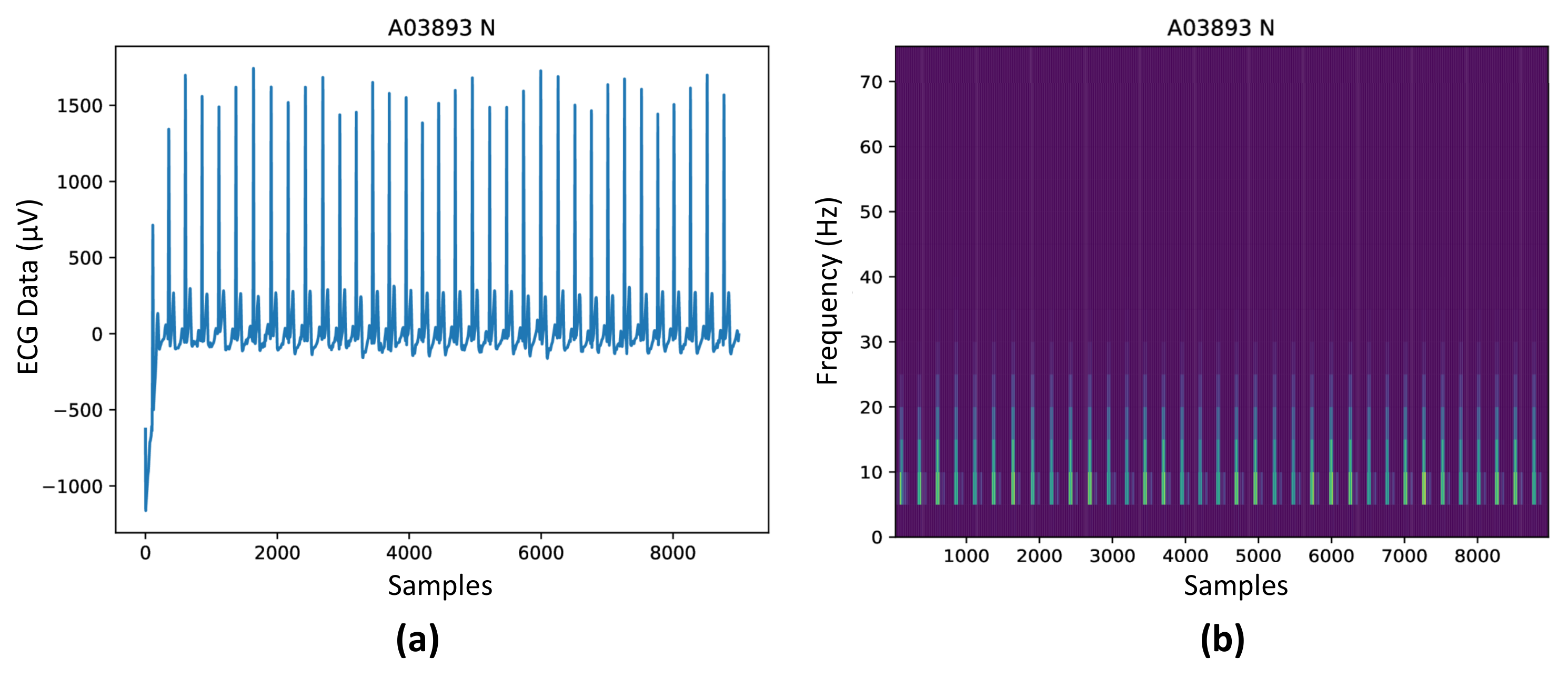}
	\caption{Preprocessing example using Record A03893 (normal class) from the public dataset of the CinC challenge: (\textbf{a}) Time series ECG signal. (\textbf{b}) Spectrogram after preprocessing.}
	\label{fig:ECGtoSpectrogram}
\end{figure}

\subsubsection{ResNet-Based AF Detector}
\label{SubSubSect:AFResNet}

The heart arrhythmia classifier specifically designed for AF detection in this work is based on the ResNet proposed by He et al. \cite{he2016deep}. The core component of the ResNet is the $ResBlock$ (denoted by $rb$ hereinafter) as illustrated at the rightmost of Figure~\ref{fig:StructuralDiagramAFDetector}. It includes two convolutional layers ($Conv1$ and $Conv2$) and a skip connection (skipping the $Conv1$ layer). By utilizing the skip connections between layers, the ResNet is able to avoid the problem of vanishing gradients or mitigate the degradation problem in DNNs. The network structure of the proposed AF detector is shown in Figure~\ref{fig:StructuralDiagramAFDetector}. There are five $ResBlocks$ (denoted by $RB$ hereinafter) in the network and each $RB$ contains five $rb$s; therefore, the network depth is \SI{52}{} layers. It is noteworthy that the number of $RB$ and the number of $rb$ in each $RB$ were carefully determined for optimal classification performance and the experimental results are detailed in Section~\ref{SubApped:OptimalAFDetectorStrcture}. Preliminary results suggested that scaling the inputs during batch normalization (denoted by $BatNorm$) could improve the classification performance\footnote{The scaling parameter in batch normalization was set to $\gamma=3$ when implementing the ResNet-based AF detector.}.

\begin{figure}[H]
	\centering
	\includegraphics[width=0.75\columnwidth]{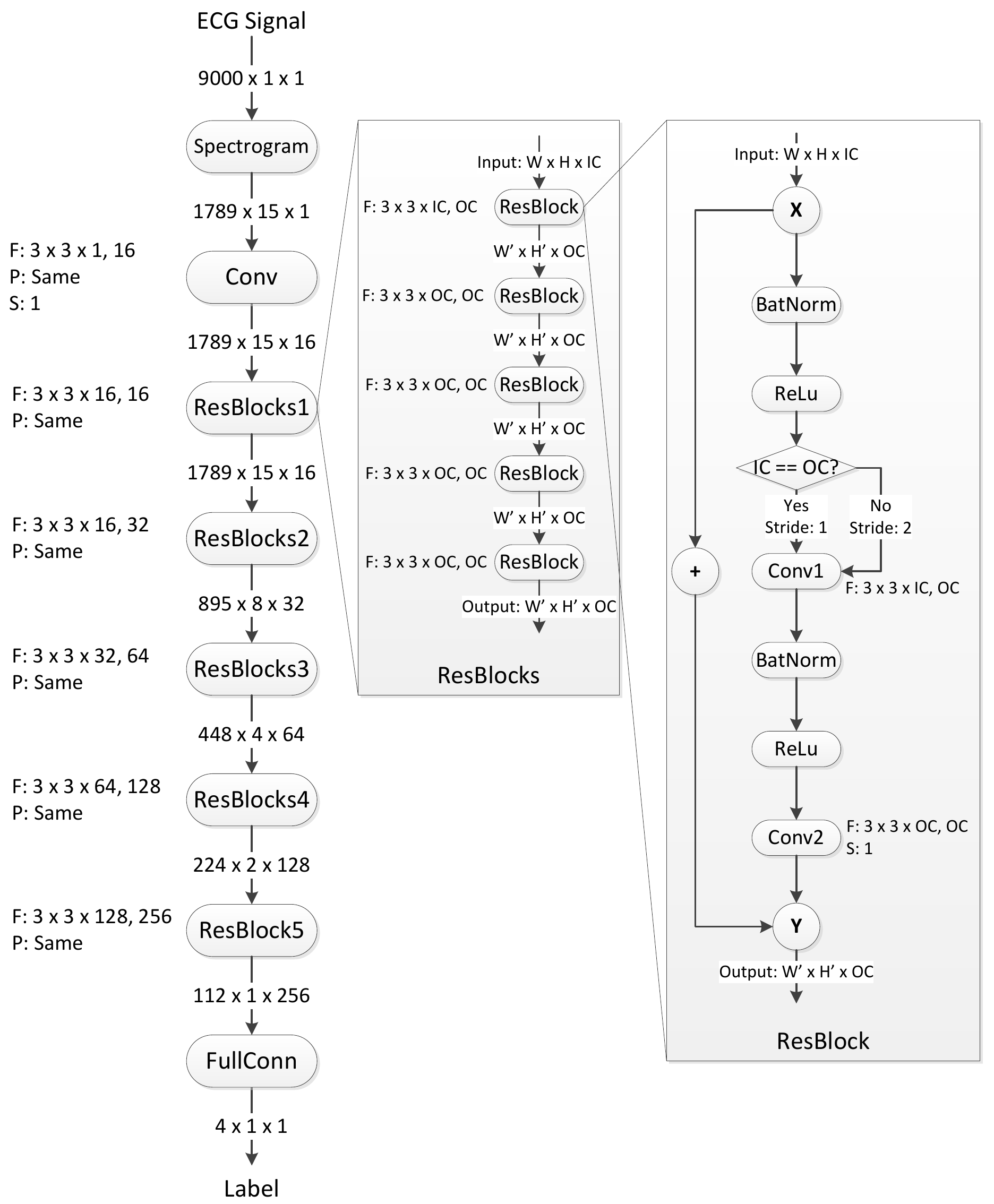}
	\caption{Structural diagram of the ResNet-based AF detector with $M=N=5$, where $N$ is the number of $ResBlocks$ ($RB$) and $M$ is the number of $ResBlock$ ($rb$) in each $RB$. The characters F, P, and S are the abbreviations of filter, padding, and stride, respectively; the IC and OC represent the number of input channels and the the number of output channels, respectively.}
	\label{fig:StructuralDiagramAFDetector}
\end{figure}

According to the structural diagram in Figure~\ref{fig:StructuralDiagramAFDetector}, each ECG record, i.e., an ECG signal with \SI{9000}{} samples, first goes through the preprocessing phase (detailed in Section~\ref{SubSubSect:AFPreprocessing}) and is converted into a 2D spectrogram with dimensions of $1789\times15$. The spectrogram is then processed by the first convolutional layer using a $3\times3$ filter with \SI{16}{} output channels. Afterwards, five $RB$s, of which each contains five $rb$s, utilizing $3\times3$ filters with doubling output channels are applied. All the $rb$s in each $RB$ employ $3\times3$ filters with identical number of input and output channels except the first one. Every $rb$ checks whether the number of input features and the number of output channels are matched. If not, a stride-2 filter, instead of a stride-1 filter, will be applied. The final layer of the proposed classifier is a fully connected layer with Softmax function; it calculates the probability distributions of the four potential classes to which the current ECG record will be classified, i.e., normal rhythm, AF rhythm, others rhythm, and noise. Utilizing the public dataset from the CinC challenge, the classification performance of this arrhythmia classifier was evaluated in Section~\ref{SubSect:PerformanceResNetAFDetector}; performance comparison between this classifier and those proposed by the first-place participants in the CinC challenge was also conducted and summarized in Section~\ref{SubSubSect:PerformanceAFDetector}.

\section{Results and Discussion}
\label{Sect:ResultsAndDiscussion}

The implementation of the personalized healthcare system monitoring users' ECG information remotely and continuously to provide prompt warnings and timely feedbacks is presented in Section~\ref{SubSect:SystemImplementation}; the limitations and the future developments are also presented. The performances of the proposed InLC lossy compression schema, which aims at boosting the battery life of the wearable ECG devices for long-term monitoring, and the ResNet-based AF detector, which classifies the heart arrhythmias from short-period single-lead ECG signals for automated AF screening/diagnosis, were evaluated in Section~\ref{SubSect:LCSchemaEvaluation} and Section~\ref{SubSect:PerformanceResNetAFDetector}, respectively; the avenues for future research including deploying the InLC schema and the AF detector in the personalized healthcare system are discussed as well.

\subsection{Implementation of the Personalized Healthcare System}
\label{SubSect:SystemImplementation}

The development of the proposed personalized healthcare system, the system architecture of which is presented in Figure~\ref{fig:EhealthSystemArchitecture}, is detailed in this section. The implementation of the wearable ECG monitoring device is first described in Section~\ref{SubSubSect:WearableECGDevice}. Following that in Section~\ref{SubSubSect:MobileApp} and Section~\ref{SubSubSect:BackEndServer} are the implementations of the mobile application and the web interface and back-end server, respectively. Finally, the limitations of the current system and the developments of future iterations are discussed in Section~\ref{SubSubSect:SystemLimitationsAndFutureWorks}.

\subsubsection{Wearable ECG Device}
\label{SubSubSect:WearableECGDevice}

The implementation of a patch-like wearable ECG monitoring device, which is lightweight and flexible, is presented in this section. The user is able to monitor his/her real-time ECG signals continuously by sticking the device, which has established a BLE connection with the user's smart phone, on the chest.

\begin{figure}[H]
	\centering
	\includegraphics[width=1\columnwidth]{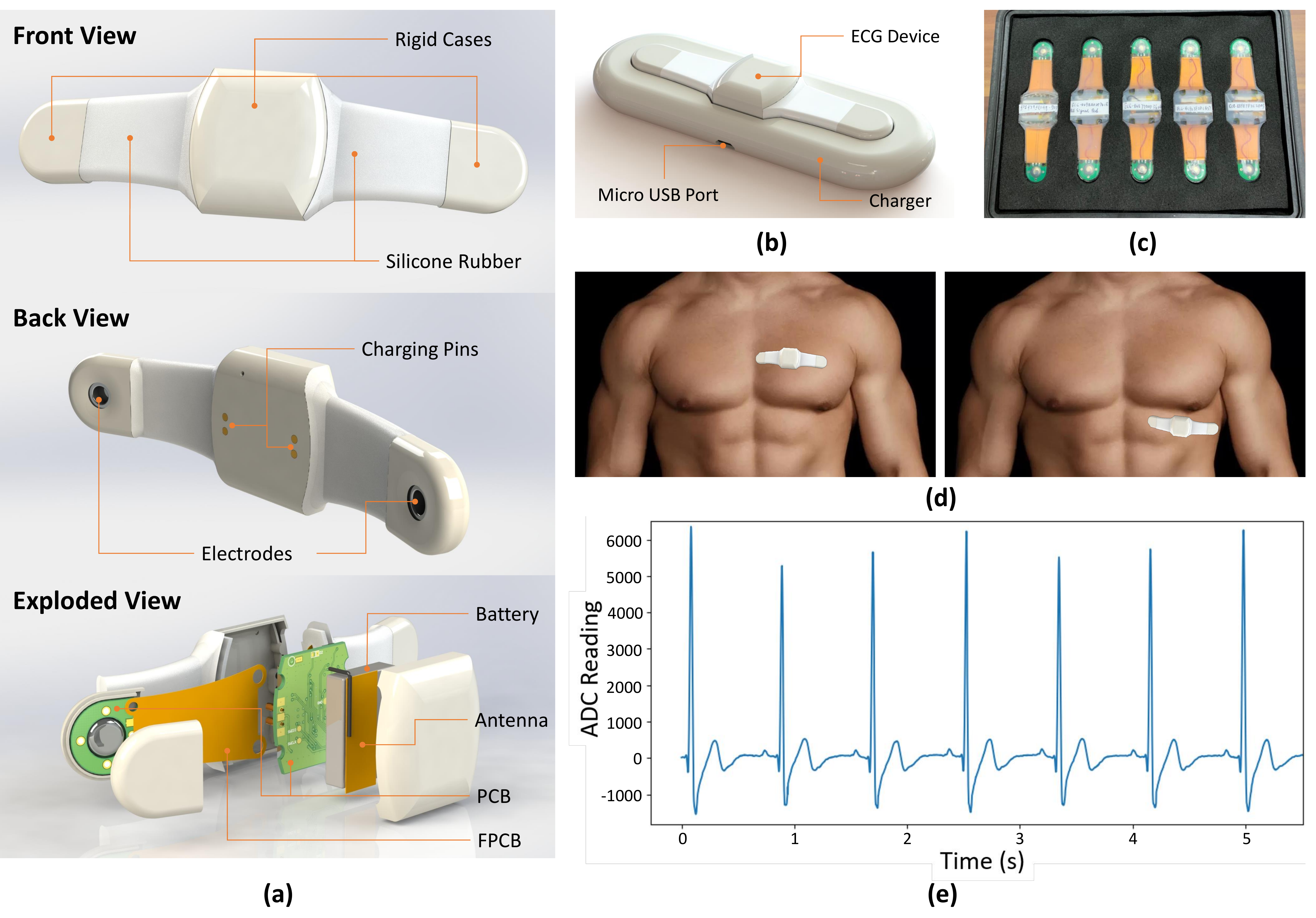}
	\caption{Illustration of the implemented wearable ECG device: (\textbf{a}) 3D drawings of the prototype. (\textbf{b}) 3D drawing of the ECG device and charger. (\textbf{c}) Five ECG device prototypes. (\textbf{d}) Placement of the device on user's chest (sticky pads are not shown). (\textbf{e}) Sample ECG data from an ECG device.}
	\label{fig:ECGDeviceDevicePlacementData}
\end{figure}

The 3D drawings of the wearable ECG device are shown in Figure~\ref{fig:ECGDeviceDevicePlacementData}a and Figure~\ref{fig:ECGDeviceDevicePlacementData}b. Five prototypes have been implemented in this work and a photo of them is illustrated in Figure~\ref{fig:ECGDeviceDevicePlacementData}c. In each wearable device, the ECG signals are acquired by the NeuroSky BMD101 analog front-end \cite{neurosky2021ultimate} with a built-in 16-bit analog-to-digital converter (ADC) sampling at \SI{512}{\hertz}. An ultra low-power ARM Cortex-M0+ micro-processor with \SI{32}{\mega\hertz} frequency, \SI{32}{\kilo\byte} Flash, and \SI{8}{\kilo\byte} SRAM is utilized to process the raw data from the analog front-end as well as establish Bluetooth communication with the smart phone using the DA14580 BLE SoC \cite{dialog2021bluetooth}. The two wings of the device are made of flexible printed circuit board (FPCB) encased in the food grade silicone rubber so that it can be attached on the user's chest conveniently and comfortably as shown in Figure~\ref{fig:ECGDeviceDevicePlacementData}d\footnote{To stick the wearable ECG device to the user's chest, two ECG sticky pads or stickers attaching to the two electrodes of the wearable device are needed. The sticky pads are not shown in Figure~\ref{fig:ECGDeviceDevicePlacementData}.}. The positions at which the ECG device should be placed were carefully selected as detailed in Section~\ref{SubApped:ECGDevicePlacementTest} and Section~\ref{SubApped:ECGSignalDisturbanceFromActivities}. A sample of ECG data collected by the wearable device, which was attached at the left position of Figure~\ref{fig:ECGDeviceDevicePlacementData}d, is illustrated in Figure~\ref{fig:ECGDeviceDevicePlacementData}e. The total cost and weight of the ECG device (without the sticky pads and charger) are \SI{35}[US\$]{} and \SI{30}{\gram}, respectively, and it has dimensions of $\SI{111.0}{\milli\meter} \times \SI{30.0}{\milli\meter} \times \SI{11.3}{\milli\meter}$. The ECG device is powered by a single Li-Po battery with total capacity of \SI{150}{\milli\ampere{}\hour}. Estimated by performing three charge-discharge cycles on the five prototypes, the ECG devices can operate for about \SI{20}{\hour} on a single charge and require about \SI{2.5}{\hour} to fully recharge.

The wearable ECG device has four operating modes, namely the sleep mode, the charge mode, the idle mode, and the run mode. The ECG device enters the sleep mode when battery is low (e.g., $\leq\SI{3.2}{\volt}$); power supplies of the BMD101 analog front-end and the DA14580 BLE SoC are shut off by the micro-processor which afterwards enters low-power mode resulting in a negligible current consumption. The only way to exit the sleep mode is by placing the ECG device on the charger as illustrated in Figure~\ref{fig:ECGDeviceDevicePlacementData}b; the device switches to the charge mode immediately and begins to charge the battery. Several neodymium magnets are utilized in the charger such that the charging pins on the device can snap into the charging port conveniently. In the charge mode, the micro-processor is running at full speed; the power supply of the DA14580 is restored and it is advertising at \SI{0.2}{\hertz}. The name of the ECG device will appear in the mobile application, and the BLE connection between the ECG device and the smart phone is initiated by the user by selecting the target ECG device listed on the mobile application. If the BLE connection has been established, the battery voltage information and the charging current and status are transmitted to the mobile application every second. The current consumption in the charge mode is omitted. Whenever the ECG device is charged but the BLE connection has not been established, it enters the idle mode, in which the BMD101 and the DA14580 are kept powering off and advertising, respectively. In the idle mode, the micro-processor is running in low-power mode resulting in a total current consumption as low as \SI{0.1}{\milli\ampere} at \SI{3.7}{\volt} supply. Once the BLE connection has been established, the ECG device switches from the idle mode to the run mode immediately. In this mode, the BMD101 is powered on and taking samples at \SI{512}{\hertz}; each sample is a 2's complement 16-bit ADC reading. The micro-processor is running at full speed to retrieve the raw ADC readings from the BMD101 and pack every 16 ADC readings into a data package before being transmitted by the DA14580. Therefore, the packed ADC readings from the ECG device are transmitted to the mobile application at \SI{32}{\hertz}. Besides, the heart rate and ECG signals quality determined by the BMD101 and the battery voltage information are transmitted to the mobile application at \SI{1}{\hertz}. The average current consumption in the run mode is around \SI{6.7}{\milli\ampere} at \SI{3.7}{\volt} supply. Once the BLE connection is terminated or no heartbeats have been detected by the BMD101 for \SI{5}{\minute}, the ECG device switches back to the idle mode for energy saving purpose.

As detailed in Section~\ref{SubApped:ECGSignalDisturbanceFromActivities}, the muscle contractions related disturbances to the ECG signals acquired by the wearable device were investigated. In addition, the intra- and inter- consistencies of the outputs from the five prototypes were also investigated in Section~\ref{SubApped:SignalComparisonBetweenDeviceAndEquipment} and Section~\ref{SubApped:SignalComparisonAmongDevices}, respectively. To verify whether or not the quality of ECG signals collected by the wearable device is sufficient for cardiac arrhythmia diagnosis, the ECG device was first evaluated against a medical equipment which provides diagnostic ECG of patients in Section~\ref{SubApped:SignalComparisonBetweenDeviceAndEquipment}. Then, 13 types of rhythms recored by the ECG device were sent to two professional cardiologists for diagnosis without discovering the exact rhythm types in Section~\ref{SubApped:DiagnosingECGSignalsFromDevice}. The evaluation results suggested that the wearable ECG device has excellent intra-consistency and acceptable inter-consistency; the quality of ECG signals from the device is sufficient for cardiac arrhythmia diagnosis, given that the rhythms have distinguishable criteria established from the Lead II.

\subsubsection{Mobile Application}
\label{SubSubSect:MobileApp}

The mobile application was built upon a cross-platform framework named React Native \cite{facebook2021react}, which allows the same pieces of code to run on both Android and iOS devices. The system architecture of the mobile application is depicted in Figure~\ref{fig:MobileAppArchi}. The screenshots taken from the mobile application running on an Android device are shown in Figure~\ref{fig:MobileAppScreeShots}.

\begin{figure}[H]	
	\centering
	\includegraphics[width=1\columnwidth]{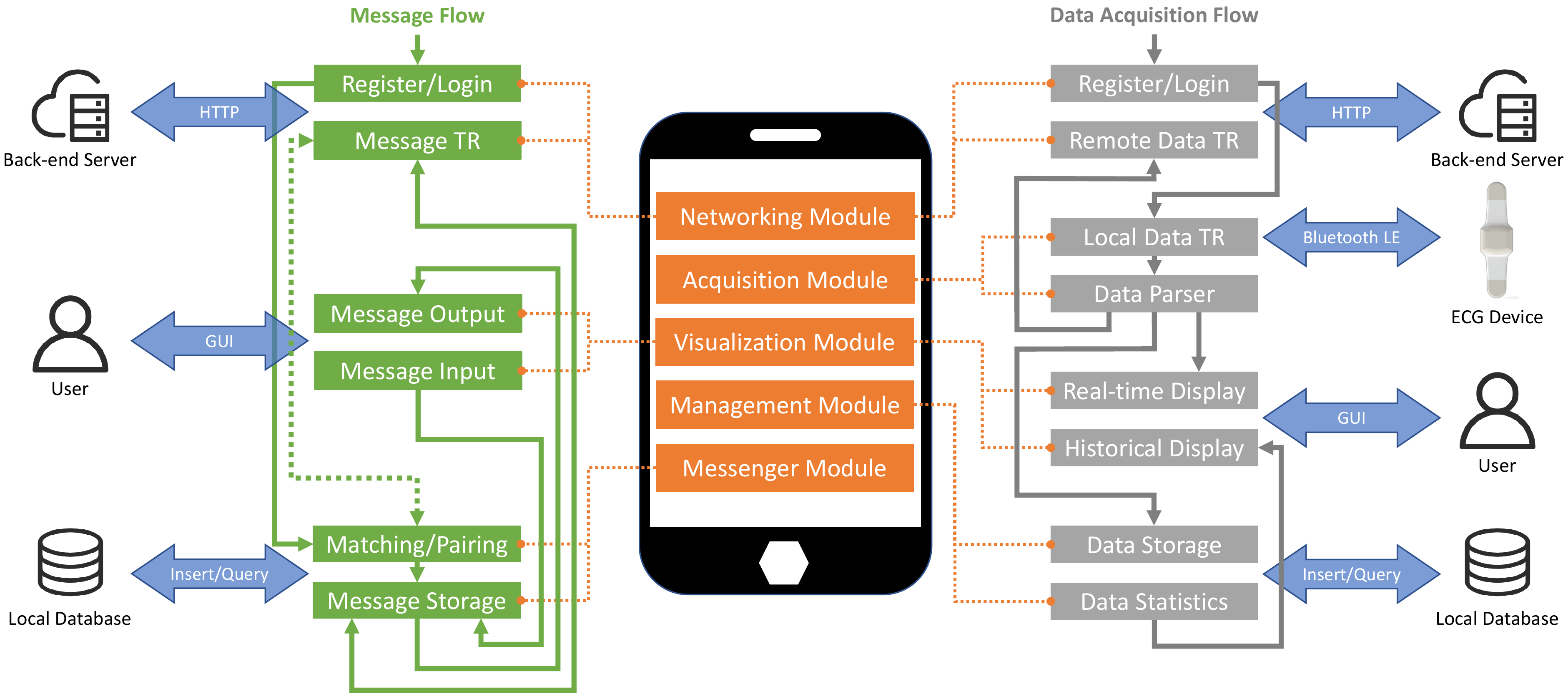}
	\caption{System architecture diagram of the mobile application. The flowcharts in green and gray represent the message flow and the data acquisition flow of the mobile application, respectively.}
	\label{fig:MobileAppArchi}
\end{figure}

\begin{figure}[H]	
	\centering
	\includegraphics[width=1\columnwidth]{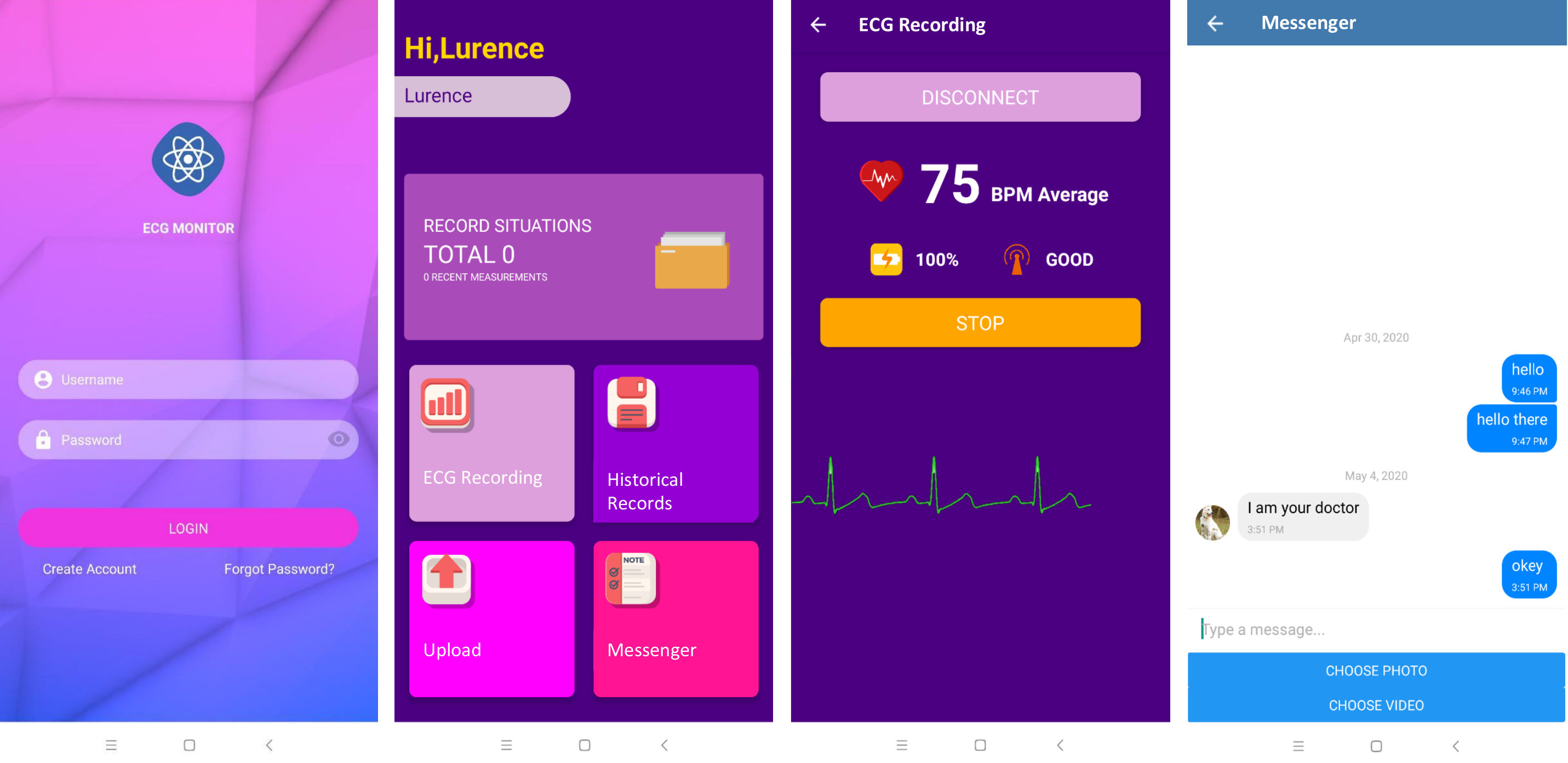}
	\caption{Screenshots of the implemented mobile application on an Android device. The screenshots from left to right are the register/login screen, the main screen, the real-time ECG screen, and the messenger screen, respectively.}
	\label{fig:MobileAppScreeShots}
\end{figure}

In the networking module, the register/login submodule provides a register and login interface to the user as illustrated in the first screenshot of Figure~\ref{fig:MobileAppScreeShots}. For registration, the user needs to input a valid e-mail address, a password, and the verification code received from the back-end server. Login authentication is required for the user to proceed. After login, the mobile application jumps to the main screen as depicted in the second screenshot of Figure~\ref{fig:MobileAppScreeShots}. The user is able to get access to his/her historical ECG records that are stored in the local Realm \cite{realm2021database} database (records fetched from the remote database of the back-end server are firstly loaded into the local database) by clicking the Historical Records tab. The historical display submodule then renders the graphical user interface (GUI) for visualizing the ECG records as well as some basic statistics such as average heart rate generated by the data statistics submodule. By clicking the Upload tab, the user can manually upload his/her previous ECG records stored in the local database (those have not been automatically uploaded during recording) to the back-end server via the remote data transmission and reception (TR) submodule.

To get the real-time ECG recordings from a wearable ECG device, the user needs to press the ECG Recording tab and the mobile application will jump to the real-time ECG screen, in which the local data TR submodule is ready to establish BLE connection with the ECG device. The BLE connection is initiated by the user by selecting the target device listed on the mobile application. The local data TR submodule also records the connected BLE slave's media access control (MAC) address in order to automatically issue reconnection. The data packages received from an ECG device are decoded by the data parser submodule to retrieve respective information, including charging current and status, battery voltage, heart rate, ECG signal quality, and ADC readings of the ECG signals. The real-time display submodule then renders the GUI for visualizing the retrieved information as shown in the third screeshot of Figure~\ref{fig:MobileAppScreeShots}. Concurrently, the retrieved information, which has been tagged with the BLE slave's MAC address and the current date and time from the smart phone, are inserted into the local database by the data storage submodule and transmitted to the back-end server by the remote data TR submodule. Note that the ECG device is transmitting the packed ADC readings to the mobile application at \SI{32}{\hertz}. To prevent overloading the mobile application and the back-end server, the real-time display submodule has a refresh rate of \SI{16}{\hertz}, while the retrieved information is inserted in the local database and uploaded to the back-end server every \SI{10}{\second}.

To communicate with the health advisor, the user needs to press the Messenger tab in the main screen and the mobile application will jump to the messenger screen as depicted in the fourth screenshot of Figure~\ref{fig:MobileAppScreeShots}. If the user have not paired with any health advisor, the matching/pairing submodule will list all the available advisors and handle the pairing request issued by the user. The paring request is transmitted to the back-end server by the message TR submodule; a pairing process is completed once the health advisor accepts the request. Now the user can communicate with the paired health advisor by inputing texts, images, and videos to the GUI rendered by the message input submodule. These inputs are stored in the local database by the message storage submodule before being transmitted to the back-end server via the message TR submodule. The message TR submodule is also responsible for fetching the messages, which were sent by the advisor, from the back-end server every \SI{2}{\second}. The fetched messages are first stored in the local database and then rendered to the user by the message output submodule. In addition, the alarms issued by the health advisor or the abnormal situations detected by the back-end server are also fetched from the server by the message TR submodule periodically; they are rendered as pop-up messages by the message output submodule to promptly warn the user.

\subsubsection{Back-end Server and Web Interface}
\label{SubSubSect:BackEndServer}

For the back-end server, all services were developed upon the Spring \cite{vmware2021spring} platform and deployed on a cloud cluster hosting two Tencent cloud instances and three AWS EC2 instances. For rapid prototyping, all communications among different cloud instances or between the back-end server and the mobile application are based on the hypertext transfer protocol (HTTP). The system architecture diagram of the back-end server is illustrated in Figure~\ref{fig:BackEndServerArchi}.

\begin{figure}[H]	
	\centering
	\includegraphics[width=0.7\columnwidth]{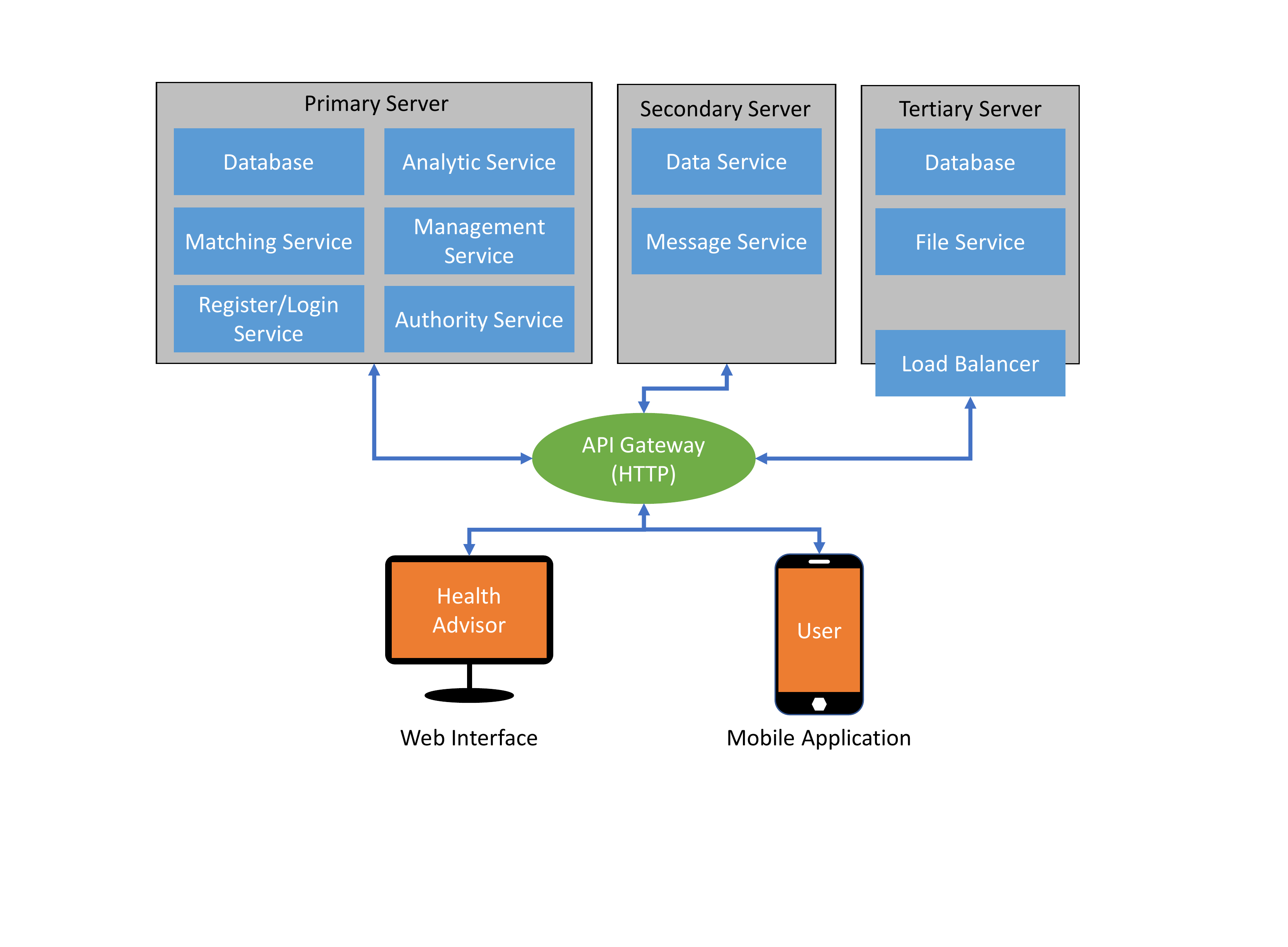}
	\caption{System architecture diagram of the back-end server. The primary and the secondary servers are respectively deployed on two Tencent cloud instances, while the tertiary server is deployed on three AWS EC2 instances.}
	\label{fig:BackEndServerArchi}
\end{figure}

While most of the back-end services and the primary MySQL \cite{mysql2021database} database are deployed on the primary server, the data and message services, and the file service and secondary MongoDB \cite{mongo2021database} database are deployed on the secondary and the tertiary servers, respectively. Such mechanism significantly relaxes the network traffic on the primary server because the data (i.e., charging current and status, battery voltage, heart rate, ECG signal quality, and ADC readings of ECG signals tagged with the wearable devices' MAC addresses and the date and time from smart phones) and messages (i.e., conversations between users and health advisors) streaming from the mobile applications are respectively handled by the data and message services that were built upon the Apache Kafka \cite{apache2017kafka} high-performance distributed streaming platform. Thanks to the distributed and scalable nature of this platform, the back-end server can be easily scaled up by deploying more data services and message services on additional instances to handle more streaming traffic. Enabled by the data service and the message service, the data and messages produced by the mobile application (i.e., producer) are first buffered in the secondary server and then consumed by the primary server (i.e., consumer), e.g., inserting them into the primary database. It is noteworthy that, whenever a message from the mobile application contains image or video, the application programming interface (API) gateway will first direct the message to the tertiary server which consists of three AWS EC2 instances. The load balancer then automatically selects a proper instance to handle the incoming message, and the file service in that instance inserts the uploaded image or video into the secondary database and returns its uniform resource location (URL) to the message service in the secondary server. Because the roles of a producer and a consumer are interchangeable, the message service in the secondary server is also responsible for fetching the messages (texts and URLs) from the primary database and retrieving the images or videos from the tertiary server (using the stored URLs). These retrieved contents are then delivered to the mobile application. Besides, the feedbacks automatically generated by the analytic service or manually issued by the health advisors via management service are delivered to the mobile application by the message service as well.

Other than the mobile application, a web interface was also developed for the users and health advisors to visualize the ECG signals acquired from the wearable devices and communicate with each other. Similar to the mobile application, a successful login of a registered account is required to interact with the web interface. The registration and login requests are handled by the register/login service; it also guarantees that there is only one concurrent login per account. After login, different permissions are granted to a user and a health advisor by the authority service. While the users are restricted to the acquired data of their own, the health advisors have unlimited access to the data from the users who have paired with them as shown in the screenshots of Figure~\ref{fig:WebInterface}. The paring and unpairing requests are handled by the matching service. In addition to communicating with the paired users, the health advisors can use the web interface to issue timely feedbacks to them via the management service.

\end{paracol}
\begin{figure}[t]
	\widefigure
	\centering
	\includegraphics[width=1\columnwidth]{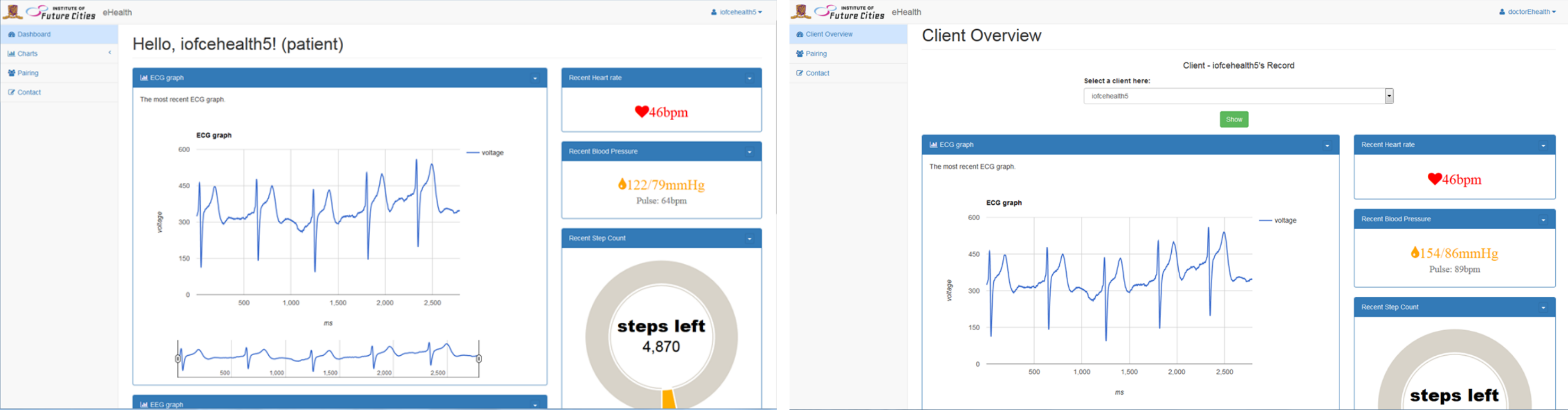}
	\caption{Screenshots of the implemented web interface. The images from left to right are the main pages of the web interface for the users and the health advisors, respectively.}
	\label{fig:WebInterface}
\end{figure}
\begin{paracol}{2}
\switchcolumn

\subsubsection{Limitations and Future Works}
\label{SubSubSect:SystemLimitationsAndFutureWorks}

Limited to what is practical, the analytic service in the back-end server is currently producing simple statistical results and the current diagnoses are provided by the health advisors. The development of automating the diagnostic procedure by adopting machine learning techniques and deep neural networks is enthusiastically awaited, especially the capability of identifying heart arrhythmias such as AF (e.g., the ResNet-based AF detector presented in Section~\ref{SubSect:AutomaticAFDetector}). For rapid prototyping, the back-end server and the mobile application are communicating via HTTP; for security reasons, the hypertext transfer protocol secure (HTTPS) is preferred in the future development. Besides, the usability of the wearable ECG device and the mobile application could be improved to prevent overwhelming the users and discouraging their initiatives. For example, to deploy the InLC lossy compression schema detailed in Section~\ref{SubSect:ECGLossy} on the wearable ECG device to boost its battery life for long-term monitoring.

\subsection{Performance Evaluation of the Lossy Compression Schemata}
\label{SubSect:LCSchemaEvaluation}

The compression efficiency and distortion of the OD, the GSVQ, the PCA-based, and the InLC schemata on the ECG signals from the PhysioNet MIT-BIH arrhythmia database \cite{moody2001the,goldberger2000physiobank} were investigated in this section. The compression efficiency and the distortion are quantified by the CR defined in Equation~\ref{eq:CR} and the PRD or RMSE defined in Equation~\ref{eq:PRD} or Equation~\ref{eq:RMSE}, respectively. The MIT-BIH database contains \SI{48}{} records of half-hour two-channel ECG signals, and each record was digitized at \SI{360}{\hertz} per channel with 11-bit resolution over a \SI{10}{\milli\volt} range. Following the experimental setups presented in Section~\ref{SubSubSect:ExperimentalSetupsLCs}, the digitized readings of each record's first channel were used to evaluate the long-term (\SI{30}{\minute} ECG signal of each record) and short-term (first \SI{1}{\minute} ECG signal of each record) performances of the aforementioned lossy compression schemata in Section~\ref{SubSubSect:AveragePerformanceLCs}. Afterwards, the performances of the InLC schemata with different configurations/setups on the first \SI{1}{\minute} ECG signals of all \SI{48}{} records were evaluated in Section~\ref{SubSubSect:AveragePerformanceInLC}. The limitations of this work as well as the future research are discussed in Section~\ref{SubSubSect:DiscussionAndFutureWorksLC}.

\subsubsection{Experimental Setups}
\label{SubSubSect:ExperimentalSetupsLCs}

In order to calculate the CR values, the number of bits used to encode the compressed signals is first defined accordingly. For the OD and the GSVQ schemata, the resolutions\footnote{Note that the parameters with 16-bit resolution are floating point numbers, and those with other resolutions are integers.} of the codeword index $i^{*}$, the original segment length $l_{t}$, the gain $g_{t}$, and the offset $o_{t}$ (for OD only) are 6-, 9-, 16-, and 16-bit, respectively. While each element of a new codeword generated in the OD is a 16-bit floating point number, the GSVQ encodes the length of the residual stream, the value of a significant point, and the distances between adjacent significant points using 9-, 11-, and 4-bit integers, respectively. Note that the GSVQ's codebook was generated from the \SI{30}{\minute} ECG signals of all \SI{48} records. The resolutions of the PCA schema's original segment length, gain, and offset are identical to those of the OD's because they utilize the same preprocessing chain. Each element of the vector $\vec{\mu}$ and the matrices $\mathbold{\Psi}$ and $\mathbold{Y}$ generated by the PCA encoder is a 16-bit floating point number. For the InLC schema, the start index $i^{*}$ and the fragment length $l^{*}$ are 10-bit integers, while the gain $g^{*}$ and the offset $o^{*}$ are 16-bit floating point numbers.

As described in Section~\ref{SubSubSect:IssuesInExistingLossy} and Section~\ref{SubSubSect:IntuitiveLossy}, the compression efficiency and distortion of the OD, the GSVQ, and the InLC schemata are controlled by their respective threshold values ($\epsilon$ or $A_{th}$) once the remaining parameters are settled, whereas the PCA schema's performance is determined by the number of normalized segments $N$ and the number of principle components $K$. For the ECG signals from the MIT-BIH database, preliminary experiments suggested that good compression efficiency and low distortion can be achieved with $N=50$ and variable $K$. The results presented in Section~\ref{SubSubSect:AveragePerformanceLCs} and Section~\ref{SubSubSect:AveragePerformanceInLC} were obtained by varying the aforementioned determining parameters (i.e., $\epsilon$, $A_{th}$, or $K$)\footnote{Specifically, the $\epsilon$ for the OD, the $A_{th}$ for the GSVQ, the $K$ for the PCA, and $\epsilon$ for the InLC were varying within the sets $\{0.02,0.04,\cdots,0.30\}$, $\{0.02,0.04,\cdots,0.30\}$, $\{15,13,\cdots,1\}$, and $\{2,4,6,8,10,15,20,30,40,50,60,70\}$, respectively.}. Since the lossy compression methods achieve excellent compression efficiency at the cost of signal distortion, a satisfactory schema should maximize the CR while maintaining low PRD or RMSE (trade-off between efficiency and distortion).

\subsubsection{Average Performance of the OD, GSVQ, PCA, and InLC Schemata}
\label{SubSubSect:AveragePerformanceLCs}

The average performance of the OD, the GSVQ, the PCA, and the InLC schemata for the \SI{30}{\minute} and the \SI{1}{\minute} ECG signals from the \SI{48}{} records of the MIT-BIH database are illustrated in Figure~\ref{fig:LCPerformanceCRPRD} and Figure~\ref{fig:LCPerformanceCRRMSE}. The compression efficiency versus distortion curves for the \SI{1}{\minute} ECG signals were labeled by the schema names and symbol ``-1''. It is apparent that the proposed InLC schema can achieve much higher CR than the others (at least doubled the CR values) under certain PRD or RMSE, for both long-period (\SI{30}{\minute}) and short-period (\SI{1}{\minute}) ECG signals. That is, the InLC surpassed the others in terms of compression efficiency and distortion.

While the InLC and the PCA schemata demonstrated better performances for the short-period ECG signals, the GSVQ's results for the short- and long-period ECG signals were nearly identical. In contrast, compared to the results for the long-period ECG signals, the OD's efficiency was significantly reduced while performing compression on the short-period ones. As discussed in Section~\ref{SubSubSect:IssuesInExistingLossy}, the OD's codebook is empty after initialization and the codeword synchronization process greatly reduced its compression efficiency. Consequently, the OD is not preferred for applications that require frequent codebook synchronization (e.g., measuring short periods of ECG occasionally). As for the GSVQ, nearly identical results for the short- and long-period ECG signals were achieved mainly because the off-line codebook was constructed utilizing all \SI{48}{} half-hour records. The improved performance of the PCA schema for short-period ECG is considered a coincidence because a long-period ECG signal was divided into short-period ones before performing PCA individually; the PCA schema's performance for long-period signals is considered more representative. According to the operating principle of the InLC schema presented in Section~\ref{SubSubSect:IntuitiveLossy}, the bank $\vec{B}_{b}$ contains the first $s_{b}$ samples of an ECG record. By intuition, the quasi-periodic feature of the ECG signal suggests that it would be more and more difficult to find a similar fragment $\vec{b}^{i}_{l}$ in $\vec{B}_{b}$ for the current fragment $\vec{f}_{l}$ with length $l$ as time goes on, and vise versa. Therefore, it is anticipated that the InLC demonstrated significantly better performance for the short-period ECG signals.

\end{paracol}
\begin{figure}[t]
	\widefigure
	\centering
	\includegraphics[width=0.92\columnwidth]{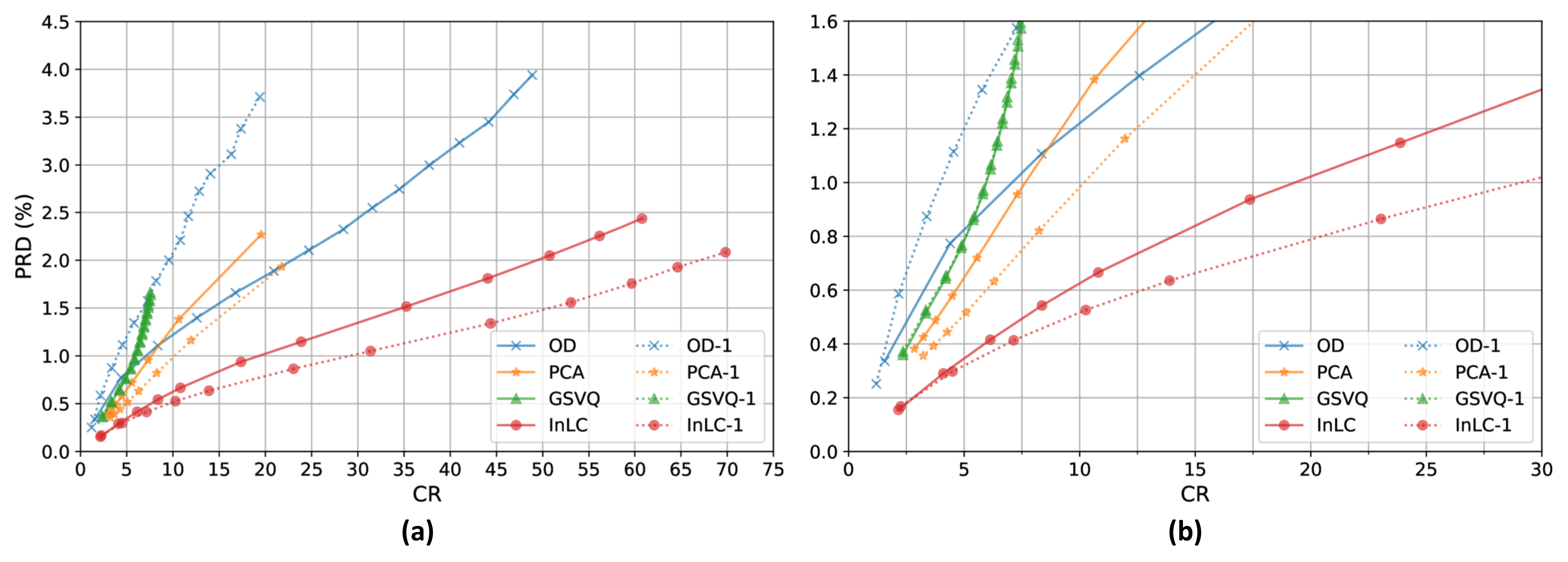}
	\caption{Average performance of the OD, PCA, GSVQ, and InLC schemata for all \SI{48}{} records in the MIT-BIH database (\SI{30}{\minute} and \SI{1}{\minute} ECG signals) with various threshold values or principle components: (\textbf{a}) Full scale CR vs. PRD. (\textbf{b}) Zoomed CR vs. PRD.}
	\label{fig:LCPerformanceCRPRD}
\end{figure}

\begin{figure}[t]
	\widefigure
	\centering
	\includegraphics[width=0.92\columnwidth]{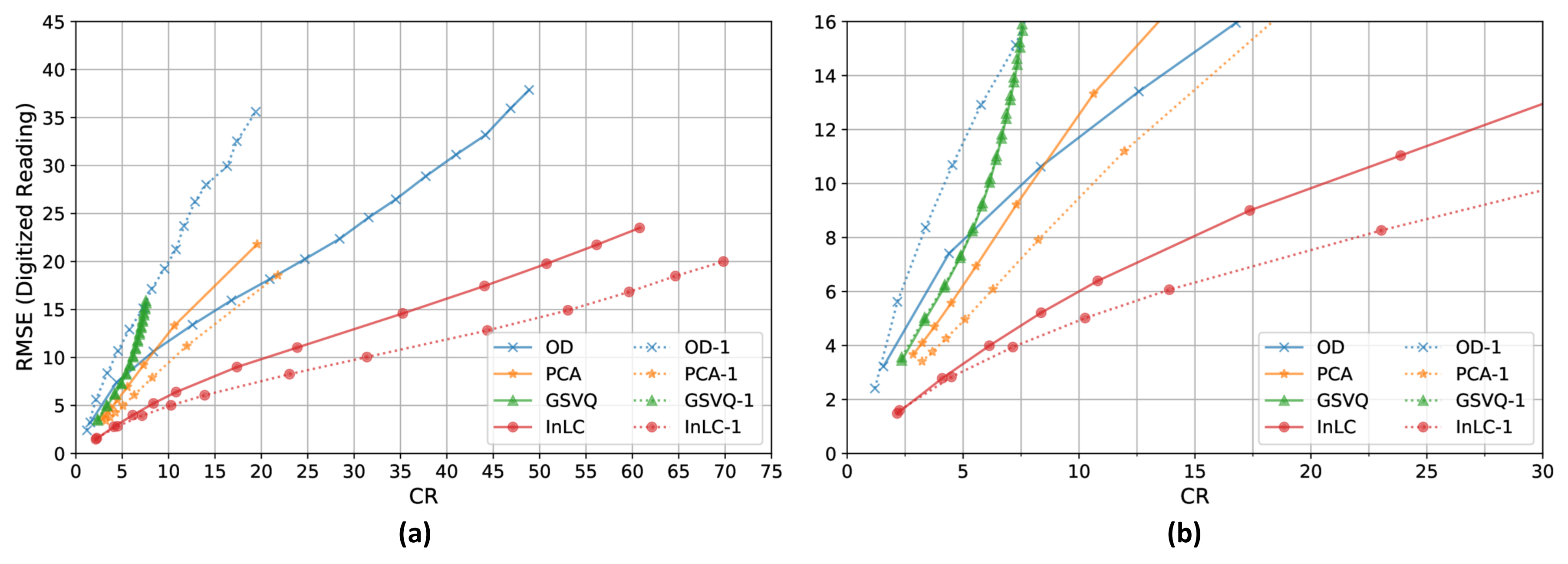}
	\caption{Average performance of the OD, PCA, GSVQ, and InLC schemata for all \SI{48}{} records in the MIT-BIH database (\SI{30}{\minute} and \SI{1}{\minute} ECG signals) with various threshold values or principle components: (\textbf{a}) Full scale CR vs. RMSE. (\textbf{b}) Zoomed CR vs. RMSE.}
	\label{fig:LCPerformanceCRRMSE}
\end{figure}
\begin{paracol}{2}
\switchcolumn

In order to improve the InLC schema's performance for long-period ECG signals, one straightforward idea is to minimize the time lag between the bank $\vec{B}_{b}$ and the current fragment $f_{l}$. Such objective can be achieved by eliminating the statement ``if $|\vec{B}_{b}|<s_{b}$'' in the Step~\ref{step4} and Step~\ref{step10} of the InLC schema's operating principle such that the bank $\vec{B}_{b}$ is updated constantly. One may notice that a near-zero distance from Equation~\ref{eq:OldDist} does not guarantee similar shapes of the two fragments $\vec{x}$ and $\vec{y}$ since the former is the necessary but insufficient condition of the latter; the current fragment $\vec{f}_{l}$ in the original ECG signal may be replaced by a distinct fragment $\hat{\vec{f}}_{l}$ in the reconstructed signal. The aforementioned phenomenon was occasionally observed with one of the fragment being a short-period straight line as generalized in Figure~\ref{fig:IllustrationStraightLine}. However, constantly updating the bank $\vec{B}_{b}$ further exacerbates such phenomenon, and the short-period curvilinear waveforms in the original ECG signal are frequently replaced by a few straight lines (hereinafter referred to as the CW-to-SL effect) in the reconstructed one as illustrated in Figure~\ref{fig:InLCStraightLineIssue}a.

\begin{figure}[H]
	\centering
	\includegraphics[width=0.98\columnwidth]{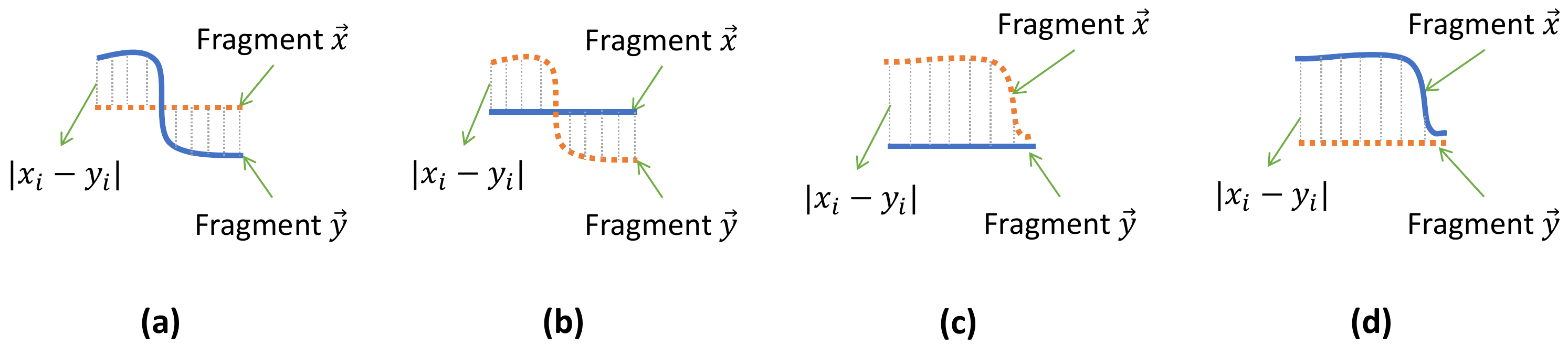}
	\caption{Illustration of the absolution differences of fragment pairs with visually different shapes but small distances from Equation~\ref{eq:OldDist}: (\textbf{a}) Z-shaped curve and straight line with intersection point. (\textbf{b}) Straight line and Z-shaped curve with intersection point. (\textbf{c}) Straight line and L-shaped curve without intersection point. (\textbf{d}) L-shaped curve and straight line without intersection point. Note that fragment pairs with horizontally/vertically mirrored Z- or L-shaped curves are not shown.}
	\label{fig:IllustrationStraightLine}
\end{figure}

To tackle such issue, a new distance function, that is, the Equation~\ref{eq:NewDist}, was proposed. The rationale of the Equation~\ref{eq:NewDist} is to prevent the occurrences of the CW-to-SL effect as generalized in Figure~\ref{fig:IllustrationStraightLine}. The condition $\sum_{i=1}^{n} \left|x_i-y_i\right| \neq \left|\sum_{i=1}^{n}(x_i-y_i)\right|$ in Equation~\ref{eq:NewDist} indicates that the fragments $\vec{x}$ and $\vec{y}$ have at least one intersection point. An infinity distance value is assigned to such fragment pair such that the occurrences as shown in Figure~\ref{fig:IllustrationStraightLine}a and Figure~\ref{fig:IllustrationStraightLine}b are eliminated. Now, only the fragment pairs without intersection will have a chance to be considered as similar depending on the threshold value $\epsilon$. Under the circumstances as shown in Figure~\ref{fig:IllustrationStraightLine}c and Figure~\ref{fig:IllustrationStraightLine}d, the distances form Equation~\ref{eq:OldDist}, i.e., $\rVert\vec{x}-\vec{y}\rVert_{\infty} - \overline{|\vec{x}-\vec{y}|}$, are negligible because the limited number of small absolute differences at the tails can not significantly lower their mean values, i.e., $\overline{|\vec{x}-\vec{y}|}$. In contrast, the values of $\overline{|\vec{x}-\vec{y}|} - \rVert\vec{x}-\vec{y}\rVert_{-\infty}$ for these fragments pairs are comparatively large. As a result, whenever a fragment pair has no intersection, utilizing the maximum value between $\rVert\vec{x}-\vec{y}\rVert_{\infty} - \overline{|\vec{x}-\vec{y}|}$ and $\overline{|\vec{x}-\vec{y}|} - \rVert\vec{x}-\vec{y}\rVert_{-\infty}$ (i.e., the second part of Equation~\ref{eq:NewDist}) as distance can prevent the occurrences as illustrated in Figure~\ref{fig:IllustrationStraightLine}c and Figure~\ref{fig:IllustrationStraightLine}d. A short period of ECG signal reconstructed from the InLC schema with constantly updated $\vec{B}_{b}$ and the new distance function is depicted in Figure~\ref{fig:InLCStraightLineIssue}b. Through visual inspection, the reconstructed ECG signal in Figure~\ref{fig:InLCStraightLineIssue}b exhibited far fewer straight lines compared to that in Figure~\ref{fig:InLCStraightLineIssue}a. However, it is anticipated that the compression efficiency of the InLC schema utilizing the new distance function would be decreased because the Equation~\ref{eq:NewDist} limits the possibilities of finding a fragment $\vec{b}^{i}_{l}$ in $\vec{B}_{b}$ that is similar to the fragment $\vec{f}_{l}$, not to mention identifying a fragment $\vec{b}^{i}_{l}$ with fairly large length $l$.

Benefited from the new distance function, a fragment pair with sufficiently small distance from the Equation~\ref{eq:NewDist} usually implies a close-to-unity gain, which is the slope from OLS regression, according to the authors' experience. Higher compression and computational efficiency of the InLC schema can be achieved by eliminating the gain $g^{*}$ (assuming unity) in the compressed information and only transmitting the new offset $o^{*}$, which is the average difference of the fragment pair. As illustrated in Figure~\ref{fig:InLCStraightLineIssue}b and Figure~\ref{fig:InLCStraightLineIssue}c, such approach will introduce slight fluctuations to the reconstructed signal that might increase the distortion. To identify the optimal configuration, the average performance of the InLC with various setups on the \SI{1}{\minute} ECG signals of all \SI{48}{} records were investigated in Section~\ref{SubSubSect:AveragePerformanceInLC}.

\subsubsection{Average Performance of the InLC with Different Configurations}
\label{SubSubSect:AveragePerformanceInLC}

The average performance of the InLC schema with various setups for the first \SI{1}{\minute} ECG signals from the \SI{48}{} records of the MIT-BIH database are illustrated in Figure~\ref{fig:InLCPerformancePRD} and Figure~\ref{fig:InLCPerformanceRMSE}. The notations ``ODF'' and ``NDF'' represent the utilization of the old (i.e., the Equation~\ref{eq:OldDist}) and the new (i.e., the Equation~\ref{eq:NewDist}) distance functions, respectively. When the bank $\vec{B}_{b}$ is updated constantly, the notation ``UP'' is used, or otherwise, the notation ``NP'' is used. While the configuration with both gain and offset is denoted by ``GO'', the one with offset only is denoted by ``OO''. Therefore, the curves denoted by ``InLC-ODF-NP-GO'' represent the compression efficiency versus distortion results from the InLC schema with Equation~\ref{eq:OldDist} distance function, static bank $\vec{B}_{b}$, and both gains and offsets, that is, the curves with notation ``InLC-1'' in Figure~\ref{fig:LCPerformanceCRPRD} and Figure~\ref{fig:LCPerformanceCRRMSE}.

\end{paracol}
\begin{figure}[t]
	\widefigure
	\centering
	\includegraphics[width=0.95\columnwidth]{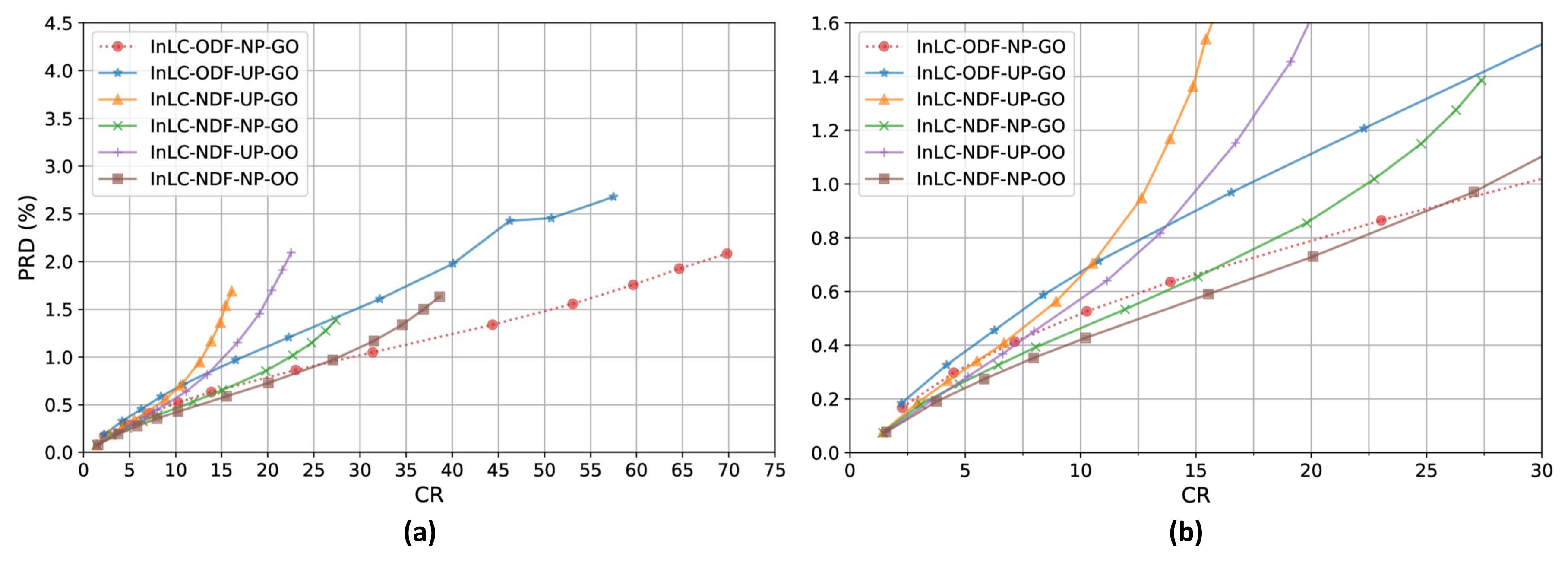}
	\caption{Average performance of the differently configured InLC schemata for all \SI{48}{} records in the MIT-BIH database (\SI{1}{\minute} ECG signals) with various threshold values: (\textbf{a}) Full scale CR vs. PRD. (\textbf{b}) Zoomed CR vs. PRD.}
	\label{fig:InLCPerformancePRD}
\end{figure}

\begin{figure}[t]
	\widefigure
	\centering
	\includegraphics[width=0.95\columnwidth]{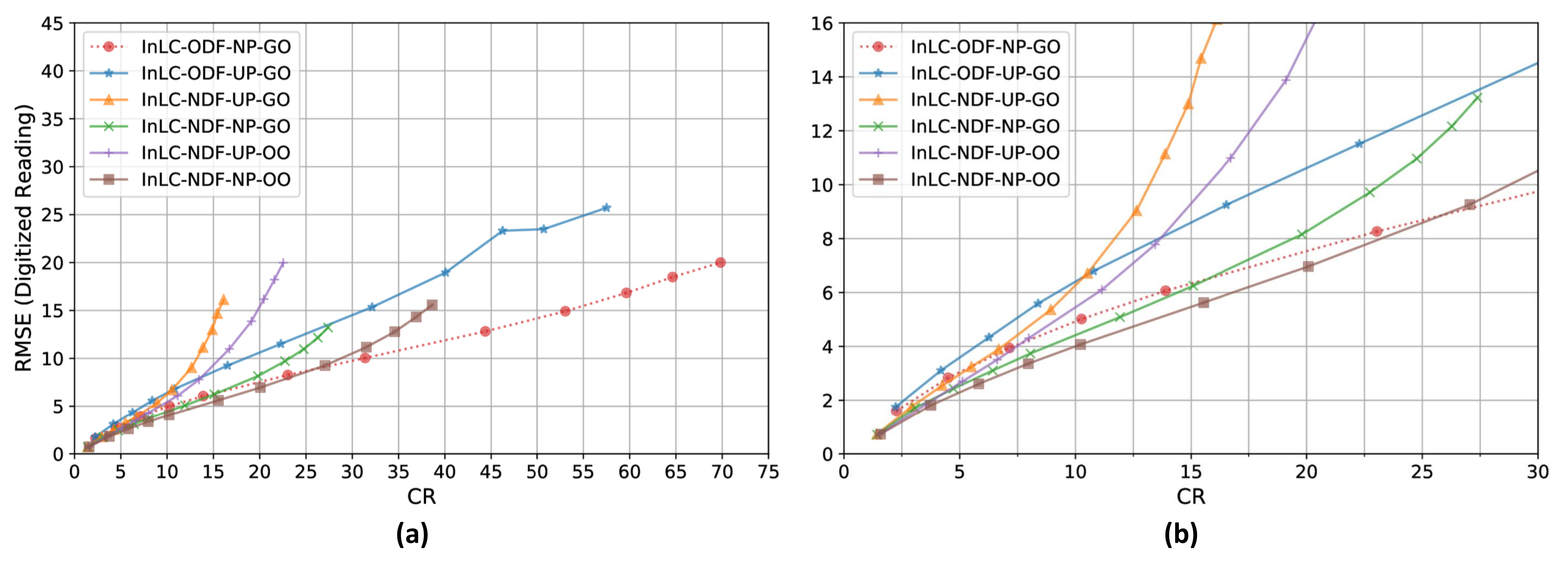}
	\caption{Average performance of the differently configured InLC schemata for all \SI{48}{} records in the MIT-BIH database (\SI{1}{\minute} ECG signals) with various threshold values: (\textbf{a}) Full scale CR vs. RMSE. (\textbf{b}) Zoomed CR vs. RMSE.}
	\label{fig:InLCPerformanceRMSE}
\end{figure}
\begin{paracol}{2}
\switchcolumn

Contrary to the speculation mentioned in Section~\ref{SubSubSect:AveragePerformanceLCs}, constantly updating the bank $\vec{B}_{b}$ by eliminating the statement ``if $|\vec{B}_{b}|<s_{b}$'' in the Step~\ref{step4} and the Step~\ref{step10} did not improve but rather impaired the compression efficiency at a certain level of distortion, especially when the acceptable distortion is comparatively high. Such conclusion was drawn by visually inspecting the CR vs. PRD and CR vs. RMSE curves in Figure~\ref{fig:InLCPerformancePRD} and Figure~\ref{fig:InLCPerformanceRMSE}. It is apparent that, given the other configurations remain identical, the curves of the InLC schemata with constantly updated $\vec{B}_{b}$ (i.e., notation ``UP'') were generally at the upper left of those with static $\vec{B}_{b}$ (i.e., notation ``NP''), i.e., lower CR at the same PRD or RMSE. However, as discussed in Section~\ref{SubSubSect:AveragePerformanceLCs}, the InLC schema with static $\vec{B}_{b}$ demonstrated better performance for the short-period ECG signals compared to that for the long-period ones. Therefore, the authors argued that higher compression efficiency at a certain level of distortion could be achieved by periodically, instead of constantly, updating the bank $\vec{B}_{b}$ of the InLC schema, e.g., re-initiating the $\vec{B}_{b}$ every \SI{1}{\minute} or \SI{5}{\minute}.

As for the pairs of CR vs. PRD or CR vs. RMSE curves from the InLC schemata with identical configurations except the distance functions, the schemata utilizing the Equation~\ref{eq:NewDist} generally resulted in lower PRD and RMSE values at low CR levels compared to those employing the Equation~\ref{eq:OldDist}. However, the PRD and RMSE values from the former schemata increased rapidly along with the CR levels and exceed those from the latter schemata at high CR levels. Such behavior is ascribed to the fact that the Equation~\ref{eq:NewDist} eliminates the occurrences of the CW-to-SL effect generalized in Figure~\ref{fig:IllustrationStraightLine} that would introduce substantial distortion, but constrains the possibilities of identifying a fairly long fragment $\vec{b}^{i}_{l}$ in $\vec{B}_{b}$ which is similar to the current fragment $\vec{f}_{l}$.

According to the empirical evidence described in Section~\ref{SubSubSect:AveragePerformanceLCs}, when utilizing the Equation~\ref{eq:NewDist} as distance function, the gain $g^{*}$ and offset $o^{*}$ (slope and intercept from OLS regression) of a fragment pair $\vec{f}^{*}$ and $\vec{b}^{*}$ can be simplified into a single offset, which is the average difference of the fragment pair, by assuming unity gain. Conceivably, better compression efficiency can be achieved with the gain eliminated. Besides, calculating the average difference of a fragment pair requires much less computing resources than performing OLS regression. Although this approach will introduce slight fluctuations to the reconstructed signal as illustrated in Figure~\ref{fig:InLCStraightLineIssue}, in which case the distortion might be increased, the InLC schemata transmitting only the new offset exhibited higher CR at the same PRD and RMSE compared to those transmitting both gain and offset (CR vs. PRD or CR vs. RMSE curve pairs with identical distance function and $\vec{B}_{b}$ setup).

\subsubsection{Discussion and Future Works}
\label{SubSubSect:DiscussionAndFutureWorksLC}

In summary, the proposed InLC schema outperformed the other three recognized lossy compression schemata, in terms of compression efficiency and distortion (at least doubled CR at the same PRD or RMSE), for the short- and long-period ECG signals from the MIT-BIH database. In addition, the InLC schema requires much less memory (three \SI{2}{\kilo\byte} buffers) and computing (distance function with inexpensive operators) resources to perform compression, making it ideal for wearable devices with limited storage and computational capacity. While maintaining a low distortion level of the InLC schema, higher compression efficiency can be achieved by employing the Equation~\ref{eq:NewDist} distance function, updating the bank $\vec{B}_{b}$ periodically (not constantly), and transmitting the new offset only (average difference of a fragment pair). Deploying the InLC schema on the wearable ECG device implemented in Section~\ref{SubSubSect:WearableECGDevice} and evaluating its energy efficiency in the field are enthusiastically awaited. Notice that the future iteration of the proposed personalized healthcare system (detailed in Section~\ref{SubSect:SystemImplementation}) will employ the deep neural networks for automated diagnosis, for example, the ResNet-based heart arrhythmia classifier proposed in Section~\ref{SubSect:AutomaticAFDetector}. In the future research, it is of paramount importance to assess the impact from a lossy compression schema in the diagnostic results. Moreover, it is also worth exploring how the deep neural networks' hidden features could improve the efficiency of a lossy compression schema without influencing the diagnostic results.

\subsection{Performance Evaluation of the ResNet-Based AF Detector}
\label{SubSect:PerformanceResNetAFDetector}

The proposed ResNet-based heart arrhythmia classifier specifically designed for AF detection was implemented using TensorFlow 1.1 and Python 2.7. The neural network models were trained and tested on a high-performance computing cluster equipped with multiple Nvidia GeForce RTX 2080 Ti graphics cards. Each card consists of a Turing\textsuperscript{TM} graphics processing unit (GPU) and \SI{11}{\giga\byte} of RAM, and was fully utilized for training a single model. In the remainder of this section, the setups for training, validation, and testing are first introduced in Section~\ref{SubSubSect:TrainValidTestSetAFDtector}. Then, three voting strategies were proposed in Section~\ref{SubSubSect:VotingAFDtector} to improve the classification performance, the $F_{1}$ measure to be specific, of the arrhythmia classifier. The final classification performance on the testing dataset (subset of the public dataset from the CinC challenge) is summarized in Section~\ref{SubSubSect:PerformanceAFDetector}; an average $F_{1}=\SI{85.10}{\percent}$ and a best $F_{1}=\SI{87.31}{\percent}$ were achieved by the ResNet-based AF detector, while the $F_{1}$ measures of the first-place participants in the CinC challenge ranged from \SIrange{78}{85}{\percent}. A variation of the proposed heart arrhythmia classifier and the future research are discussed in Section~\ref{SubSubSect:DiscussionFutureWorkAFDetector}.

\subsubsection{Training, Validation, and Testing Setups}
\label{SubSubSect:TrainValidTestSetAFDtector}

Referring to Section~\ref{SubSubSect:AFDatasetAndMetrics}, the public dataset from the CinC challenge consists of \SI{5977}{} records of \SI{30}{\second} ECG signals. These ECG records were randomly divided into two subsets, namely the training dataset with \SI{5500}{} records and the testing dataset with \SI{477}{} records, while maintaining classification profiles similar to the entire public dataset as listed in Table~\ref{Tab:AFDatasetProfile}. In the training dataset, the numbers of records of the normal rhythm, AF rhythm, others rhythm, and noise were \SI{3400}{} (\SI{61.8}{\percent}), \SI{460}{} (\SI{8.4}{\percent}), \SI{1530}{} (\SI{27.8}{\percent}), and \SI{110}{} (\SI{2.0}{\percent}), respectively. The corresponding numbers of the testing dataset were \SI{278}{} (\SI{58.3}{\percent}), \SI{39}{} (\SI{8.2}{\percent}), \SI{145}{} (\SI{30.4}{\percent}), and \SI{15}{} (\SI{3.1}{\percent}), respectively.

In the training phase, the 10-fold cross-validation method was adopted; each fold of ECG records and the training dataset yielded identical classification profile. Additionally, the mini-batch stochastic gradient descent with momentum (SGDM) optimizer was utilized for training the neural network models; the training processes were terminated when the validation loss was settled. Preliminary results showed that the models using \SI{64}{}-record mini-batches with classification profile similar to that of the training dataset\footnote{In this work, the size of a mini-batch was empirically determined to achieve good trade-off between training time and classification performance; each 64-record mini-batch contained \SI{40}{} normal classes (\SI{62.5}{\percent}), \SI{5}{} AF classes (\SI{7.8}{\percent}), \SI{18}{} others classes (\SI{28.1}{\percent}), and \SI{1}{} noise class (\SI{1.6}{\percent}).} generally resulted in better classification performance on the testing dataset compared to those using mini-batches with random classification profiles. It is expected because the randomly selected mini-batches usually contained few or even no ECG records of the AF or the noise classes due to the limited number of these rhythm classes in the training dataset, in which case the trained models would be biased.

\subsubsection{Voting Strategies}
\label{SubSubSect:VotingAFDtector}

In order to fully utilize the information provided by the training dataset, three voting strategies were applied on the predicted classes from the ten models, which were trained in a 10-fold cross-validation experiment, when classifying the ECG records in the testing dataset. These strategies are stated as follows:
\begin{itemize}
	\item Strategy 1: Take the predicted class with maximum occurrence;

	\item Strategy 2: Take the predicted class with maximum probability;

	\item Strategy 3: Take the predicted class with maximum occurrence if there exists only one; otherwise, take the predicted class with maximum probability.
\end{itemize}

\subsubsection{Classification Performance on Testing Dataset}
\label{SubSubSect:PerformanceAFDetector}

The classification performances, including the saturated top-1 accuracies and $F_{1}$ measures of the training, validation, and testing datasets, of a 10-fold cross-validation experiment are listed in Table~\ref{Tab:AFDetectorFirst10FoldResult}. The mean values of the saturated testing top-1 accuracy and $F_{1}$ measure were $\overline{A_{1}}=\SI{86.12}{\percent}$ and $\overline{F_{1}}=\SI{82.71}{\percent}$, respectively, while their variations were negligible ($CV\leq\SI{1.0}{\percent}$). The negligible $CV$ values suggested that these results were representative for estimating the true performance of the proposed AF detector. Applying the three voting strategies (detailed in Section~\ref{SubSubSect:VotingAFDtector}) on the predicted classes from the ten models in a 10-fold cross-validation experiment generally improved the saturated testing $A_{1}$ and $F_{1}$ compared to their mean values. As presented in Table~\ref{Tab:AFDetectorFirst10FoldStrategies}, the maximum improvement was achieved while employing the third strategy with final saturated testing $A_{1}=\SI{87.00}{\percent}$ and $F_{1}=\SI{83.35}{\percent}$.

The 10-fold cross-validation experiment had been repeated for five times in this work, and the training and testing datasets in each experiment were randomly initiated according to the setups stated in Section~\ref{SubSubSect:TrainValidTestSetAFDtector}. The results are presented in Section~\ref{SubApped:AFDetectorPerformanceFiveExperiments}. In accordance with Table~\ref{Tab:AFDetectorFive10FoldResult}, the mean values of the saturated testing top-1 accuracies and $F_{1}$ measures\footnote{The values of $\overline{A_{1}}$ and $\overline{F_{1}}$ listed in Table~\ref{Tab:AFDetectorFive10FoldResult} were the mean values achieved by averaging the values of $A_{1}$ and $F_{1}$ from the ten folds.} over these five 10-fold experiments were $\overline{A_{1}}=\SI{87.00}{\percent}$ and $\overline{F_{1}}=\SI{83.69}{\percent}$, respectively. Their variations were also negligible with $CV<\SI{1}{\percent}$. Consequently, it can be concluded that, without any voting strategy, the classification performance of the ResNet-based AF detector was $A_{1}=\SI{87.00}{\percent}$ or $F_{1}=\SI{83.69}{\percent}$. However, as presented in Table~\ref{Tab:AFDetectorFive10FoldStrategies}, applying the voting strategies in each 10-fold experiment generally improved the final classification performance on the testing dataset, and the final values of testing $A_{1}$ and $F_{1}$ from experiments utilizing the third voting strategy were usually the largest. Overall, the average classification performance of the ResNet-based AF detector utilizing the third voting strategy was $A_{1}=\SI{88.05}{\percent}$ or $F_{1}=\SI{85.10}{\percent}$ ($CV<\SI{1.7}{\percent}$), while the best performance achieved in this work was $A_{1}=\SI{88.89}{\percent}$ or $F_{1}=\SI{87.31}{\percent}$ (second 10-fold experiment). As listed in Table~\ref{Tab:AFDetectorPerformanceComparison}, the average $F_{1}$ measure ($F_{1}=\SI{85.10}{\percent}$) achieved in this work was \SIrange{1.3}{9.1}{\percent} larger than those obtained by the first-place participants in the CinC challenge, except the third approach ($F_{1}=\SI{85}{\percent}$) proposed by Teijeiro et al. \cite{teijeiro2017arrhythmia}. Notice that the aforementioned approach utilized three different classifiers to predict an ECG record; therefore, its complexity is much higher than the ResNet-based AF detector. A simple classifier is more preferable as it unlocks the potential of being deployed on the edge devices, e.g., the smart phones and the wearable ECG monitoring devices, such that the off-line classification and the prompt pre-diagnosis are achievable.

\begin{specialtable}[H]
\caption{Comparison of classification performance between the ResNet-based AF detector and those approaches proposed by the first-place participants of the CinC challenge.}
\label{Tab:AFDetectorPerformanceComparison}
\small
\begin{tabular}{cccccc}
\toprule
& \multicolumn{5}{c}{\textbf{Summary of Classification Performance (\%)}} \\
\cmidrule[0.6pt]{2-6}
& \textbf{$F_{1}(N)$} & \textbf{$F_{1}(A)$} & \textbf{$F_{1}(O)$} & \textbf{$F_{1}(\sim)$} & \textbf{$F_{1}$} \\
\midrule
\textbf{Datta et al.} \cite{datta2017identifying}
	& 90.95 & 79.78 & 77.19 & -- & 82.64 \\
\midrule
\textbf{Zabihi et al.} \cite{zabihi2017detection}
	& 90.87 & 83.51 & 73.41 & 50.42 & 83 \\
\midrule
\textbf{Mahajan et al.} \cite{mahajan2017cardiac}
	& 91 & 74 & 70 & -- & 78 \\
\midrule
\multirow{3}{*}{\textbf{Teijeiro et al.} \cite{teijeiro2017arrhythmia}}
	& -- & -- & -- & -- & 83 \\
	& -- & -- & -- & -- & 83 \\
	& -- & -- & -- & -- & \textbf{85} \\
\midrule
\multirow{2}{*}{\textbf{Hong et al.} \cite{hong2017encase}}
	& 92 & 84 & 74 & -- & 83 \\
	& 92 & 85 & 74 & -- & 84 \\
\midrule
\textbf{This Work (Best)}
	& \textbf{93.49} & \textbf{88.10} & \textbf{80.34} & \textbf{54.55} & \textbf{87.31} \\
\textbf{{This Work} (Average)}
	& \textbf{93.03} & 83.78 & \textbf{78.47} & \textbf{65.48} & \textbf{85.10} \\
\bottomrule
\end{tabular}
\\
\footnotesize
Note that the standard deviations of the average $F_{1}(N)$, $F_{1}(A)$, $F_{1}(O)$, and $F_{1}(\sim)$ from this work were \SI{0.69}{\percent}, \SI{3.08}{\percent}, \SI{1.47}{\percent}, and \SI{13.25}{\percent}, respectively. The classification performances of the approaches proposed by the first-place participants in the challenge were obtained using the public dataset with labels in version 2 \cite{clifford2017af}.
\end{specialtable}

\subsubsection{Discussion and Future Works}
\label{SubSubSect:DiscussionFutureWorkAFDetector}

Utilizing the \SI{30}{\second} ECG records in the public dataset of the CinC challenge, the ResNet-based AF detector proposed in this work achieved an average and a best $F_{1}$ measures of $F_{1}=\SI{85.10}{\percent}$ and $F_{1}=\SI{87.31}{\percent}$, respectively. Compared to the first-place participants in the CinC challenge, this work obtained the highest $F_{1}$ measure with one single classifier (\SIrange{1.3}{9.1}{\percent} improvement of average $F_{1}$). Although only the \SI{30}{\second} records (\SI{70}{\percent} of the public dataset) were used for training and testing the ResNet, the performance results are considered representative as utilizing more information during training usually results in better performance given that the inputs are not biased.

Since the core component of the proposed AF detector is a ResNet, any techniques that can transform the time series ECG signals into 2D images are applicable for replacing the STFT preprocessing block illustrated in Figure~\ref{fig:AFDetectorStructureDiagram}. A new heart arrhythmia classifier based on the ResNet is then established. For example, the recurrence plot is a powerful tool proposed by Eckmann et al. \cite{eckmann1995recurrence} to characterize the time constancy of dynamical systems. Based on the recurrence plot, a number of applied and theoretical research has been conducted \cite{zhou2007cluster,liu2020fractal}. The preliminary experimental results showed that utilizing the recurrence plots of the time series ECG signals as the ResNet's inputs produced promising results. An example recurrence plot is depicted in Figure~\ref{fig:RecurrencePlotExample}; the element at position $(i,j)$ of the recurrence plot is the difference\footnote{The absolute $L^{\infty}$ norm between two ECG segments was used in the preliminary experiment.} between the ECG segments with length $l$ at indexes $i$ and $j$. Empirically, improved classification performance of the ResNet was achievable by eliminating the ECG signals' DC components and high-frequency noises before transforming into recurrence plots. In the preliminary experiment, the empirical mode decomposition (EMD) was employed to filter the ECG signals. It is known as an effective method to capture the meaningful components, defined as the intrinsic mode functions (IMFs), from a given time series signal \cite{huang1998empirical}. Its effectiveness has been demonstrated in many applied studies \cite{zhou2010empirical,zhou2013empirical,leung2019integration,jiang2021long}. With proper filtering on the \SI{30}{\second} ECG signals and appropriate downsizing on the converted recurrence plots\footnote{The filtered ECG signal contained the third to the fifth IMFs from the EMD; the recurrence plot was downsized by dividing the ECG signal into segments with length $l=45$ and calculating the differences between segments.}, the new arrhythmia classifier's preliminary classification performance\footnote{The Adam optimizer was used in the preliminary experiment; better performance is anticipated with fine-tuned SGDM optimizer and conversion parameters (filter, segment length, etc.).} was $A_{1}=\SI{83.82}{\percent}$ or $F_{1}=\SI{80.21}{\percent}$. The authors anticipate that higher accuracy is achievable in classifying an ECG record by combing the predicted results from both heart arrhythmia classifiers. An automated AF detector integrating both classifiers and deploying such AF detector in the personalized healthcare system implemented in Section~\ref{SubSect:SystemImplementation} are enthusiastically awaited. Nevertheless, as presented in Section~\ref{SubSubSect:DiscussionAndFutureWorksLC}, the mutual influence between the performance of a lossy compression schema (i.e., the InLC proposed in Section~\ref{SubSubSect:IntuitiveLossy}) and the diagnostic results from a deep neural network (i.e., the ResNet-based heart arrhythmia classifier) is worth investigating in the future research.

\begin{figure}[H]
	\centering
	\includegraphics[width=0.9\columnwidth]{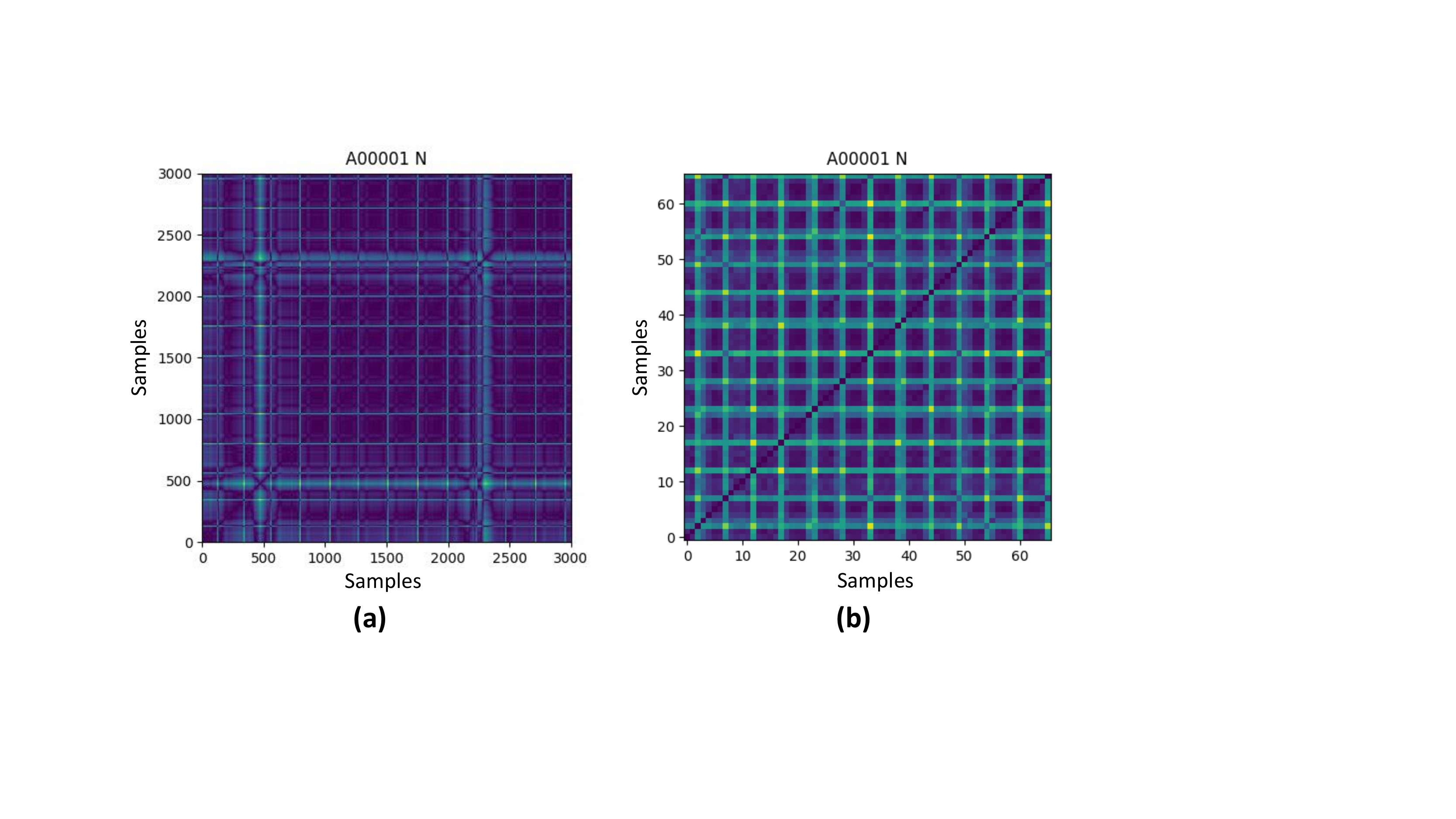}
	\caption{Recurrence plot example using Record A00001 (normal class) from the public dataset of the CinC challenge: (\textbf{a}) Converted from the first \SI{10}{\second} of the original ECG signal with $l=1$. (\textbf{b}) Converted from the first \SI{10}{\second} of the filtered ECG signal with $l=45$.}
	\label{fig:RecurrencePlotExample}
\end{figure}

\section{Conclusions}
\label{Sect:Conclusion}

Extended form an earlier IoT-based e-health system, which aimed at studying the association between HRV and the risks of stroke, CVDs, etc., a personalized healthcare system embodying a wearable ECG monitoring device for data acquisition, a mobile application for data visualization and relay, and a back-end server for data management and analytics, was developed in this manuscript. This system provides a convenient way for the users to keep track of their health conditions using wearable devices and smart phones. They are continuously monitored by the back-end server which provides timely health warnings and feedbacks. It also enables the health advisors to monitor their paired users' ECG information remotely and communicate with them for further diagnosis and interventions. The wearable ECG devices have been evaluated against a medical equipment and demonstrated excellent intra-consistency ($\overline{CV_{RMS}}\approx\SI{5.5}{\percent}$), acceptable inter-consistency ($\overline{CV_{RMS}}\approx\SI{12.1}{\percent}$), and negligible RR-interval errors ($ARE<\SI{1.4}{\percent}$). Empirically confirmed by two professional cardiologists, the quality of ECG signals acquired by the wearable ECG device is sufficient for cardiac arrhythmia diagnosis given that the rhythms have distinguishable criteria established from the Lead II.

To boost the battery life of the wearable ECG device for continuous and long-term monitoring, a lossy compression schema named InLC was proposed in this manuscript. Higher energy efficiency is achieved by compressing the acquired ECG signals before sending to the smart phones. The InLC schema outperformed the other three recognized lossy compression schemata in terms of compression efficiency and distortion (at least $2\times$ the CR at a certain PRD or RMSE) for the ECG signals from the MIT-BIH database. Compared with the recognized schemata, it eliminates the use of R-peak detector for segmenting the ECG signals and requires much less memory (three \SI{2}{\kilo\byte} buffers) and computing (inexpensive distance and offset functions) resources to perform compression. Besides, the intrinsic time lag of the InLC schema is within a few seconds (determined by $s_{f}$). Overall, it is ideal for the wearable ECG devices with hardware constrains and application with real-time requirement. Implementing the InLC schema on the wearable ECG device and investigating its field performances are enthusiastically awaited.

To enable the automated diagnostic/screening capability in the personalized healthcare system, a ResNet-based heart arrhythmia classifier specifically designed for AF detection was developed. This AF detector is capable of identifying the arrhythmia type of a short-period single-lead ECG signal with high sensitivity and accuracy, which is the key for continuous monitoring. For the \SI{30}{\second} ECG records from the CinC challenge, the average and the best testing $F_{1}$ measures of the AF detector were $F_{1}=\SI{85.10}{\percent}$ and $F_{1}=\SI{87.31}{\percent}$, respectively, while the $F_{1}$ measures obtained by the first-place participants in the challenge ranged from \SIrange{78}{85}{\percent}. It is anticipated that, with such heart arrhythmia classifier being deployed on the personalized healthcare system, timely warnings and pre-diagnoses are accessible to the users and health advisors round the clock. Further diagnosis and intervention from the informed health advisors are achievable in a more labor- and time-efficient way, such that the users could take early treatments before the development of any inalterable life-threatening conditions/diseases. The avenues for future research including improving the classification performance of the AF detector, implementing the AF detector in the proposed personalized healthcare system, as well as investigating the mutual influence between the performance of the InLC schema and the diagnostic results from the AF detector, are also discussed.

\vspace{6pt} 



\authorcontributions{Conceptualization, K.S. Woo, K.S. Leung, Y. Leung, P.W. Lee; methodology, W.Y. Yi, P.F. Liu, J.M. Chen, and S.L. Lo; software, W.Y. Yi, P.F. Liu, J.M. Chen, and S.L. Lo; validation, W.Y. Yi, P.F. Liu, S.L. Lo, Y.F. Chan, and Y. Zhou; formal analysis, W.Y. Yi, P.F. Liu, and Y. Zhou; investigation, W.Y. Yi, P.F. Liu, and Y.F. Chan; resources, K.S. Woo, K.S. Leung, Y. Leung, and P.W. Lee; data curation, W.Y. Yi and P.F. Liu; writing---original draft preparation, W.Y. Yi and P.F. Liu; writing---review and editing, S.L. Lo, Y.F. Chan, Y. Zhou, K.S. Leung, Y. Leung, K.S. Woo, and P.W. Lee; visualization, W.Y. Yi and P.F. Liu; supervision, K.S. Leung, Y. Leung, and K.S. Woo; project administration, K.S. Leung, Y. Leung, and K.S. Woo; funding acquisition, K.S. Woo, Y. Leung, and K.S. Leung. All authors have read and agreed to the published version of the manuscript.}

\funding{This research was funded by the Faculty Strategic Development Fund of The Faculty of Social Science of The Chinese University of hong Kong. It was also partially sponsored by the Dr Stanley Ho Medical Development Foundation for Health Promotion Activities and Related Research.}

\acknowledgments{The authors would like to thank Mr Xian-Yang Peng and Mr Xin-Lin Wu, Department of Computer Science and Engineering of The Chinese University of Hong Kong, for developing the mobile application and back-end server for the personalized healthcare system. We also showed our gratitude to Ms Ada Yu, Department of Medicine and Therapeutics of The Chinese University of Hong Kong, who provided great assistances during the evaluation of the wearable ECG device. We further acknowledged Mr Wei Li and Mr Atif Khurshid, Department of Computer Science and Engineering of The Chinese University of Hong Kong, for implementing the preliminary version of the ResNet-based AF detector.}

\conflictsofinterest{The authors declare no conflict of interest.}


\clearpage
\abbreviations{The following abbreviations are used in this manuscript:\\
\noindent 
\begin{tabular}{@{}ll}
IoT 	& Internet of things \\
MEMS 	& micro-electro-mechanical system \\
ECG 	& electrocardiography \\
AF 		& atrial fibrillation \\
CVD 	& cardiovascular disease \\
WHO 	& World Health Organization \\
NN 		& neural network \\
EEG 	& electroencephalography \\
HRV 	& heart rate variability \\
BLE 	& Bluetooth low energy \\

AZTEC 	& amplitude zone time epoch coding \\
TP 		& tunning point \\
CORTES 	& coordinate reduction time encoding system \\
SAPA 	& scan-along polygonal approximation \\
LTC 	& lightweight temporal compression \\
KLT 	& Karhunen-Lo\`eve transform \\
FFT	 	& fast Fourier transform \\
DWT 	& discrete wavelet transform \\
DCT 	& discrete cosine transform \\
FWT 	& fast Walsh transform \\
PCA 	& principal component analysis \\
OD 		& on-line dictionary \\
GSVQ 	& gain-shape vector quantization \\
SOMP 	& simultaneous orthogonal matching pursuit \\
BSBL 	& block sparse Bayesian learning \\
AE 		& autoencoder \\
DC 		& direct current \\
DyWT 	& dyadic wavelet transform \\
LBG 	& Linde-Buzo-Gray \\
InLC 	& intuitive lossy compression \\
OLS 	& ordinary least-squares \\

CR 		& compression ratio \\
PRD 	& percentage root-mean-square difference \\
PRDN 	& percentage root-mean-square difference normalized \\
RMSE 	& root-mean-square error \\
RMSEP 	& root-mean-square error peak-to-peak \\
SNR 	& signal-to-noise ratio \\
MAE 	& maximum amplitude error \\
CC 		& cross correlation \\
CDI 	& clinical distortion index \\
WDD 	& weighted diagnosis distortion \\
DDM 	& diagnostic distortion measure \\
QS 		& quality score \\

DNN 	& deep neural network \\ 
CNN 	& convolutional neural network \\
CRNN 	& convolutional recurrent neural network \\
LSTM 	& long-short term memory \\
ResNet 	& deep residual network \\
STFT 	& short-time Fourier transform \\ 
\end{tabular}

\begin{tabular}{@{}ll}
ADC 	& analog-to-digital converter \\
MHz 	& mega hertz \\
KB 		& kilobyte \\
SRAM 	& static random access memory \\
SoC 	& system on a chip \\
PCB 	& printed circuit board \\
FPCB 	& flexible printed circuit board \\
Li-Po 	& lithium-polymer \\
GUI 	& graphical user interface \\
TR 		& transmission and reception \\
MAC 	& media access control \\
HTTP 	& hypertext transfer protocol \\
API 	& application programming interface \\
URL 	& uniform resource location \\
HTTPS 	& hypertext transfer protocol secure \\

GPU		& graphics processing unit \\
SGDM 	& stochastic gradient descent with momentum \\
IMF 	& intrinsic mode function \\
EMD 	& empirical mode decomposition \\

bpm 	& beats per minute \\
RL 		& right leg \\
RA 		& right arm \\
LL 		& left leg \\
RC 		& resistor-capacitor \\
SD 		& standard deviation \\
ARE 	& absolute relative error \\
CV 		& coefficient of variation \\
RMS 	& root-mean-square \\
AFL 	& atrial flutter \\
VF 		& ventricular fibrillation \\
LBBB 	& left bundle branch block \\
RBBB 	& right bundle branch block \\
AV 		& atrioventricular \\
PVC 	& premature Ventricular contraction \\
SVT 	& supraventricular tachycardia \\
VT 		& ventricular tachycardia \\

Adam 	& adaptive moment estimation \\
FS 		& full-scale \\
HS 		& half-scale \\
EXP 	& experiment \\
\end{tabular}}

\clearpage
\appendixtitles{yes} 
\appendixstart
\appendix

\section{Test and Evaluation Results of the Implemented ECG Device}
\label{Apped:EvaluationResultsOfECGDevice}

\subsection{Placement Test of the ECG Device}
\label{SubApped:ECGDevicePlacementTest}

In this experiment, the implemented ECG device was placed at seven different positions on three normal subjects' chests (two male and one female adults) and the acquired ECG signals were inspected visually. These positions were empirically selected covering different areas of the subject's chest. The intended purpose of this experiment is to identify the optimal position for the ECG device such that the P-wave, the QRS-complex, and the T-wave of a normal sinus rhythm can be clearly observed from the acquired ECG signals. The ECG of a heart in normal sinus rhythm showing standard P-wave, QRS-complex, and T-wave is depicted in Figure~\ref{fig:ECGBeat}. For illustration purpose, only the ECG signals from one of the male adults are presented in Figure~\ref{fig:ECGPlacement1}; the ECG signals from the other two adults manifested similar characteristics.

\begin{figure}[H]
	\centering
	\includegraphics[width=0.7\columnwidth]{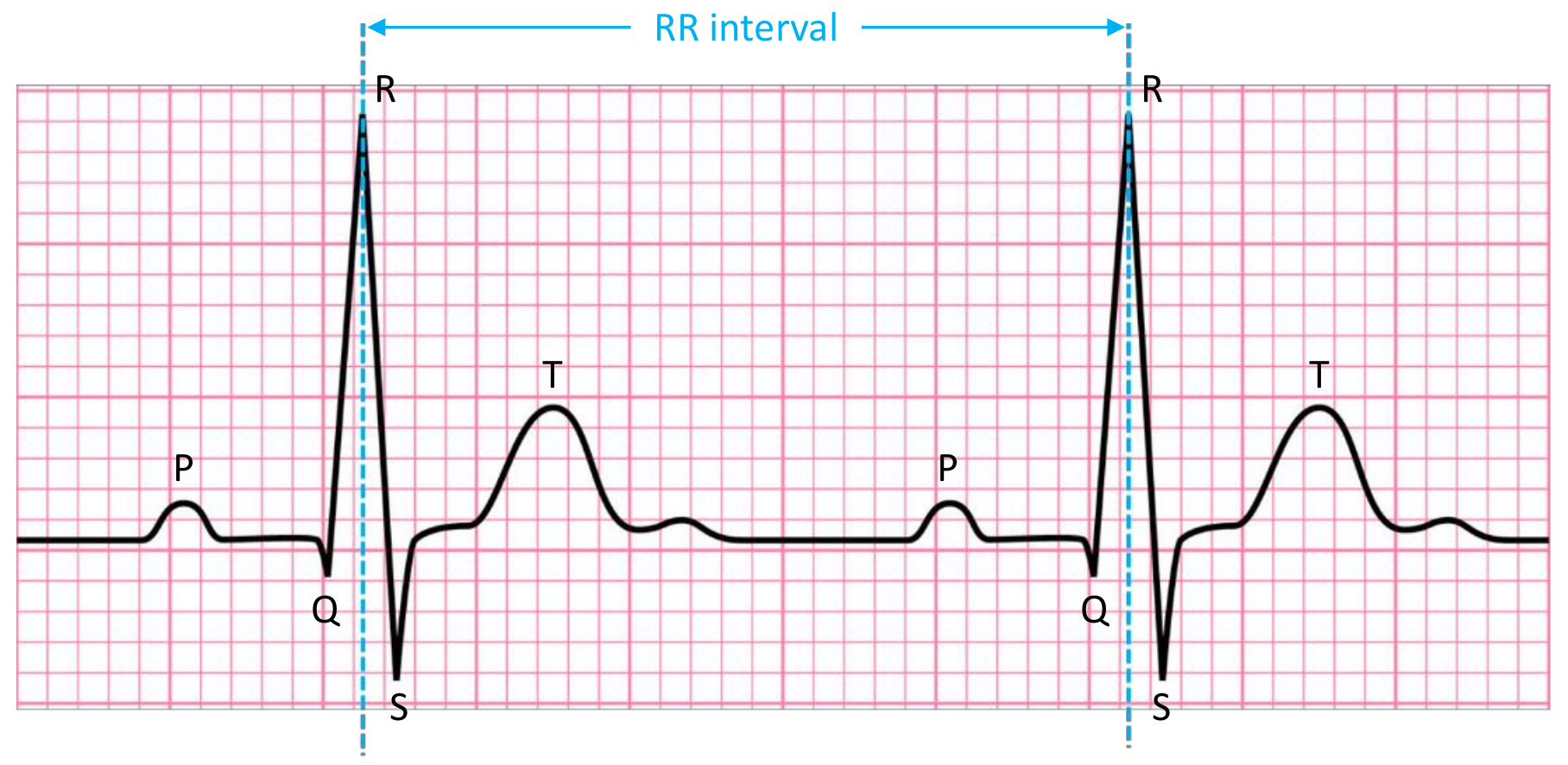}
	\caption{ECG of a heart in normal sinus rhythm showing standard P-wave, QRS-complex, and T-wave. The time interval between two adjacent R-peaks is represented as RR-interval.}
	\label{fig:ECGBeat}
\end{figure}

As illustrated in Figure~\ref{fig:ECGPlacement1}, no P-wave and T-wave can be observed from the ECG signals collected at Position 2 and Position 6. In contrast, the ECG signals from the remaining positions clearly exhibit the P-waves, QRS-complexes, and T-waves. These characteristics are essential for diagnosis of cardiac arrhythmia, e.g., detecting AF from the ECG signals. Given that the ECG signals from Position 1 have a waveform closest to the normal sinus rhythm as depicted in Figure~\ref{fig:ECGBeat}, it is considered to be the optimal placement position for the wearable ECG device. Notice that the R-peaks in the ECG signals from Position 2 have the largest magnitude among all positions. This position would be a good candidate when investigating the HRV only, because the enormous magnitude of the R-peaks makes distinguishing adjacent heartbeats much easier, e.g., using simple thresholding method.

It is noteworthy that better quality of ECG signals is achievable with proper skin preparation. For example, as illustrated in Figure~\ref{fig:ECGPlacement1Cont}a, slightly sanding the skin with the Red Dot\textsuperscript{TM} Trace Pre as well as adding the cardiogram conductive gel on the sticky pads before attaching the ECG device at Position 1 of subject's chest resulted in larger magnitude of R-peaks and more visually distinguishable P-waves compared to the ECG signals in Figure~\ref{fig:ECGPlacement1}a. In addition to attaching the wearable device on the chest, ECG signals can also be retrieved from a subject by putting his/her two thumbs on the device's electrodes (sticky pads in between) as indicated in Figure~\ref{fig:ECGPlacement1Cont}b. Although the retrieved ECG signals were a bit noisy, the R-peaks were still visually identifiable; it is sufficient for short-term HRV investigation.

In summary, attaching the wearable ECG device at Position 1 of a normal subject's chest produced an ECG waveform similar to the normal sinus rhythm with standard P-wave, QRS-complex, and T-wave. This position is selected as the optimal placement for the wearable device and should be strictly followed when collecting ECG records from the recruited subjects. Skin preparation should be performed whenever applicable.

\begin{figure}[H]
	\centering
	\includegraphics[width=0.95\columnwidth]{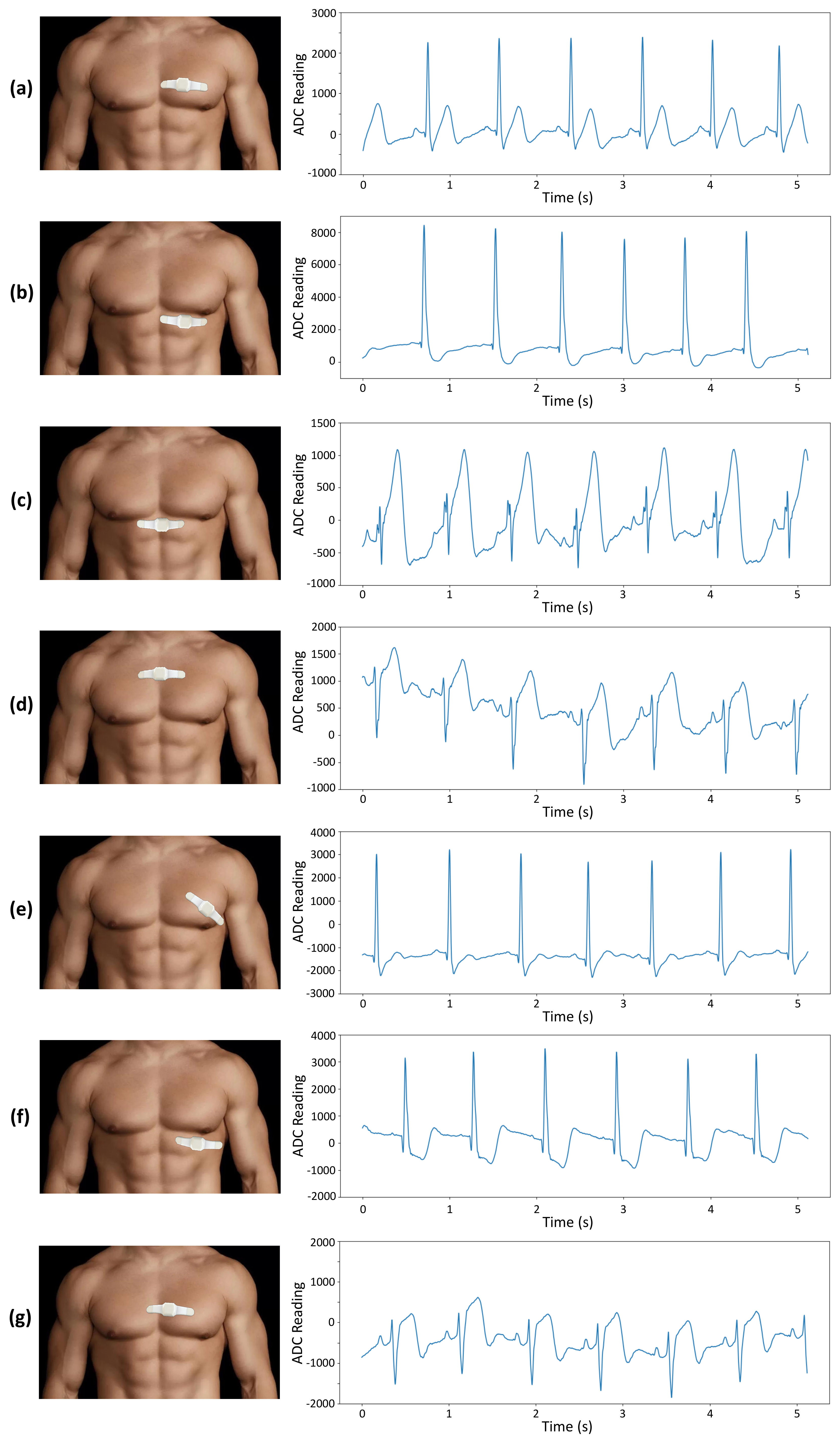}
	\caption{Placement test of the ECG device at: (\textbf{a}) Position 1. (\textbf{b}) Position 2. (\textbf{c}) Position 3. (\textbf{d}) Position 4. (\textbf{e}) Position 5. (\textbf{f}) Position 6. (\textbf{g}) Position 7. Images on the left are illustrating the device's positions on the subject's chest and graphs on the right are the ECG signals collected. The sticky pads are not shown.}
	\label{fig:ECGPlacement1}
\end{figure}

\begin{figure}[H]
	\centering
	\includegraphics[width=0.95\columnwidth]{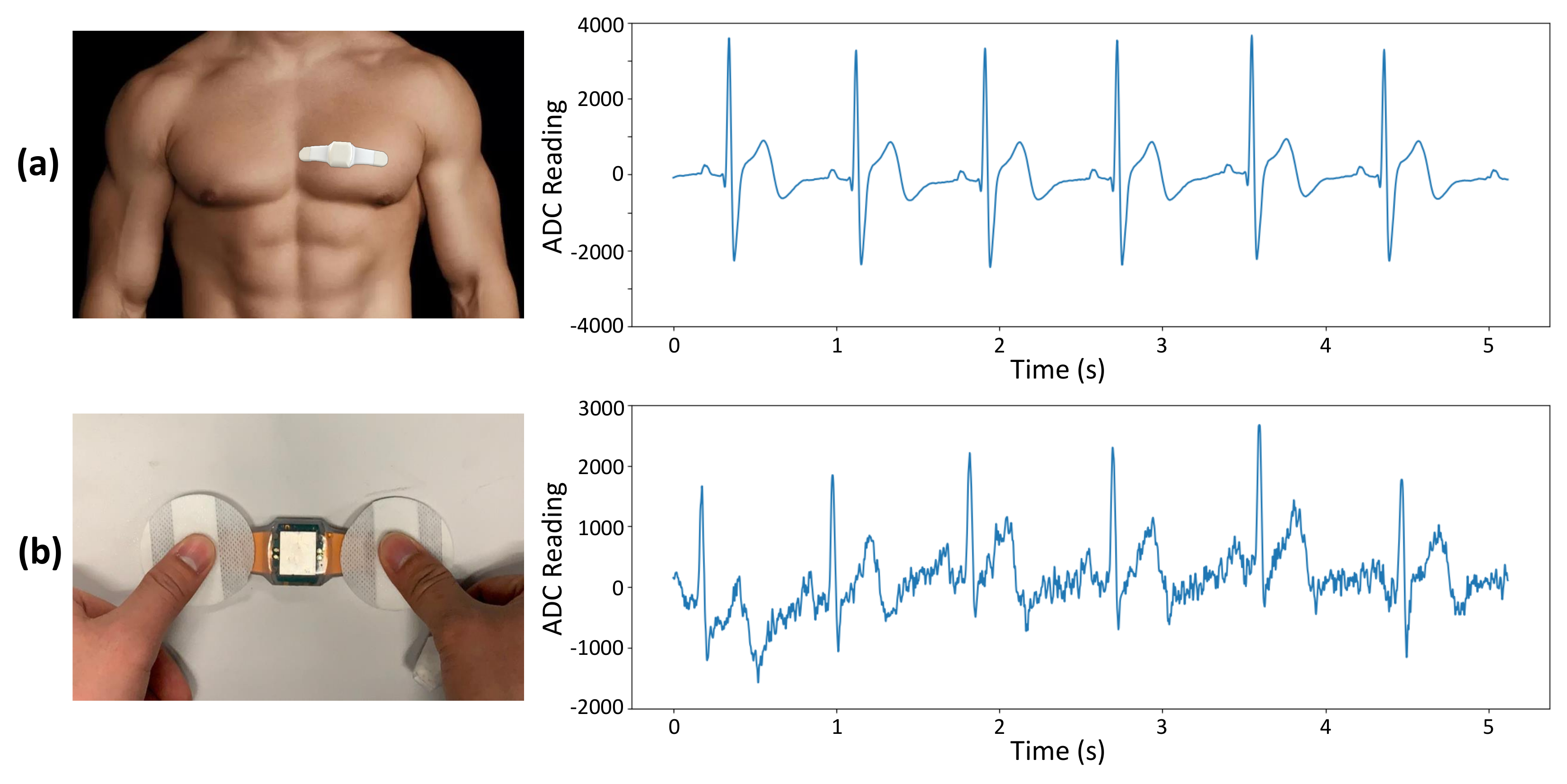}
	\caption{Placement test of the ECG device: (\textbf{a}) Position 1 with skin preparation (sticky pads are not shown). (\textbf{b}) The subject's two thumbs are placed on the sticky pads that are connected to the ECG device's electrodes.}
	\label{fig:ECGPlacement1Cont}
\end{figure}

\subsection{ECG Signal Disturbance from Subject Activities}
\label{SubApped:ECGSignalDisturbanceFromActivities}

It is well established that electrical signals will be generated in muscles during the contraction representing neuromuscular activities \cite{reaz2006techniques}. Such kind of signals will contaminate the acquired ECG signals \cite{clifford2006ecg,gacek2011ecg}. In this experiment, the muscle contractions related disturbances to the ECG signals acquired by the wearable device at Position 1 (see Figure~\ref{fig:ECGPlacement1}a) were investigated. Since the Position 1 is right on top of the pectoralis major, it is anticipated that the ECG signals acquired at this position will be heavily influenced by the subject activities. Therefore, the ECG signals acquired at Position 2 (see Figure~\ref{fig:ECGPlacement1}b), at which fewer major muscles are involved, were also investigated. As stated in Section~\ref{SubApped:ECGDevicePlacementTest}, the ECG signals acquired at this position manifest R-peaks with enormous magnitude but only provide the RR-interval information for HRV analysis.

The wearable device was attached at Position 1 and Position 2 of a male adult to recored the ECG signals. With the device attached at each position, the subject was asked to perform the following activities:
\begin{enumerate}
\item	Sitting at rest; 
\item	Walking;
\item	Typing/Writing;
\item	Jogging;
\item	Performing push-ups;
\item	Squatting;
\item	Performing chest flies;
\item	Swing arms;
\item	Wriggling;
\item	Performing sit-ups.
\end{enumerate}
A \SI{30}{\second} of ECG record was acquired for each activity at each position. For the sake of clarity, only \SI{5}{\second} of ECG signals of each record were inspected in this section.

Considering the ECG records acquired while sitting at rest as reference, it is apparent that the signals at Position 1 (graphs on the left columns of Figure~\ref{fig:ECGDuringActivities} and Figure~\ref{fig:ECGDuringActivitiesCont}) were significantly influenced by most of the activities except walking and typing/writing; the ECG signals during performing push-ups, chest flies, and sit-ups became so contaminated that even the R-peaks can hardly be identified. Although the noise components of the contaminated ECG signals can be filtered out utilizing various signal processing techniques \cite{asgari2017novel,hossain2020denoising}, it is out of the scope of this manuscript. Contrary to the Position 1, the ECG signals acquired at Position 2 (graphs on the right columns of Figure~\ref{fig:ECGDuringActivities} and Figure~\ref{fig:ECGDuringActivitiesCont}) exhibited clearly identifiable R-peaks for all the activities. Approved by two professional cardiologists, when recording subjects' RR-interval information for HRV analysis, the Position 2 is better than the Position 1 in terms of activities attributed noise immunity.

To sum up, the ECG signals acquired at Position 1 manifested clear P-waves, QRS-complexes, and T-waves while the subject was at rest or performing light activities; the signals were significantly contaminated during intensive activities. Clear R-peaks were identifiable from the ECG signals acquired at Position 2 during all the selected activities; the signals were sufficient for HRV analysis only.

\end{paracol}
\appendix
\begin{figure}[!t]	
	\widefigure
	\centering
	\includegraphics[width=0.95\columnwidth]{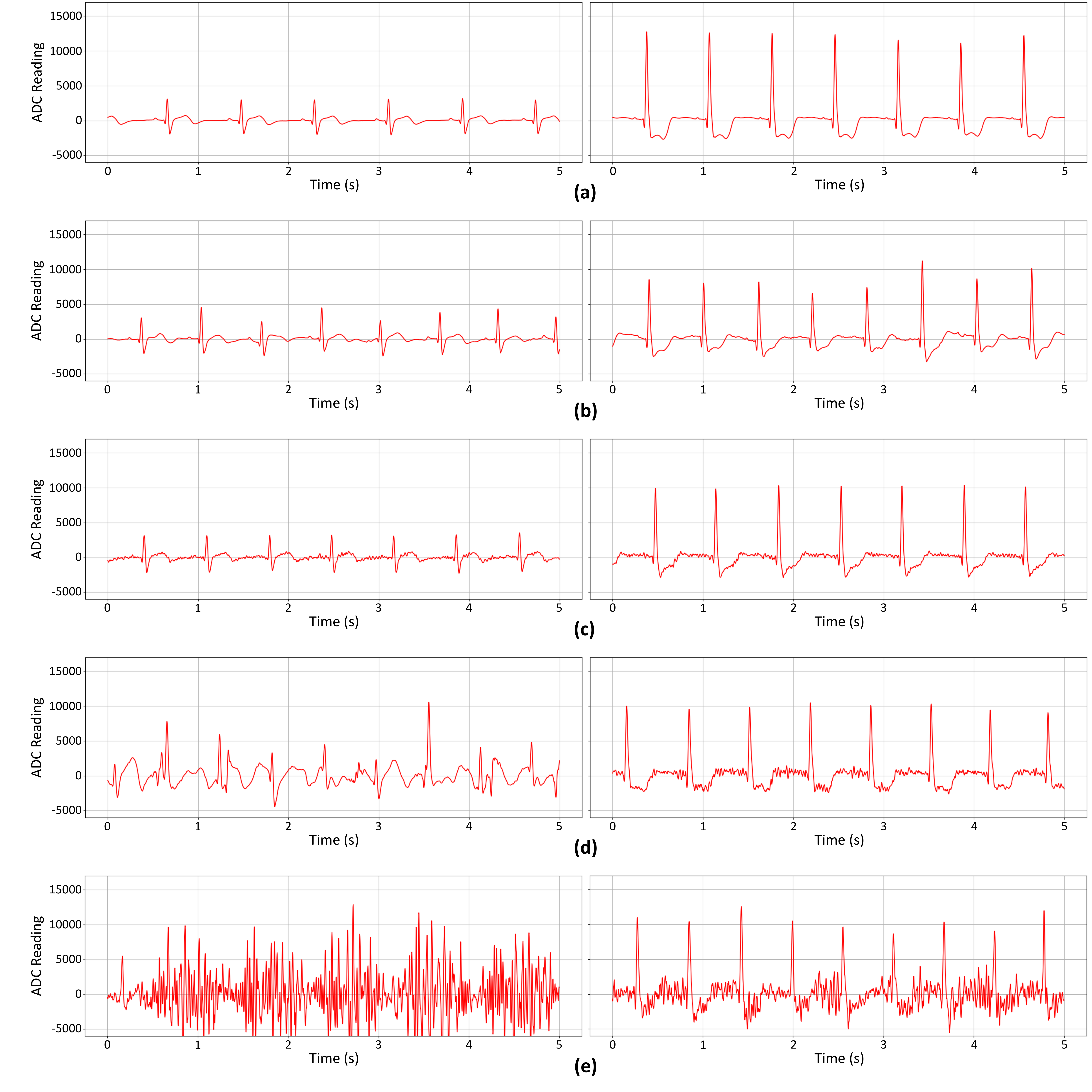}
	\caption{\SI{5}{\second} ECG signals collected by the wearable device at Position 1 (left) and Position 2 (right) when subject was performing different activities: (\textbf{a}) Sitting at rest. (\textbf{b}) Walking. (\textbf{c}) Typing/Writing. (\textbf{d}) Jogging. (\textbf{e}) Performing push-ups.}
	\label{fig:ECGDuringActivities}
\end{figure}

\begin{figure}[!t]	
	\widefigure
	\centering
	\includegraphics[width=0.95\columnwidth]{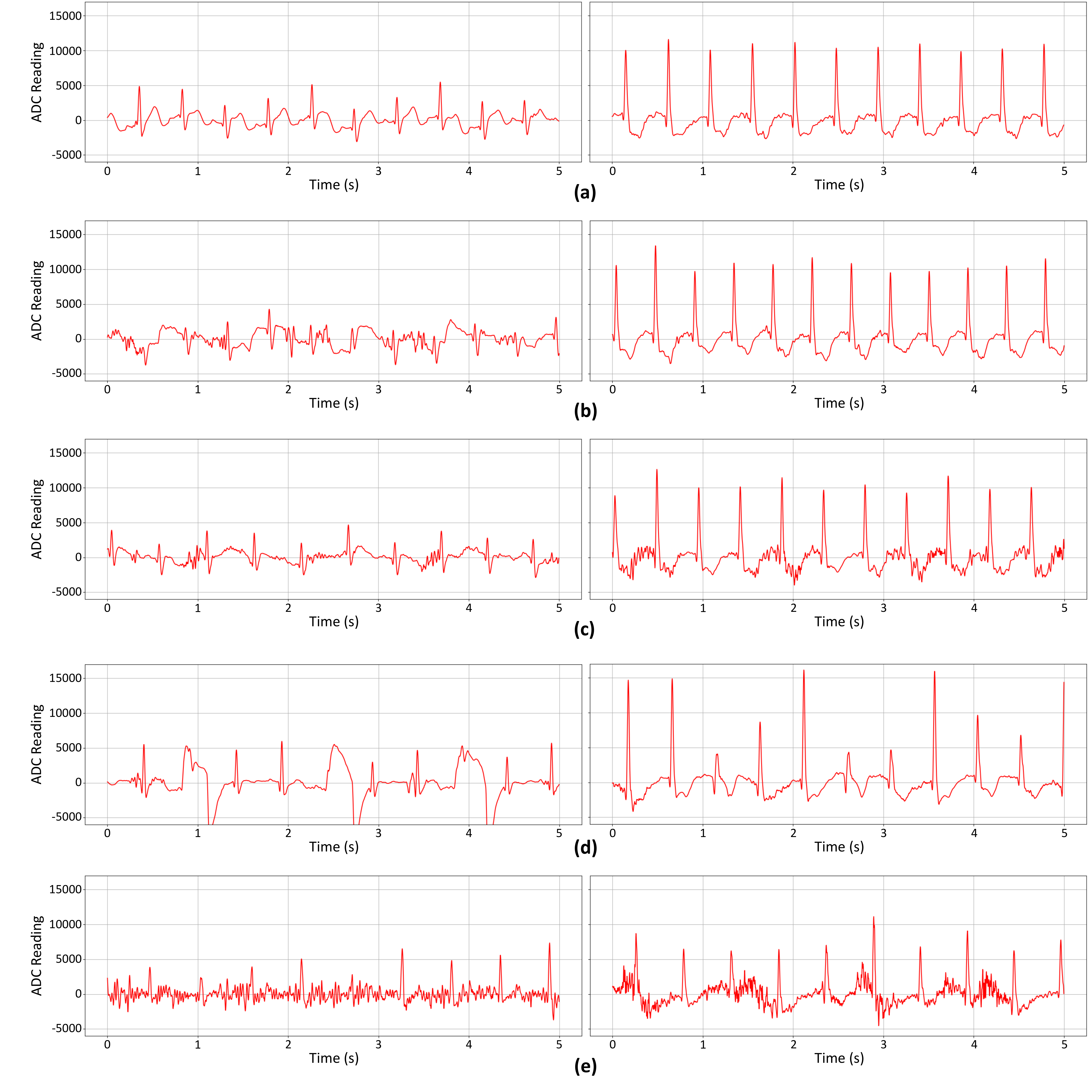}
	\caption{\SI{5}{\second} ECG signals collected by the wearable device at Position 1 (left) and Position 2 (right) when subject was performing different activities: (\textbf{a}) Squatting. (\textbf{b}) Performing chest flies. (\textbf{c}) Swing arms. (\textbf{d}) Wriggling. (\textbf{e}) Performing sit-ups.}
	\label{fig:ECGDuringActivitiesCont}
\end{figure}
\begin{paracol}{2}
\switchcolumn

\clearpage
\subsection{Comparison of ECG Signals from the Wearable Device and a Medical Equipment}
\label{SubApped:SignalComparisonBetweenDeviceAndEquipment}

In order to verify whether or not the quality of ECG signals collected by the wearable device is sufficient for cardiac arrhythmia diagnosis, its outputs were compared against that of the Philips PageWriter Trim II ECG machine\footnote{The sampling rate of the ECG machine is \SI{500}{\hertz}.} through visual inspection. This medical grade machine is used in the Prince of Wales Hospital in Hong Kong to provide diagnostic ECG of patients. All of the ECG waveforms collected by the wearable device and the medical machine were generated by the Fluke ProSim2 Vital Sign Simulator \cite{fluke2021patient} which is able to output normal sinus rhythms and arrhythmias at 12-lead configuration (referenced to right leg electrode, i.e., RL) with user adjustable ECG rate and amplitude.

The ECG simulator was configured to generate normal sinus rhythm waveforms at \SI{60}{}, \SI{80}{}, and \SI{100}{} beats per minutes (bpm) with \SI{1}{\milli\volt} amplitude. In addition, a sinusoid performance waveform at \SI{5}{\hertz} within \SIrange{0}{1}{\milli\volt} was also generated for comparison. For each configuration, the ECG simulator was first connected to the ECG device and then the ECG machine; about \SI{10}{\second} of data of each waveform were collected. The waveforms generated by the ECG simulator under specific configuration were assumed to be identical throughout the experiment.

Due to regulation of the hospital, the direct digital outputs from the ECG machine were not accessible but the printed paper sheets from the machine. These paper sheets containing the 12-lead ECG graphs were first converted into digital images using a scanner. Then, the Engauge Digitizer \cite{mitchell2021engauge} was utilized to recover the data points from those graphs. It is noteworthy that only the Lead II ECG graphs on the paper sheets were digitized because the specified amplitudes are for Lead II \cite{fluke2021prosim}, in which the P-waves, QRS-complexes, and the T-waves can be clearly observed, and the ECG signals acquired by the wearable device at Position 1 have a waveform (see Figure~\ref{fig:ECGPlacement1}a and Figure~\ref{fig:ECGPlacement1Cont}a) similar to that of a standard Lead II ECG. Conceivably, the two electrodes of the wearable device were connected to the right arm (RA) and left leg (LL) electrodes of the simulator.

The outputs of various waveforms measured by the ECG device (ADC readings) and the ECG machine (digitized data points) were shifted up or down such that the final outputs have a zero baseline. In addition, the ADC readings from the ECG device were converted into voltages in order to compare with the outputs from the ECG machine. Because the wearable device and the medical machine were not taking measurements simultaneously, their outputs were aligned with the first peaks or the first R-peaks of the acquired waveforms. The R-peaks were determined by the fast QRS detection method introduced by Elgendi \cite{elgendi2013fast} and confirmed via visual inspection. Finally, the acquired waveforms from the ECG device and the ECG machine are presented in Figure~\ref{fig:ComparisonECGDeviceMachine}. For clarity's sake, only five consecutive cycles of the sinusoid performance waveforms and three consecutive heartbeats of the normal sinus rhythm waveforms are presented.

For the sinusoid performance waveforms, it is apparent that the outputs from both the device and the machine had similar amplitude. However, their maximum magnitudes were about \SI{0.95}{\milli\volt} while the nominal amplitude accuracy of the ECG simulator is $\pm\SI{2}{\percent}$ (on Lead II). Regarding the shape of the acquired waveforms, the outputs from the device were closer to a sinusoid waveform than that from the machine. The deformation of the latter is mainly ascribed to the digitizing process which approximated the scanned ECG graphs via spline interpolation and manual adjustments.

For the normal sinus rhythm waveforms with various ECG rates, the magnitudes of the R-peaks acquired by the ECG device were significantly lower than that acquired by the ECG machine. While the nadirs of the Q-waves acquired by the ECG device were generally higher than that from the ECG machine, the nadirs of the S-waves from both of them were similar. The lower R-peaks and higher Q-nadirs of the waveforms from the ECG device were probably attributed to the BMD101's built-in band-pass filter (cut-off frequencies from \SIrange{0.5}{103}{\hertz}) and the on-board RC low-pass filter (cut-off frequency about \SI{32}{\hertz}) at the BMD101's inputs that significantly attenuated the high frequency components of the ECG waveforms. The P-waves and T-waves acquired by the ECG devices were typically above and below that acquired by the ECG machine, respectively. Besides, while the data points within the TP-intervals (between the tail of current T-wave and the head of next P-wave) from the ECG machine formed a set of flat lines, the data points within the TP-intervals from the ECG device were ascending. The exact causes require further investigation. Although the shapes of the normal sinus rhythm waveforms measured by the ECG device and the ECG machine were different at certain levels, the time of occurrences of the corresponding P-waves, QRS-complexes, and T-waves were well aligned. To further investigate the quality of ECG signals from the wearable device, various types of rhythms generated by the simulator were recorded by the ECG device and diagnosed by two professional cardiologists in Section~\ref{SubApped:DiagnosingECGSignalsFromDevice}.

\begin{figure}[H]	
	\centering
	\includegraphics[width=0.85\columnwidth]{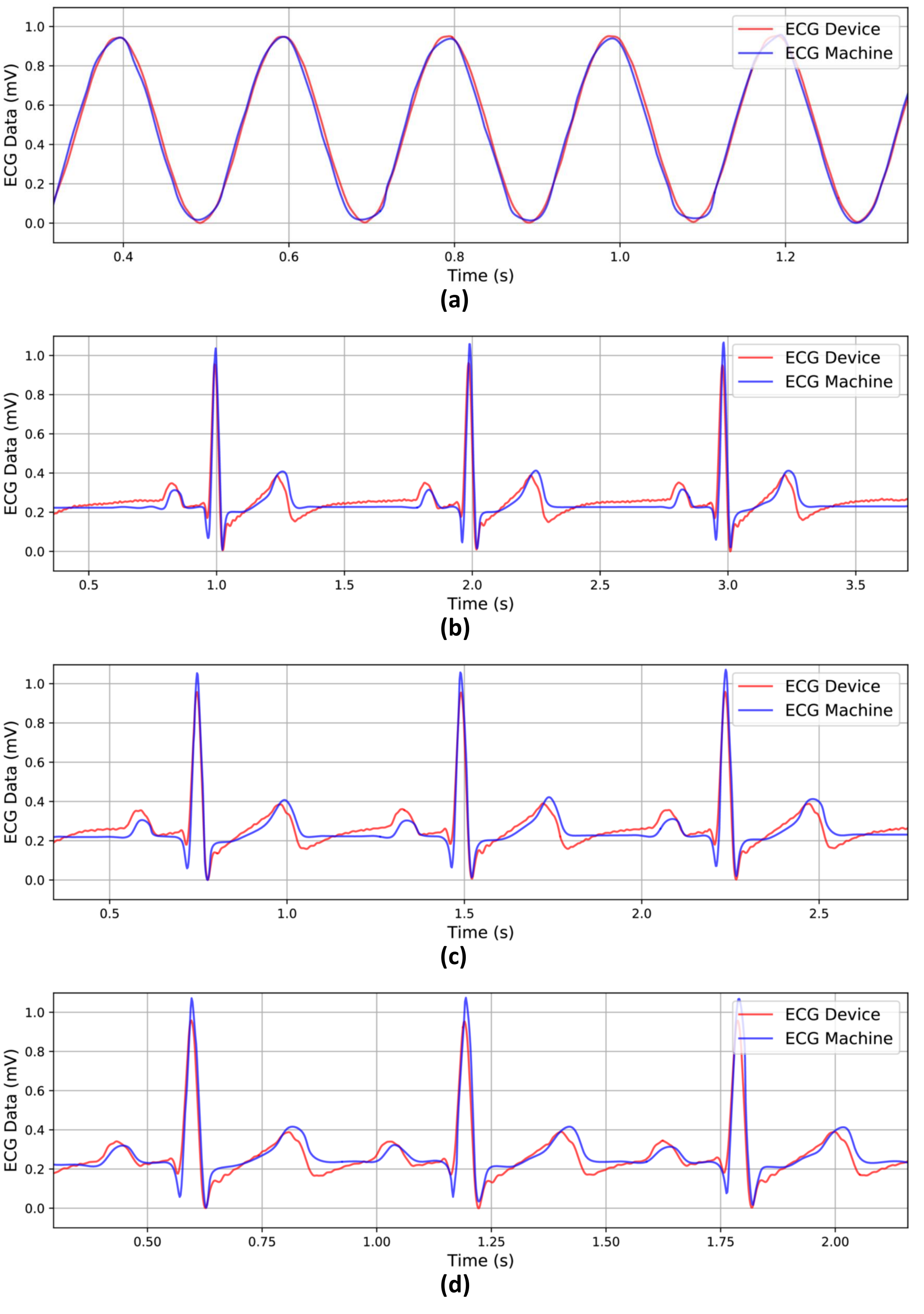}
	\caption{Comparison of waveforms (ECG simulator generated with \SI{1}{\milli\volt} amplitude) acquired by the ECG device (from the ADC) and the ECG machine (from the digitizing software): (\textbf{a}) \SI{5}{\hertz} sinusoid performance waveform. (\textbf{b}) Normal sinus rhythm waveform at \SI{60}{\bpm}. (\textbf{c}) Normal sinus rhythm waveform at \SI{80}{\bpm}. (\textbf{d}) Normal sinus rhythm waveform at \SI{100}{\bpm}. For the sake of clarity, only 5 consecutive cycles of the sinusoid performance waveforms and 3 consecutive heartbeats of the normal sinus rhythm waveforms are shown.}
	\label{fig:ComparisonECGDeviceMachine}
\end{figure}

The intra-consistency (or beat-wise consistency) of the outputs from the wearable device and the medical machine was also investigated in this section. For the sinusoid performance waveforms, a beat is defined by the data points between two adjacent peaks; for the normal sinus rhythm waveforms, an ECG beat contains the data points within a particular RR-interval as shown in Figure~\ref{fig:ECGBeat}. Note that an ECG beat has a different shape compared to a normal heartbeat which has difficulty in identifying the exact start- and end-time. All the beats extracted from the outputs of a particular waveform are illustrated in the first rows of Figure~\ref{fig:BeatConsiTestSinusoid}, Figure~\ref{fig:BeatConsiTest60bpm}, Figure~\ref{fig:BeatConsiTest80bpm}, and Figure~\ref{fig:BeatConsiTest100bpm}, where the symbol $N$ represents the number of beats extracted; these beats were visually overlapping with each other and no essential difference among beats was observed.

To quantify the intra-consistency, for those beats extracted from a particular waveform, the coefficient of variation (CV) value at each sampling instant was calculated\footnote{A CV value at a certain sampling instant will be calculated only when there are at least three data points at that instant.}. These CV values are illustrated in the second rows of Figure~\ref{fig:BeatConsiTestSinusoid}, Figure~\ref{fig:BeatConsiTest60bpm}, Figure~\ref{fig:BeatConsiTest80bpm}, and Figure~\ref{fig:BeatConsiTest100bpm}, and the $CV_{RMS}$ value at the upper right corner of each graph represents the root-mean-square of all CV values. Ideally, the $CV_{RMS}$ value is equal to zero. For every waveform generated by the simulator, the intra-consistency of beats, quantified as $CV_{RMS}$, from the ECG machine was significantly larger than that from the ECG device. Apart from the deformation attributed to the digitizing process, the lower Q-nadirs of the ECG beats from the medical machine made a significant contribution to the larger $CV_{RMS}$ values\footnote{The CV value at a certain sampling instant is exceedingly large if the magnitude of measurement at that instant is significantly small.} when compared to the ECG device. It is apparent that the $CV_{RMS}$ values of the ECG beats from the wearable device increased along with the ECG rates. The major reason is that an ECG waveform with higher ECG rate implies ECG beats with shorter TP-intervals during which the CV values were comparatively small. Overall, when measuring the normal sinus rhythm waveforms, the intra-consistency, quantified as mean $CV_{RMS}$, of the outputs from the wearable device was around \SI{5.5}{\percent} which could be considered negligible.

In conclusion, the normal sinus rhythm waveforms measured by the ECG device and the ECG machine were different at certain levels, especially the magnitudes of the Q-nadirs and R-peaks, and the shapes of the T-waves and TP-intervals. However, the time of occurrences of their corresponding P-waves, QRS-complexes, and T-waves were well aligned. The intra-consistency of the outputs from the medical machine was significantly worse than that from the wearable device primarily due to the deformation introduced by the digitizing process. When measuring the normal sinus rhythm waveform at a certain ECG rate, the outputs from the wearable device were considered consistent across ECG beats ($\overline{CV_{RMS}} \approx \SI{5.5}{\percent}$).

\begin{figure}[!t]	
	\centering
	\includegraphics[width=0.9\columnwidth]{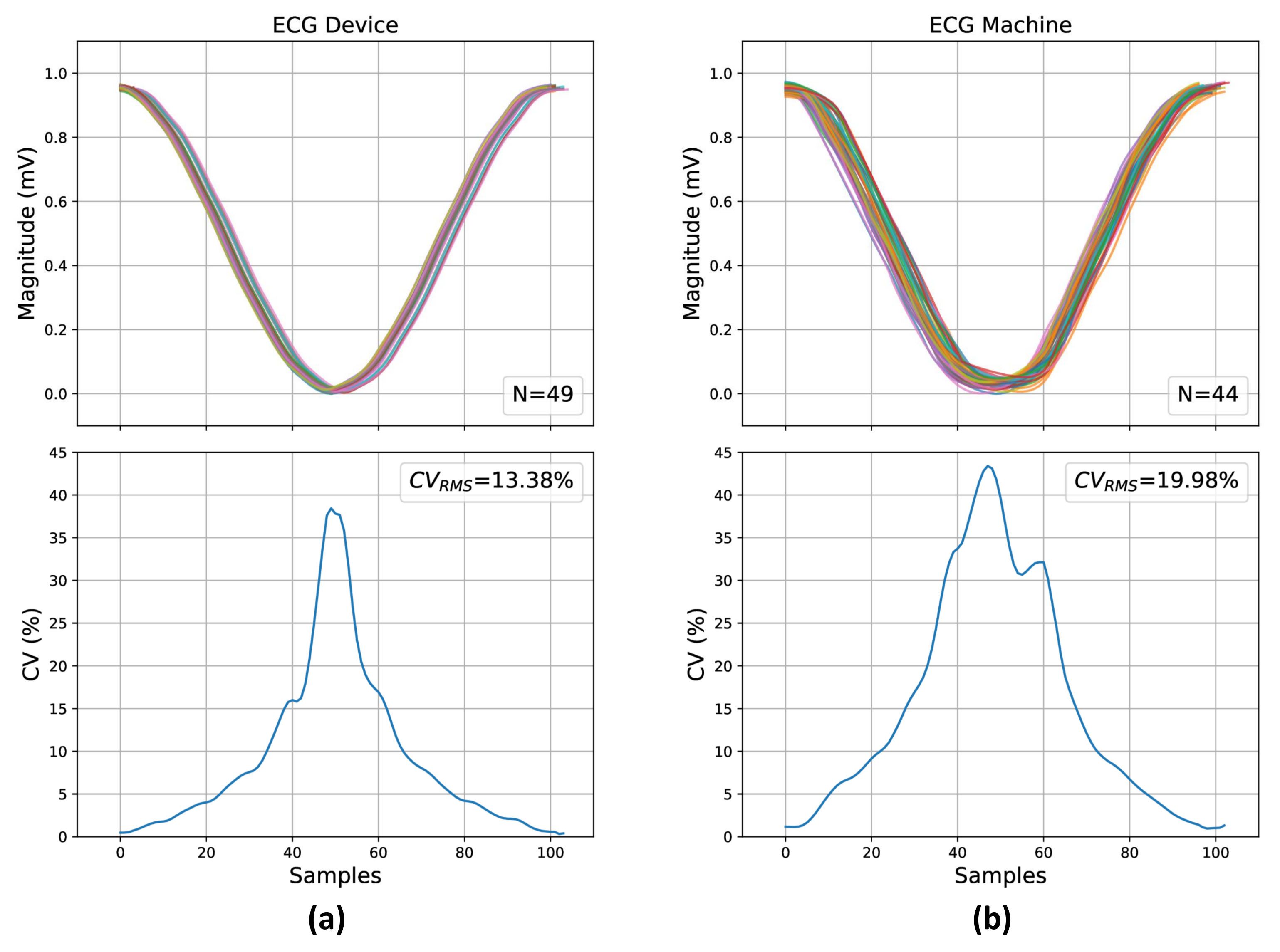}
	\caption{Consistency test of the \SI{5}{\hertz} sinusoid performance waveform (with \SI{1}{\milli\volt} amplitude) acquired by: (\textbf{a}) the ECG device. (\textbf{b}) the ECG machine. The graphs in the first row contain the cosine cycles (defined by the data points between two adjacent peaks) retrieved from the acquired waveforms, and the symbol $N$ represents the number of cycles. The graphs in the second row illustrate the $CV$ values calculated at each sampling instant.}
	\label{fig:BeatConsiTestSinusoid}
\end{figure}

\begin{figure}[!t]	
	\centering
	\includegraphics[width=0.9\columnwidth]{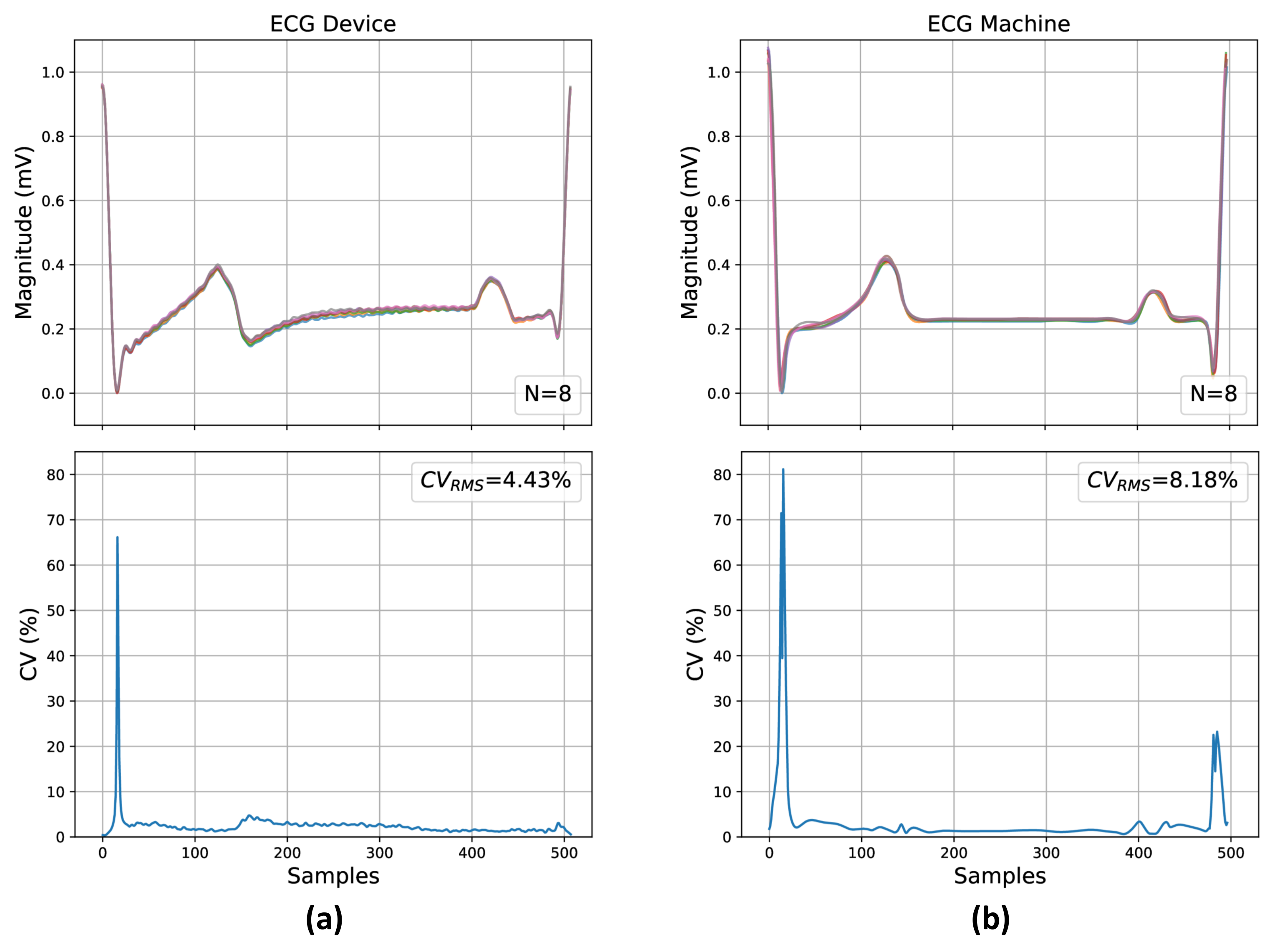}
	\caption{Consistency test of the \SI{60}{\bpm} normal sinus rhythm waveform (with \SI{1}{\milli\volt} amplitude) acquired by: (\textbf{a}) the ECG device. (\textbf{b}) the ECG machine. The graphs in the first row contain the ECG beats (defined by the data points between two adjacent R-peaks) retrieved from the acquired waveforms, and the symbol $N$ represents the number of ECG beats. The graphs in the second row illustrate the $CV$ values calculated at each sampling instant.}
	\label{fig:BeatConsiTest60bpm}
\end{figure}

\begin{figure}[!t]	
	\centering
	\includegraphics[width=0.9\columnwidth]{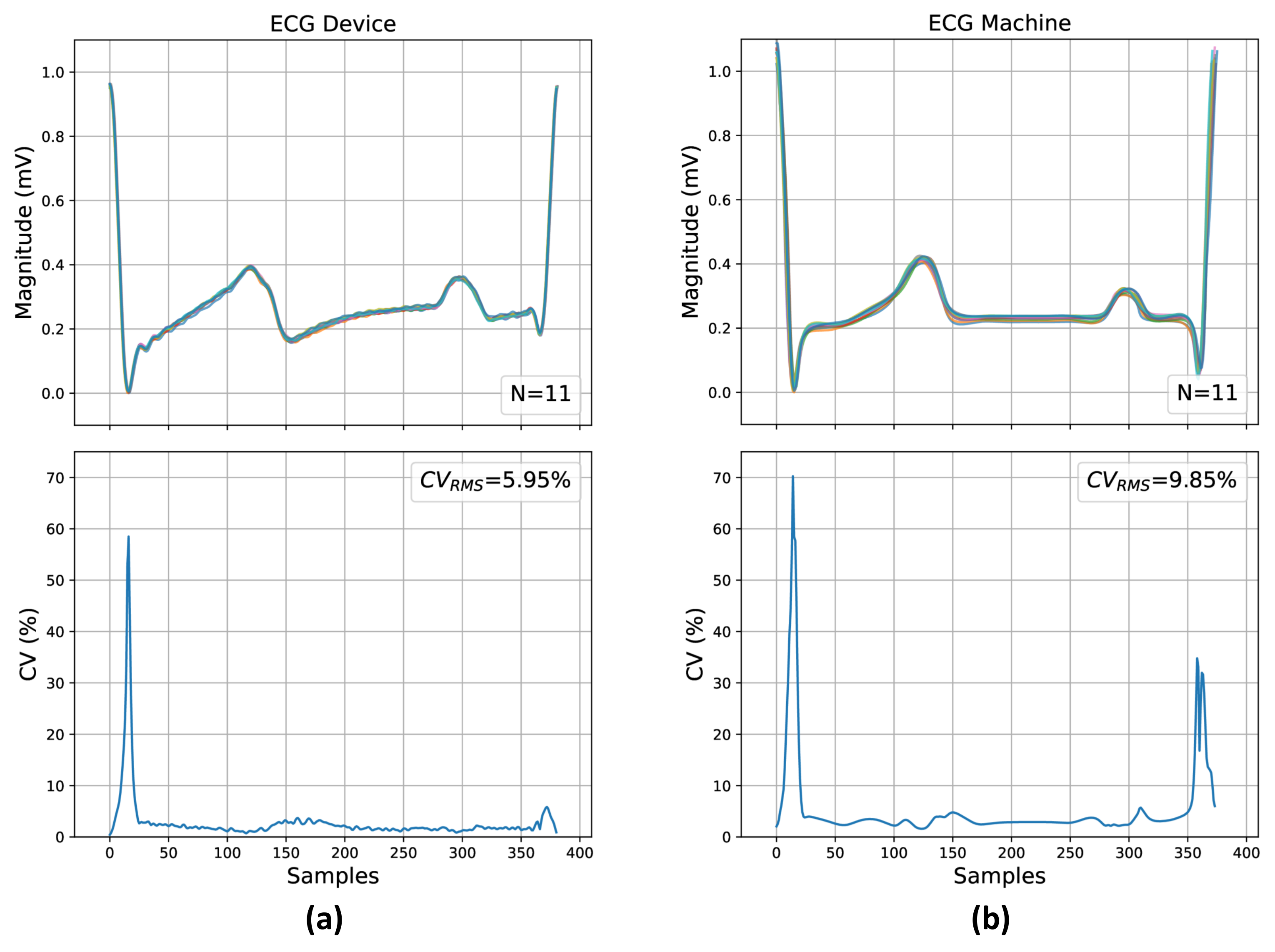}
	\caption{Consistency test of the \SI{80}{\bpm} normal sinus rhythm waveform (with \SI{1}{\milli\volt} amplitude) acquired by: (\textbf{a}) the ECG device. (\textbf{b}) the ECG machine. The graphs in the first row contain the ECG beats (defined by the data points between two adjacent R-peaks) retrieved from the acquired waveforms, and the symbol $N$ represents the number of ECG beats. The graphs in the second row illustrate the $CV$ values calculated at each sampling instant.}
	\label{fig:BeatConsiTest80bpm}
\end{figure}

\begin{figure}[!t]	
	\centering
	\includegraphics[width=0.9\columnwidth]{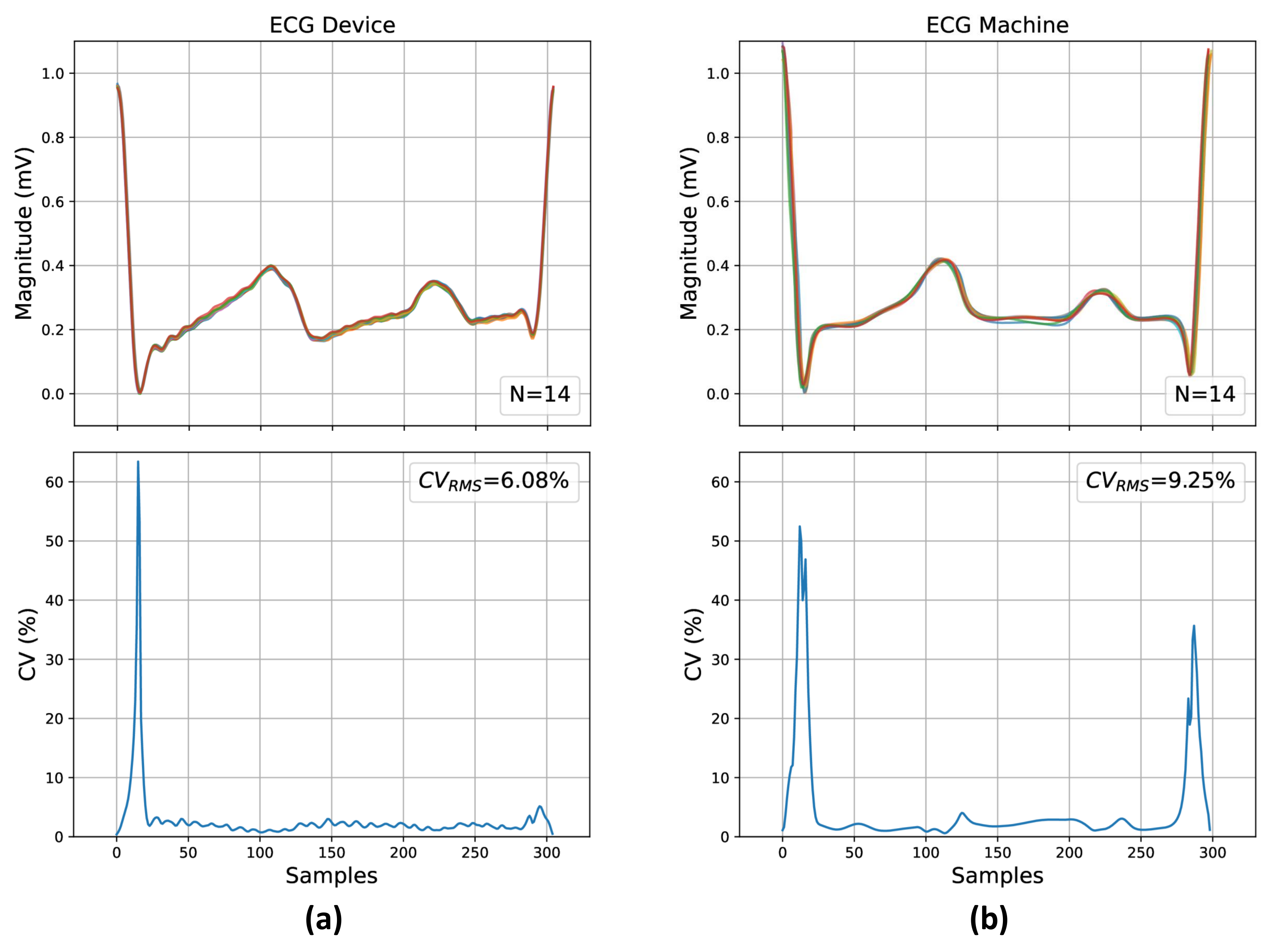}
	\caption{Consistency test of the \SI{100}{\bpm} normal sinus rhythm waveform (with \SI{1}{\milli\volt} amplitude) acquired by: (\textbf{a}) the ECG device. (\textbf{b}) the ECG machine. The graphs in the first row contain the ECG beats (defined by the data points between two adjacent R-peaks) retrieved from the acquired waveforms, and the symbol $N$ represents the number of ECG beats. The graphs in the second row illustrate the $CV$ values calculated at each sampling instant.}
	\label{fig:BeatConsiTest100bpm}
\end{figure}

\clearpage
\subsection{Comparison of ECG Signals from Five Implemented ECG Devices}
\label{SubApped:SignalComparisonAmongDevices}

Five wearable ECG devices have been implemented and the inter-consistency (or unit-wise consistency) of the outputs from these devices were investigated in this section. These devices were, one after the other, connected to the RA and LL electrodes of the ECG simulator which was configured to generate normal sinus rhythm waveforms at \SI{60}{bpm}, \SI{80}{bpm}, \SI{100}{bpm}, and \SI{120}{bpm} with \SI{1}{\milli\volt} amplitude. For every ECG device, about \SI{10}{\minute} of data of each waveform were collected. The waveforms generated by the ECG simulator under specific configuration were assumed to be identical throughout the experiment.

Similar to the experiment presented in Section~\ref{SubApped:SignalComparisonBetweenDeviceAndEquipment}, the ADC outputs of various waveforms measured by the five wearable devices were first respectively shifted up with zero baselines and then converted into voltages. Because the wearable devices were not taking measurements simultaneously, their outputs of the normal sinus rhythm waveform at certain ECG rate were aligned with the first R-peaks. The R-peaks were determined by the fast QRS detection method introduced by Elgendi \cite{elgendi2013fast} and confirmed via visual inspection. As depicted in the first column of Figure~\ref{fig:UnitWiseConsiTest}, the ECG beats (defined by the data points between two adjacent R-peaks) retrieved from the waveform measured by a wearable device were visually overlapping with each other. The intra-consistency among ECG beats of a wearable device, quantified as $CV_{RMS}$, was no larger than \SI{5.2}{\percent} and could be considered negligible. However, compared to the ECG beats of a wearable device, the ECG beats among these five devices manifested worse consistency.

Taking the beat-wise average of the ECG beats from a wearable device, the representative ECG beat of a wearable device when measuring a normal sinus rhythm waveform at certain ECG rate was achieved. As illustrated in the second column of Figure~\ref{fig:UnitWiseConsiTest}, the representative ECG beats of the Device 5 were generally away from those of the other devices, especially for those data points within the SQ-intervals. To quantify the inter-consistency, for the five representative ECG beats extracted from the waveforms at certain ECG rate, the CV value at each sampling instant was calculated. These CV values are illustrated in the third column of Figure~\ref{fig:UnitWiseConsiTest}, and the root-mean-square of the CV values in each graph, denoted by $CV_{RMS}$, is presented at the upper left corner. Apparently, the local maximum CV values were introduced during the intervals between the Q-waves and the R-peaks. As the magnitudes of data points in these intervals were increasing rapidly, slight time-shifts of the representative ECG beats, which are shown in the second column of Figure~\ref{fig:UnitWiseConsiTest}, will resulted in huge CV values. Because the ECG beats of a wearable device demonstrated excellent consistency (see the first column of Figure~\ref{fig:UnitWiseConsiTest}), the time-shifts among the representative ECG beats from different wearable devices were unlikely attributed to the performance of the ECG simulator but the timing accuracy of the BMD101 analog front-ends' internal RC oscillators. Overall, when measuring the normal sinus rhythm waveforms, the inter-consistency, quantified as mean $CV_{RMS}$, of the outputs from the five wearable devices, was around \SI{12.1}{\percent}. It has been empirically confirmed by two professional cardiologists that such level of variations in the ECG beats from different wearable devices will not influence the diagnostic results.

\end{paracol}
\appendix
\begin{figure}[t]
	\widefigure
	\centering
	\includegraphics[width=0.95\columnwidth]{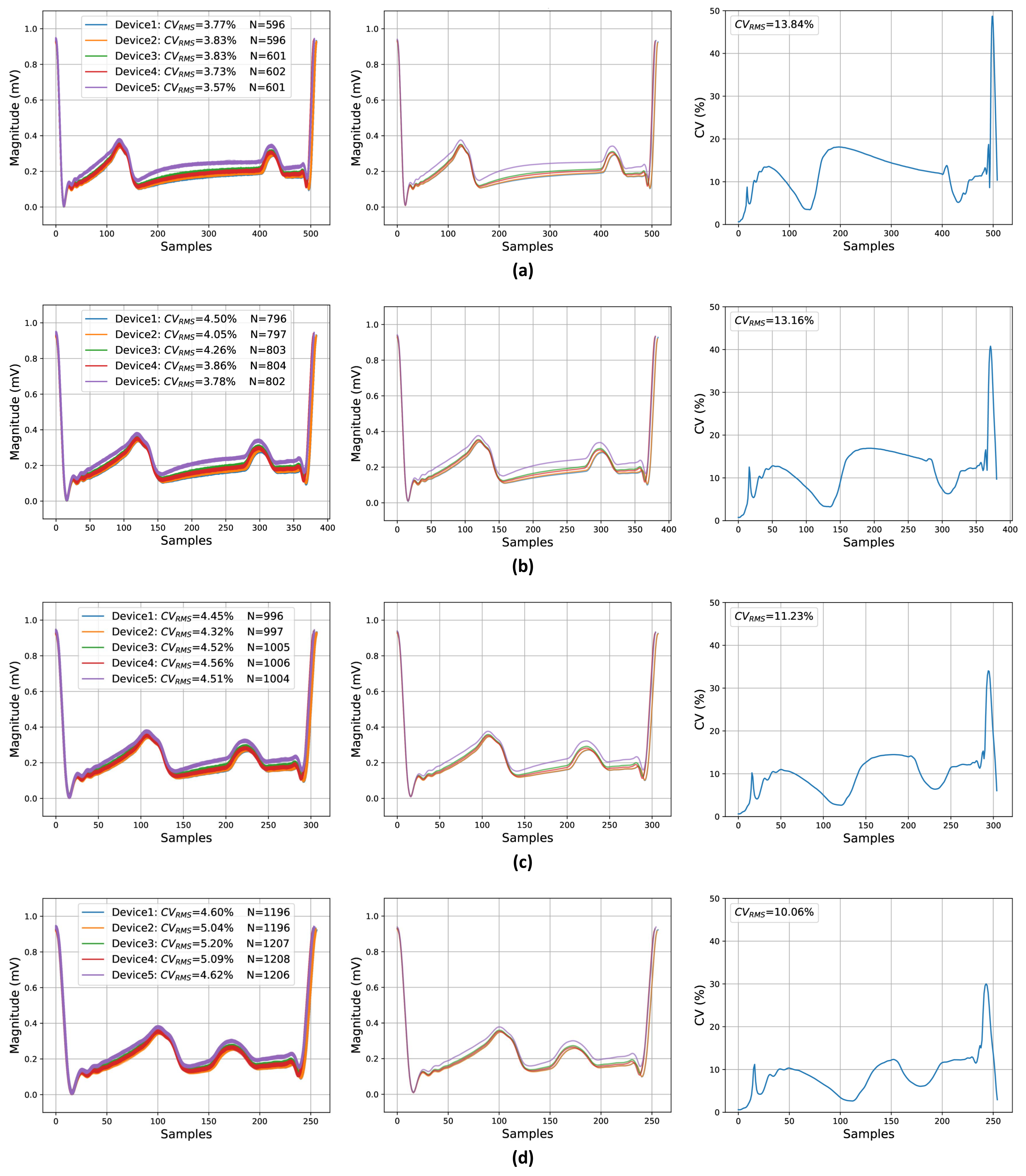}
	\caption{Consistency test of the normal sinus rhythm waveforms (with \SI{1}{\milli\volt} amplitude) acquired by the 5 wearable devices: (\textbf{a}) ECG rate at \SI{60}{\bpm}. (\textbf{b}) ECG rate at \SI{80}{\bpm}. (\textbf{c}) ECG rate at \SI{100}{\bpm}. (\textbf{d}) ECG rate at \SI{120}{\bpm}. The graphs in the first column contain the ECG beats (defined by the data points between two adjacent R-peaks) retrieved from the acquired waveform of each wearable device, and the symbol $N$ represents the number of ECG beats. The graphs in the second column illustrate the representative ECG beats (beat-wise average of the ECG beats acquired by a wearable device) of the wearable devices. In the third column, the $CV$ values in each graph are calculated from the corresponding representative ECG beats shown in the second column.}
	\label{fig:UnitWiseConsiTest}
\end{figure}
\begin{paracol}{2}
\switchcolumn

In addition, for the normal sinus rhythm waveforms at each ECG rate, the time differences between the nominal RR-interval (calculated from the configured ECG rate of the simulator) and those RR-intervals calculated from the ECG beats retrieved from the wearable devices' outputs were analyzed in this section. Note that a RR-interval is the time interval between two adjacent R-peaks, and the nominal rate accuracy of the ECG simulator is $\pm\SI{1}{\percent}$. Therefore, the nominal RR-intervals of the normal sinus rhythm waveforms at \SI{60}{\bpm}, \SI{80}{\bpm}, \SI{100}{\bpm}, and \SI{120}{\bpm} ECG rates are around \SI{1.00\pm0.01}{\second}, \SI{0.75\pm0.0075}{\second}, \SI{0.60\pm0.006}{\second}, and \SI{0.50\pm0.005}{\second}, respectively. The basic statistics and the error bars (one sample standard deviation) of the RR-intervals calculated from the ECG beats acquired by the five wearable devices are presented in Table~\ref{Tab:BasicStatisticsRR} and Figure~\ref{fig:RRIntervalErrorBar}, respectively.

\clearpage

\begin{specialtable}[H]
\caption{Basic statistics of RR-intervals calculated from the ECG beats of the 5 wearable devices.}
\label{Tab:BasicStatisticsRR}
\small
\begin{tabular}{cccccccc}
\toprule
& \textbf{ECG Rate} & \textbf{Device 1} & \textbf{Device 2} & \textbf{Device 3} & \textbf{Device 4} & \textbf{Device 5} \\
\midrule
						& \SI{60}{bpm}  & 0.9997 & 0.9994 & 0.9906 & 0.9897 & 0.9918 \\
\textbf{Sample Mean} 	& \SI{80}{bpm}  & 0.7498 & 0.7496 & 0.7429 & 0.7422 & 0.7439 \\
\textbf{(s)}			& \SI{100}{bpm} & 0.5998 & 0.5997 & 0.5944 & 0.5938 & 0.5951 \\
						& \SI{120}{bpm} & 0.4998 & 0.4997 & 0.4953 & 0.4948 & 0.4959 \\
\midrule
						& \SI{60}{bpm}  & 0.0007 & 0.0009 & 0.0008 & 0.0009 & 0.0008 \\
\textbf{Sample SD} 		& \SI{80}{bpm}  & 0.0007 & 0.0008 & 0.0009 & 0.0004 & 0.0006 \\
\textbf{(s)}			& \SI{100}{bpm} & 0.0006 & 0.0005 & 0.0009 & 0.0006 & 0.0009 \\
						& \SI{120}{bpm} & 0.0006 & 0.0007 & 0.0010 & 0.0009 & 0.0006 \\
\midrule
						& \SI{60}{bpm}  & 0.20 & 0.20 & 1.17 & 1.17 & 0.98 \\
\textbf{Maximum ARE}	& \SI{80}{bpm}  & 0.26 & 0.26 & 1.04 & 1.30 & 1.04 \\
\textbf{(\%)} 			& \SI{100}{bpm} & 0.39 & 0.39 & 1.04 & 1.37 & 1.04 \\
						& \SI{120}{bpm} & 0.39 & 0.39 & 1.17 & 1.17 & 1.17 \\
\bottomrule
\end{tabular}
\\
\footnotesize
SD and ARE are the abbreviations of the standard deviation and the absolute relative error, respectively.
\end{specialtable}

\end{paracol}
\appendix
\begin{figure}[H]
	\widefigure
	\centering
	\includegraphics[width=0.98\columnwidth]{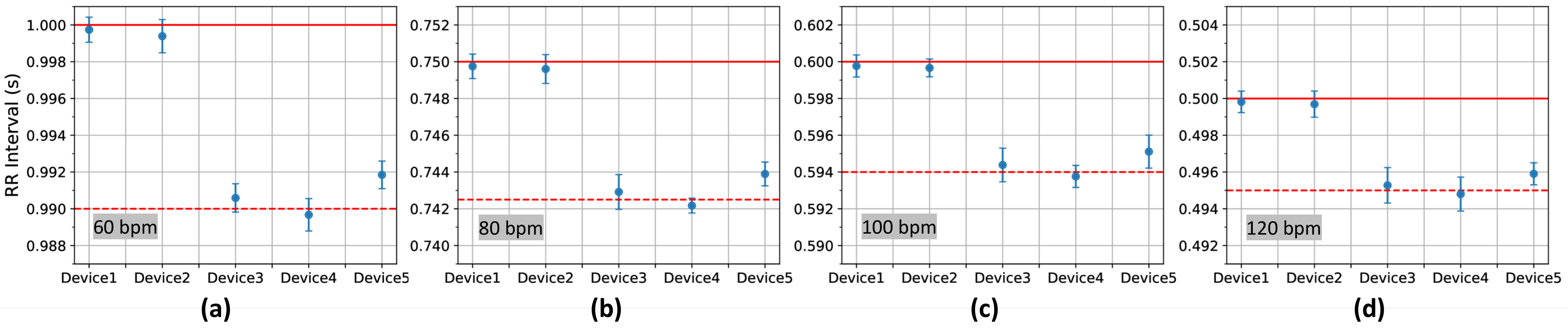}
	\caption{Error bars (one sample SD) of the RR-intervals calculated from the ECG beats (defined by the data points between two adjacent R-peaks) of the 5 wearable devices when measuring normal sinus rhythm waveforms with ECG rate at: (\textbf{a}) \SI{60}{\bpm}. (\textbf{b}) \SI{80}{\bpm}. (\textbf{c}) \SI{100}{\bpm}. (\textbf{d}) \SI{120}{\bpm}. The solid red lines represent the ideal nominal RR-interval values at respective ECG rates; the dashed red lines represent the lower bounds of the nominal RR-intervals, determined by the nominal rate accuracy of the ECG simulator, at respective ECG rates.}
	\label{fig:RRIntervalErrorBar}
\end{figure}
\begin{paracol}{2}
\switchcolumn

As listed in Table~\ref{Tab:BasicStatisticsRR} and illustrated in Figure~\ref{fig:RRIntervalErrorBar}, the sample means of RR-intervals calculated from the ECG devices' outputs were always lower than the ideal nominal values at respective ECG rates. The RR-intervals of the Device 3, 4, 5 were typically close to the lower-bounds of the nominal values, while those of the Device 1 and 2 were slightly lower than the ideal nominal values. The sample standard deviation (SD) of RR-intervals of the Device 3 and the absolute relative error (ARE) of RR-intervals of the Device 4 were usually the largest at each ECG rate, respectively. In Table~\ref{Tab:BasicStatisticsRR}, the maximum ARE between the ideal nominal RR-intervals and the calculated ones of the wearable devices were lower than \SI{1.4}{\percent}. Although the nominal rate accuracy of the ECG simulator is $\pm\SI{1}{\percent}$, the authors speculated that the time differences between the nominal RR-intervals and those calculated from the ECG devices' outputs were primarily attributed to the varying timing accuracies of internal RC oscillators among the BMD101 analog front-ends, given the excellent intra-consistency of the outputs from a wearable device (see the first column of Figure~\ref{fig:UnitWiseConsiTest}). It is noteworthy that, since the BMD101 analog front-end is sampling at \SI{512}{\hertz} and $\pm\SI{1}{}$ sample differences in the ECG beats retrieved from the wearable devices' outputs are inevitable, the RR-intervals calculated from the ECG beats at certain ECG rate have intrinsic time differences around $\pm\SI{0.002}{\second}$, i.e., $\pm\SI{0.2}{\percent}$ to $\pm\SI{0.4}{\percent}$ at \SIrange{60}{120}{\bpm} ECG rates. As empirically confirmed by two professional cardiologists, the \SI{1.4}{\percent} maximum ARE in the wearable devices' calculated RR-intervals could be considered negligible.

In summary, when measuring the normal sinus rhythm waveforms at \SIrange{60}{120}{\bpm}, the inter-consistency, quantified as mean $CV_{RMS}$, of the outputs from the five wearable devices, was around \SI{12.1}{\percent}; the maximum ARE between the ideal nominal RR-intervals and the calculated ones of the wearable devices was lower than \SI{1.4}{\percent}. Both of them were speculated to be primarily attributed to the timing accuracy of the BMD101 analog front-ends' internal RC oscillators, instead of the nominal rate accuracy of the ECG simulator. It has been empirically confirmed by two professional cardiologists that such level of variations ($\overline{CV_{RMS}} \approx \SI{12.1}{\percent}$) in the ECG beats from different wearable devices would not influence the diagnostic results, and the AREs in the wearable devices' calculated RR-intervals ($<\SI{1.4}{\percent}$) could be considered negligible.

\subsection{Diagnosing ECG Signals Collected by the Wearable Device}
\label{SubApped:DiagnosingECGSignalsFromDevice}

To further investigate whether or not the quality of ECG signals collected by a wearable device is sufficient for cardiac arrhythmia diagnosis, 13 types of rhythms generated by the ECG simulator were recorded by the wearable ECG device. The device was connected to the RA and LL electrodes (Lead II) of the simulator. Each record was trimmed and contained \SI{8}{\second} of ECG signals as shown in Figure~\ref{fig:DifferentWaveformsFromSimulator}. These processed ECG records were sent to two professional cardiologists for diagnosis without discovering the exact rhythm types. The 13 types of rhythms generated by the ECG simulator were:
\begin{enumerate}
\item	Asystole rhythm; 
\item	Atrial flutter (AFL) rhythm;
\item	Coarse atrial fibrillation (AF) rhythm;
\item	Coarse ventricular fibrillation (VF) rhythm;
\item	Left bundle branch block (LBBB) rhythm;
\item	Normal sinus rhythm;
\item	Normal sinus rhythm with \SI{50}{\hertz} noise;
\item	Normal sinus rhythm with muscle contraction noise;
\item	Right bundle branch block (RBBB) rhythm;
\item	Second-degree atrioventricular (AV) block rhythm;
\item	Premature ventricular contraction (PVC) rhythm from left ventricular;
\item 	Supraventricular tachycardia (SVT) rhythm;
\item 	Ventricular tachycardia (VT) rhythm.
\end{enumerate}

Both cardiologists had correctly labeled 11 out of 13 (about \SI{85}{\percent}) types of rhythms. The LBBB (Rhythm 5) and the RBBB (Rhythm 9) were incorrectly labeled as the RBBB rhythm and the normal sinus rhythm, respectively. The misclassification of the LBBB and the RBBB collected by the ECG device was arguable given that the Rhythm 5 (LBBB) was visually similar to a typical RBBB rhythm on Lead II \cite{ecg2021right}, and the Rhythm 9 (RBBB) was nearly identical to a normal sinus rhythm on Lead II but had deeper S-waves and larger T-waves. Notice that the specific criteria for distinguishing the LBBB and the RBBB, or the normal sinus rhythm and the LBBB or RBBB are not established from the Lead II but the other leads, especially the Lead V1 \cite{rezaie2013bundle,scherbak2019left,harkness2018right}. Therefore, additional information is needed to correctly classify the LBBB and RBBB rhythms collected by the ECG device.

In conclusion, the quality of ECG signals collected by a wearable ECG device is sufficient for cardiac arrhythmia diagnosis, and the acquired ECG rhythms, including normal ones and arrhythmias, can be correctly classified by a cardiologist given that the rhythms have distinguishable criteria established from the Lead II.

\begin{figure}[H]
	\centering
	\includegraphics[width=0.9\columnwidth]{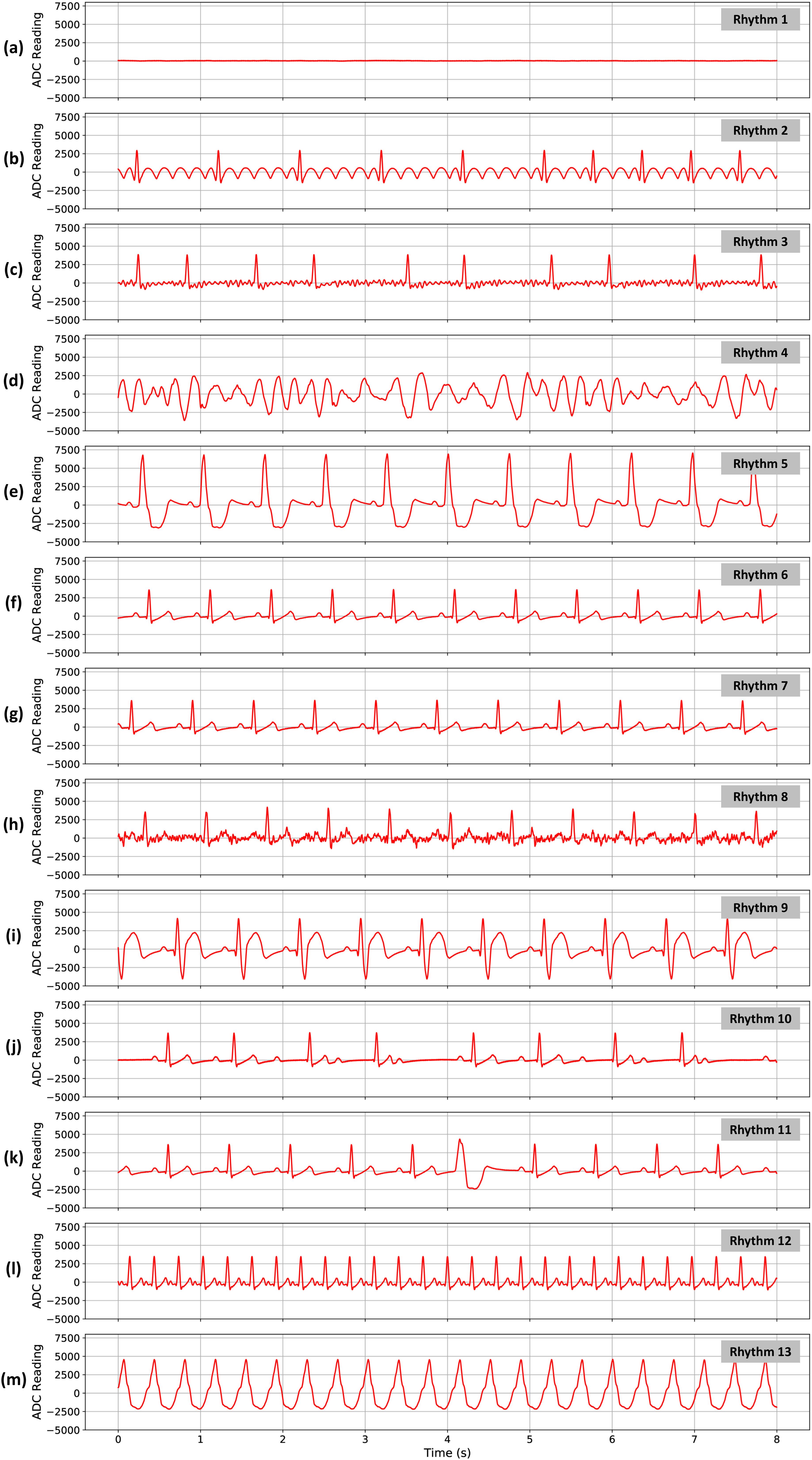}
	\caption{\SI{8}{\second} of ECG signals collected by the wearable device when the simulator was generating: (\textbf{a}) Asystole rhythm. (\textbf{b}) AFL rhythm. (\textbf{c}) Coarse AF rhythm. (\textbf{d}) Coarse VF rhythm. (\textbf{e}) LBBB rhythm. (\textbf{f}) Normal sinus rhythm. (\textbf{g}) Normal sinus rhythm with \SI{50}{\hertz} noise. (\textbf{h}) Normal sinus rhythm with muscle contraction noise. (\textbf{i}) RBBB rhythm. (\textbf{j}) Second-degree AV block rhythm. (\textbf{k}) PVC rhythm from left ventricular. (\textbf{l}) SVT rhythm. (\textbf{m}) VT rhythm.}
	\label{fig:DifferentWaveformsFromSimulator}
\end{figure}

\setcounter{figure}{0}
\section{Experimental Results of the InLC Schema}
\label{Apped:ExperimentalResultsInLC}

\subsection{Monotonicity of the Distance Functions in InLC Schema}
\label{SubApped:MonotonicityInLCDistFunct}

In order to improve the computational efficiency of the InLC schema, the length $l_{nxt}$ of next fragment $\vec{f}_{l_{nxt}}$ is determined by the binary search approach as presented in Section~\ref{SubSubSect:IntuitiveLossy}. Note that the binary search approach is only applicable for monotonic function, while both of the proposed distance functions (i.e., Equation~\ref{eq:OldDist} and Equation~\ref{eq:NewDist}) are mathematically non-monotonic. However, preliminary empirical results suggested that the distances calculated from Equation~\ref{eq:OldDist} and Equation~\ref{eq:NewDist} (with infinity distances eliminated) are typically monotonically increasing along with length $l$ of fragment $\vec{f}_{l}$. An example is illustrated in Figure~\ref{fig:IllustrationInLCDist} ($s_{f}=s_{b}=l_{max}=1024$ and $l_{min}=10$); the samples in $\vec{B}_{b}$ and $\vec{B}_{f}$ are the consecutive ECG samples arbitrarily retrieved from the Record 100 of the MIT-BIH database. The horizontal axes of Figure~\ref{fig:IllustrationInLCDist} represent the length $l$ of fragment $\vec{f}_{l}$ in $\vec{B}_{f}$, and the vertical axes represent the minimum and average distances between a fragment $\vec{f}_{l}$ and every fragment $\vec{b}^{i}_{l}$ in $\vec{B}_{b}$ with start index $i$.

\end{paracol}
\appendix
\renewcommand\thefigure{B\arabic{figure}}
\begin{figure}[H]
	\widefigure
	\centering
	\includegraphics[width=0.95\columnwidth]{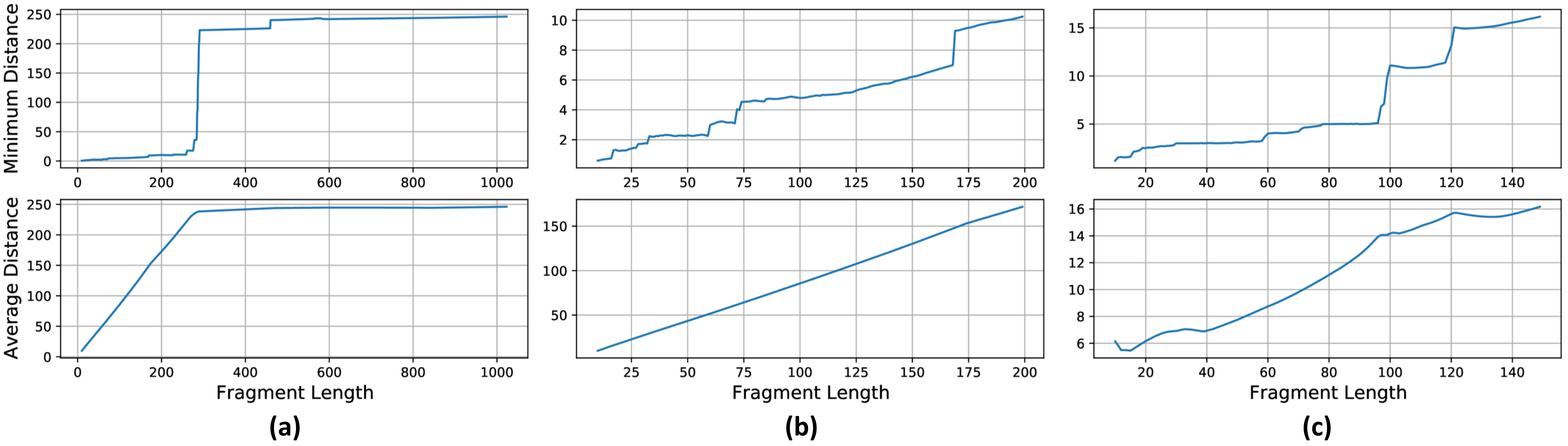}
	\caption{Illustration of the monotonicity of the proposed distance functions along with the length $l$ of fragment $f_{l}$ ($f_{l}$ is extracted from the Record 100 of MIT-BIH database with arbitrary start index): (\textbf{a}) Distances from Equation~\ref{eq:OldDist} versus $l$ ($l\leq1024$). (\textbf{b}) Distances from Equation~\ref{eq:OldDist} versus $l$ ($l\leq200$). (\textbf{c}) Distances from Equation~\ref{eq:NewDist} versus $l$ ($l\leq1024$) with infinity distances eliminated.}
	\label{fig:IllustrationInLCDist}
\end{figure}
\begin{paracol}{2}
\switchcolumn

\subsection{Reconstructed Signals from the InLC Schema with Different Setups}
\label{SubApped:ReconSigInLCDiffSetups}

As stated in Section~\ref{SubSubSect:AveragePerformanceLCs}, the short-period curvilinear waveforms in the original ECG signal are likely being replaced by a few straight lines in the reconstructed one (known as CW-to-SL effect) due to the deficiency of the InLC schema's distance function, i.e., the Equation~\ref{eq:OldDist}, especially when the bank $\vec{B}_{b}$ is updated constantly. Using a short period of ECG signal from the Record 100 of the MIT-BIH database as an example, the aforementioned phenomenon is depicted in Figure~\ref{fig:InLCStraightLineIssue}a. Excluding the P-waves, QRS-complexes, and T-waves with significant fluctuations, those remaining consecutive ECG samples in the original ECG signal were apparently replaced by a few straight lines in the reconstruct one. The CW-to-SL effect is generalized in Figure~\ref{fig:IllustrationStraightLine}. Empirically, the occurrences of such phenomenon can be effectively eliminated by further constraining the distance function as defined in the Equation~\ref{eq:NewDist} according to the rationale presented in Section~\ref{SubSubSect:AveragePerformanceLCs}. The example result of the InLC schema with constantly updated $\vec{B}_{b}$ and the Equation~\ref{eq:NewDist} as distance function is illustrated in Figure~\ref{fig:InLCStraightLineIssue}b.

Based on the authors' experience, besides eliminating the CW-to-SL effect, a fragment pair $\vec{x}$ and $\vec{y}$ with sufficiently small distance value calculated from the Equation~\ref{eq:NewDist} usually implies a close-to-unity gain (i.e., the slope of OLS regression). Consequently, the InLC schema can achieve higher compression and computational efficiency at the cost of slightly increased distortion by eliminating the gain $g^{*}$ (assuming unity) in the compressed information and only transmitting the new offset $o^{*}$, which is the average difference between the fragments $\vec{x}$ and $\vec{y}$. The example result is shown in Figure~\ref{fig:InLCStraightLineIssue}c.

\begin{figure}[H]
	\centering
	\includegraphics[width=0.85\columnwidth]{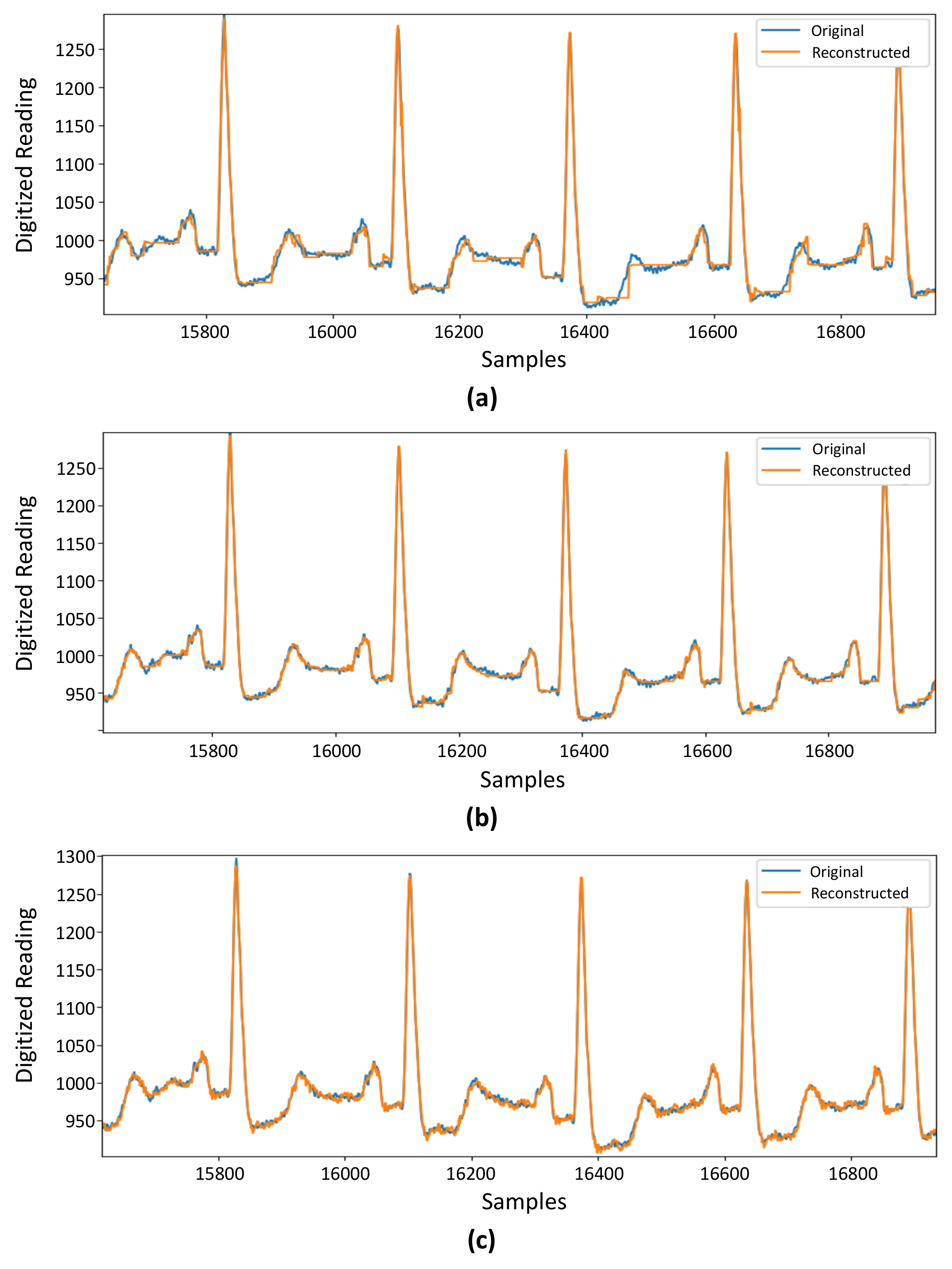}
	\caption{Examples of the original (blue) and reconstructed (orange) ECG signals (Record 100 from the MIT-BIH database) for the InLC schema (threshold $\epsilon=10.0$) with different configurations: (\textbf{a}) InLC with constantly updated $\vec{B}_{b}$, Equation~\ref{eq:OldDist} distance function, and both gain and offset. (\textbf{b}) InLC with constantly updated $\vec{B}_{b}$, Equation~\ref{eq:NewDist} distance function, and both gain and offset. (\textbf{c}) InLC with constantly updated $\vec{B}_{b}$, Equation~\ref{eq:NewDist} distance function, and only offset.}
	\label{fig:InLCStraightLineIssue}
\end{figure}

\setcounter{figure}{0}
\renewcommand\thefigure{C\arabic{figure}}
\section{Experimental Results of the ResNet-Based AF Detector}
\label{Apped:ExperimentalResultsAFDetector}

\subsection{Determining the Width and Overlapping of the Tukey Window}
\label{SubApped:WidthOverlappingOfTukey}

As presented in Section~\ref{SubSubSect:AFPreprocessing}, a Tukey window with width $W$ and overlapping $O$ was employed in the STFT to convert the time series ECG signals into 2D spectrograms. A conversion example using window with $W=60$ and $O=30$ is depicted in Figure~\ref{fig:IllustrationWindowSTFT}. Notice that most of the upper areas (i.e., those components with higher frequencies) of the spectrogram are in purple (i.e., negligible magnitudes). It is expected considering that the ECG signals in the public dataset have a nominal frequency band from \SIrange{0.5}{40}{\hertz} (refer to Section~\ref{SubSubSect:AFDatasetAndMetrics}) and the high frequency components of their FFT results have negligible magnitudes as illustrated in Figure~\ref{fig:FourTypeSignalFFT}. The authors speculated that eliminating the high frequency components, e.g., frequency components over \SI{75}{\hertz}, will not influence the classification performance of the ResNet-based AF detector.

\begin{figure}[H]
	\centering
	\includegraphics[width=0.95\columnwidth]{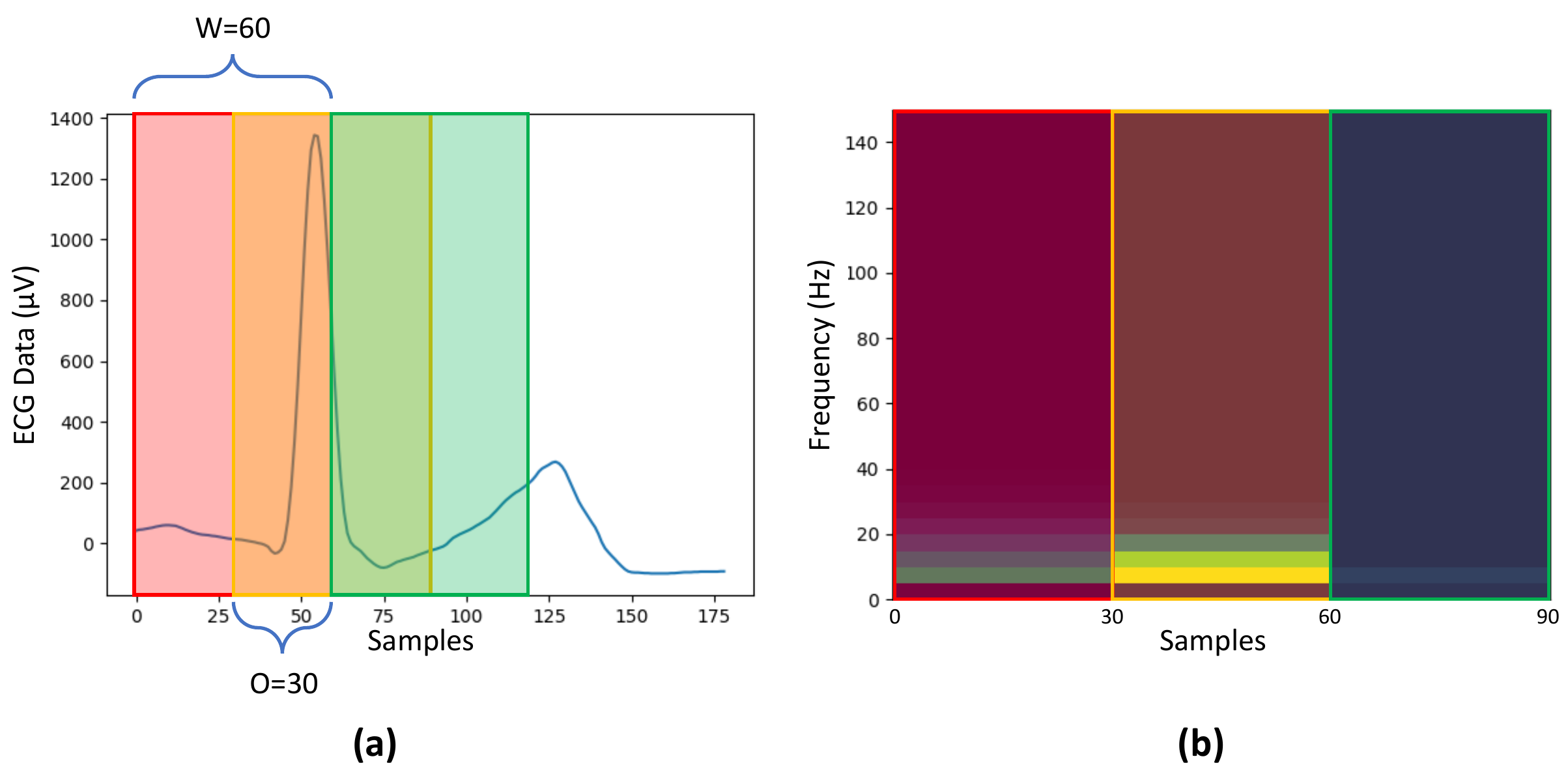}
	\caption{Illustration of converting the time series ECG signal into a spectrogram using Tukey windowed STFT with $W=60$ and $O=30$ samples: (\textbf{a}) Time series ECG signal. (\textbf{b}) Converted spectrogram of which each column highlighted in red, yellow, or green contains the STFT result (i.e., magnitudes of frequency components, which are represented using different colors) of the ECG samples beneath the window with corresponding color.}
	\label{fig:IllustrationWindowSTFT}
\end{figure}

\begin{figure}[H]
	\centering
	\includegraphics[width=0.95\columnwidth]{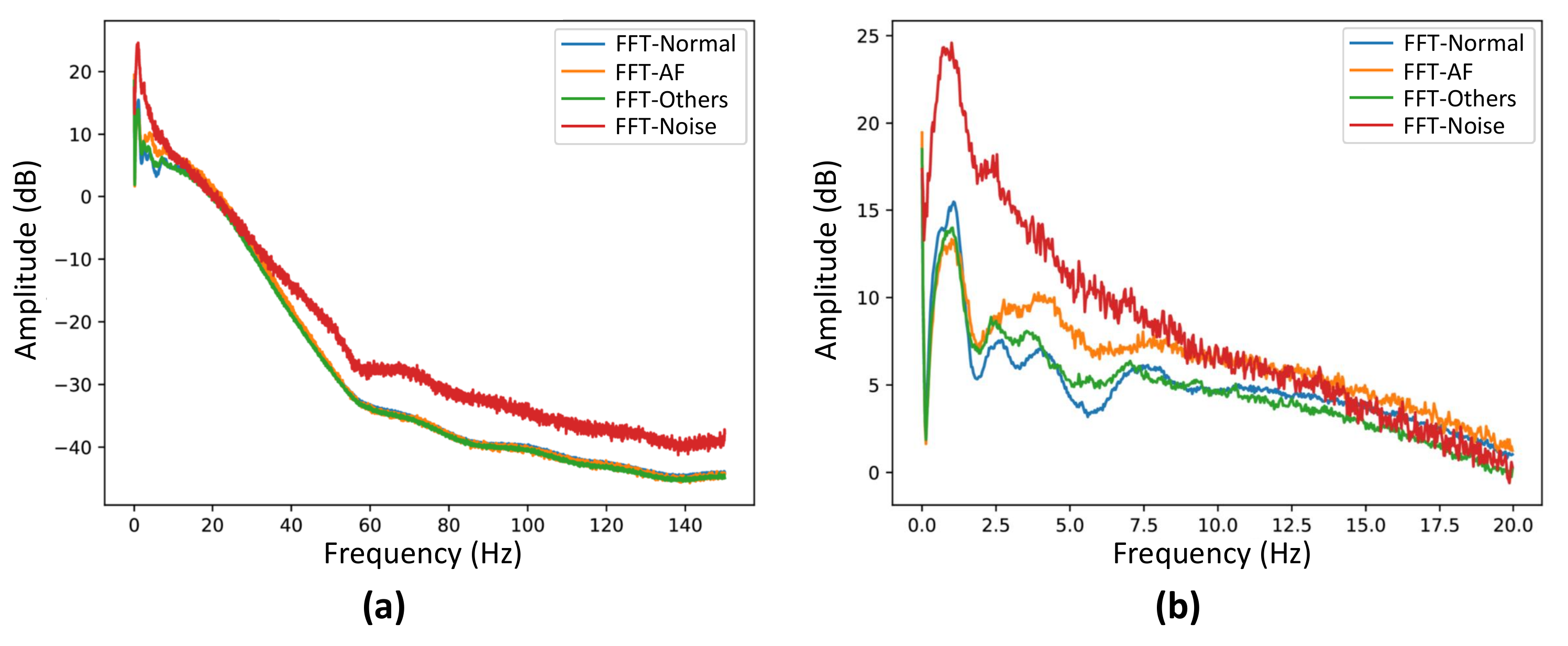}
	\caption{Average FFT results of the four classes of ECG signals (\SI{30}{\second} records) from the public dataset of the CinC challenge: (\text{a}) Frequency range from \SIrange{0}{150}{\hertz} (full scale). (\text{b}) Frequency range from \SIrange{0}{20}{\hertz} (zoomed).}
	\label{fig:FourTypeSignalFFT}
\end{figure}

To test the aforementioned speculation and determine the optimal configuration for the Tukey window, the network structure with $M=5$ and $N=3$ was initially adopted in the AF detector\footnote{The details of the ResNet-based AF detector's structure are presented in Section~\ref{SubSubSect:AFResNet} and Section~\ref{SubApped:OptimalAFDetectorStrcture}. The parameters $M=5$ and $N=3$ were empirically selected before the optimal network structure has been determined.} and the adaptive moment estimation (Adam) optimizer was used in the training phase\footnote{\label{Fn:AdamOptimizer}The Adam optimizer was utilized because it requires little tuning on the neural network's hyper-parameters that heavily influence the model's classification performance.}. Each model with particular input setting (full-scale or half-scale spectrograms, or spectrograms generated by employing Tukey windows with various $W$ and $O$) was trained for three times following the training and validation setups\footnote{\label{Fn:Adam3TimesExp}Within each of the three experiments, the first fold of ECG records was used for validation, while the remaining ones were used for training.} stated in Section~\ref{SubSubSect:TrainValidTestSetAFDtector}. The training processes were terminated when the number of iterations reached \SI{3000}{} or \SI{5000}{}. The smoothed validation top-1 accuracies and $F_{1}$ measures during training\footnote{Each model with particular input setting was trained for three times and the smoothed values are the moving averages (\SI{200}{} samples moving window) of the mean top-1 accuracies or $F_{1}$ measures.} are depicted in Figure~\ref{fig:SmoothResultsHalfFullSpectrograms}, Figure~\ref{fig:SmoothResultsVaringOSpectrograms}, and Figure~\ref{fig:SmoothResultsVaringWSpectrograms}. The labels in each figure denote the input settings; for example, the label ``Tukey-60-30-HS'' represents that the inputs of the AF detector were the lower half (components with frequencies no larger than \SI{75}{\hertz}) of the spectrograms generated by employing a Tukey window with width $W=60$ samples and overlapping $O=30$ samples.

\begin{figure}[H]
	\centering
	\includegraphics[width=0.97\columnwidth]{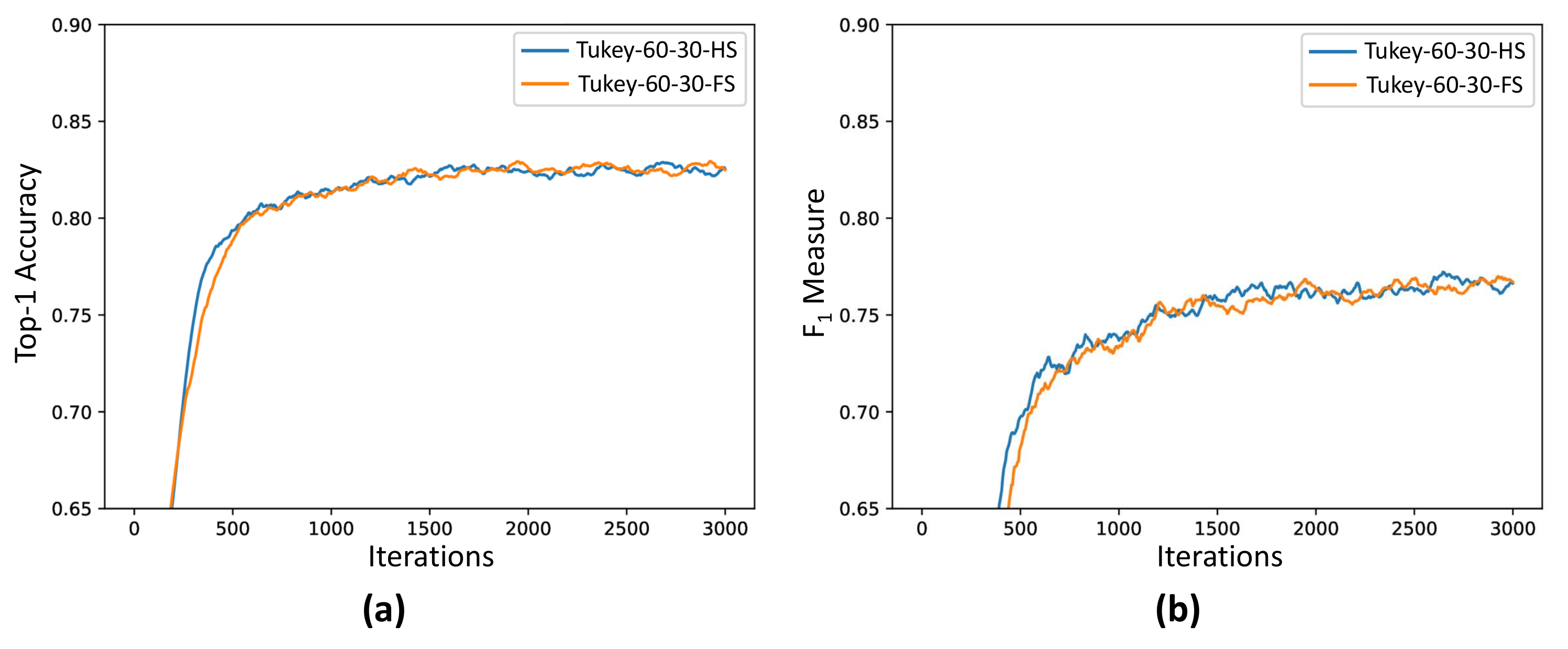}
	\caption{Classification performance of the ResNet-based AF detector ($M=5$ and $N=3$) using half-scale (HS) or full-scale (FS) spectrograms as inputs: (\textbf{a}) Smoothed validation top-1 accuracies. (\textbf{a}) Smoothed validation $F_{1}$ measures.}
	\label{fig:SmoothResultsHalfFullSpectrograms}
\end{figure}

\begin{figure}[H]
	\centering
	\includegraphics[width=0.97\columnwidth]{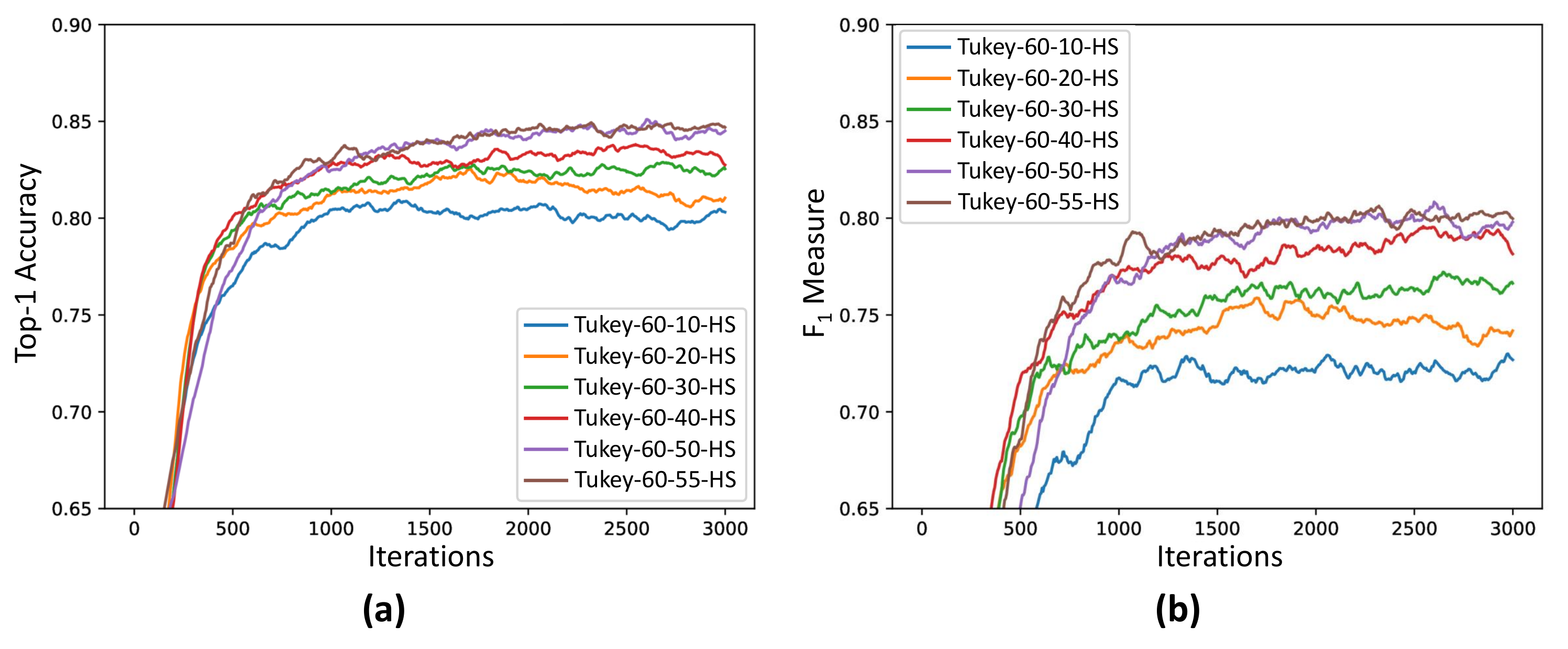}
	\caption{Classification performance of the ResNet-based AF detector ($M=5$ and $N=3$) using half-scale spectrograms, generated by employing Tukey windows with $W=60$ and various $O$, as inputs: (\textbf{a}) Smoothed validation top-1 accuracies. (\textbf{a}) Smoothed validation $F_{1}$ measures.}
	\label{fig:SmoothResultsVaringOSpectrograms}
\end{figure}

\begin{figure}[H]
	\centering
	\includegraphics[width=0.97\columnwidth]{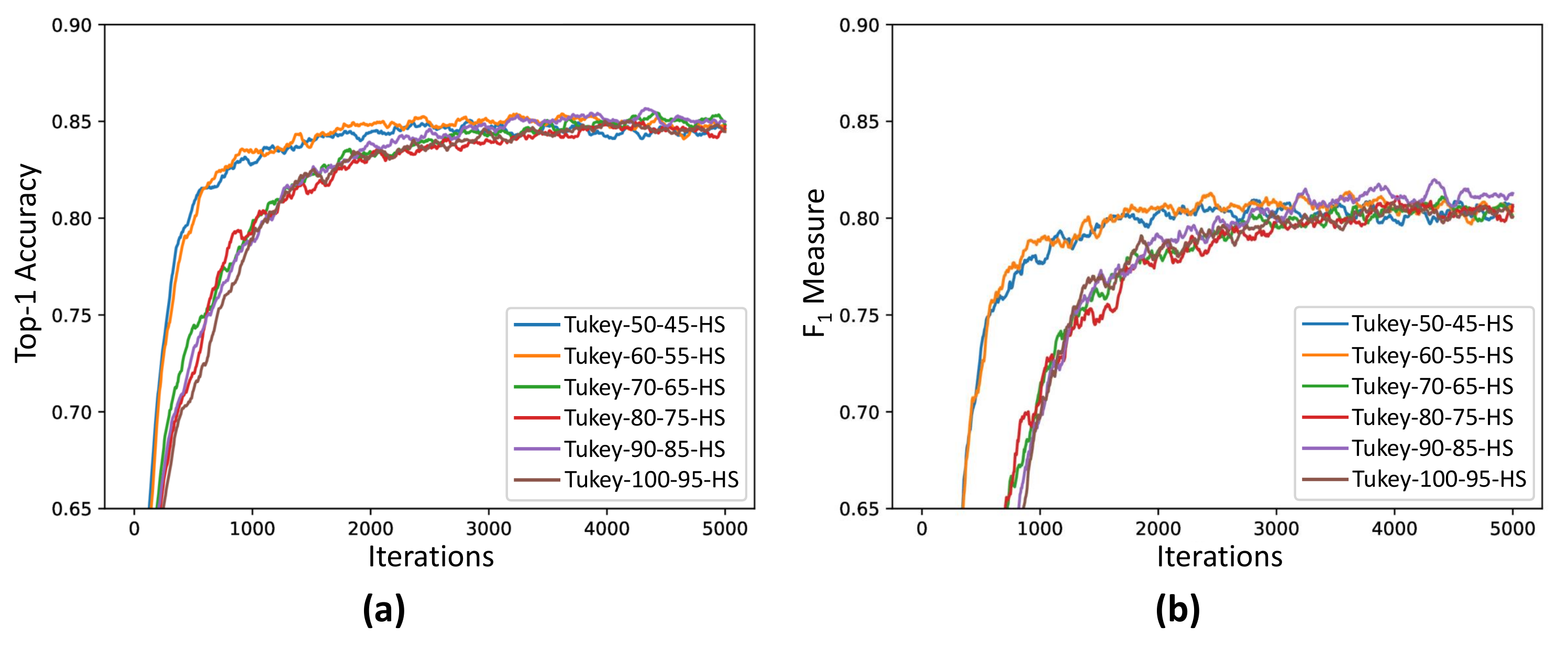}
	\caption{Classification performance of the ResNet-based AF detector ($M=5$ and $N=3$) using half-scale spectrograms, generated by employing Tukey windows with \SIrange{90}{95}{\percent} of overlapping and various $W$, as inputs: (\textbf{a}) Smoothed validation top-1 accuracies. (\textbf{a}) Smoothed validation $F_{1}$ measures.}
	\label{fig:SmoothResultsVaringWSpectrograms}
\end{figure}

As illustrated in Figure~\ref{fig:SmoothResultsHalfFullSpectrograms}, eliminating the frequency components at the upper halves (frequencies above \SI{75}{\hertz}) of the spectrograms showed negligible influence on the AF detector's saturated validation top-1 accuracy and $F_{1}$ measure but slightly boosted the convergence rate. Further considering that using spectrograms with smaller sizes as inputs generally results in a shorter training time and lower GPU memory requirement that unlocks the potential of increasing the network depth, the half-scale spectrograms were utilized as inputs for the AF detector in this work.

As depicted in Figure~\ref{fig:SmoothResultsVaringOSpectrograms} and Figure~\ref{fig:SmoothResultsVaringWSpectrograms}, the saturated validation top-1 accuracies and $F_{1}$ measures of the models employing Tukey windows with different $W$ and $O$ pairs were generally increased along with the growth of the overlapping percentages (i.e., $100 \cdot O / W$), while the width $W$ had a marginal impact on these values. Additionally, the purple and the brown curves in Figure~\ref{fig:SmoothResultsVaringOSpectrograms} suggested that increasing the overlapping percentage would not further improve the AF detector's classification performance once it was over \SI{83}{\percent} (i.e., $W=60$ and $O=50$). Therefore, overlapping percentages higher than \SI{92}{\percent} (i.e., $W=60$ and $O=55$), which imply larger sizes of converted spectrograms, were omitted here. Although the model employing a Tukey window with $W=90$ and $O=85$ (overlapping percentage about \SI{94}{\percent}) could achieve a slightly higher saturated $F_{1}$ measure compared to those employing windows with similar overlapping percentages (from \SIrange{90}{95}{\percent}) but smaller $W$, the blue and the orange curves in Figure~\ref{fig:SmoothResultsVaringWSpectrograms} indicated that models with smaller $W$ yielded faster convergence rates. Besides, if the overlapping percentage has been settled, a Tukey window with smaller $W$ usually implies a converted spectrogram with decreased size, which is preferred as less GPU memory is required. 

Eventually, the Tukey window with $W=60$ and $O=55$ was empirically selected because the model employing such window in the preprocessing phase could achieve satisfactory saturated validation top-1 accuracy and $F_{1}$ measure with a fast convergence rate. The upper half of each one of the spectrograms converted by performing Tukey windowed STFT on the time series ECG signals was eliminated to reduce it size without affecting the classification performance of the ResNet-based AF detector.

\subsection{Determining the Optimal Structure for the ResNet-Based AF Detector}
\label{SubApped:OptimalAFDetectorStrcture}

The simplified structural diagram of the ResNet-based AF detector, which utilized a Tukey window with $W=60$ samples and $O=55$ samples in the preprocessing phase, is presented in Figure~\ref{fig:AFDetectorStructureDiagram}. While the STFT preprocessing block, the first convolutional layer, and the fully connected layer are intrinsic, the structure highlighted in the orange dashed rectangle with rounded corners is configurable. Such configurable structure contains $N$ $ResBlocks$ (denoted by $RB$) and each $RB$ is constructed by $M$ $ResBlock$ (denoted by $rb$). The internal structure of a $rb$, in which two convolutional layers and a skip connection are utilized, is illustrated at the rightmost of Figure~\ref{fig:StructuralDiagramAFDetector}. Conceivably, the network depth of the proposed AF detector is $2 \cdot MN + 2$ layers.

\begin{figure}[H]
	\centering
	\includegraphics[width=0.9\columnwidth]{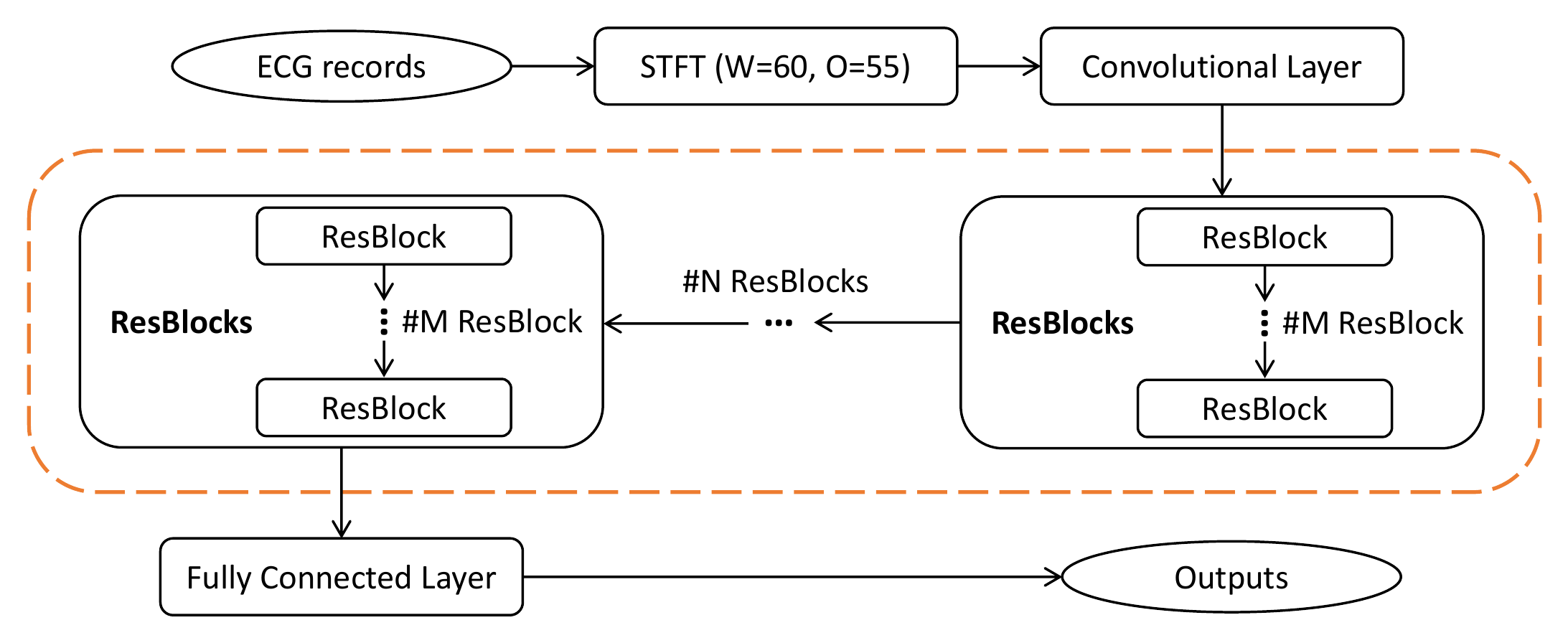}
	\caption{Abstract diagram illustrating the structure of the ResNet-based AF detector.}
	\label{fig:AFDetectorStructureDiagram}
\end{figure}

To investigate the classification performance of the AF detector with various pairs of $M$ and $N$, where $M\in\{1,\cdots,8\}$ and $N\in\{2,\cdots,7\}$, the Adam optimizer, which requires little tuning on the network's hyper-parameters\textsuperscript{\ref{Fn:AdamOptimizer}}, was utilized in the training phase. Each model with particular pair of $M$ and $N$ values was trained for three times following the training and validation setups\textsuperscript{\ref{Fn:Adam3TimesExp}} detailed in Section~\ref{SubSubSect:TrainValidTestSetAFDtector}, and the training processes were terminated when the number of iterations reached 5000. Taking the mean of the first \SI{20}{} largest saturated values (i.e., validation top-1 accuracies or $F_{1}$ measures) as the final result, the classification performance of each model with particular pair of $M$ and $N$ is summarized in Figure~\ref{fig:AFDetectorPerformanceDifferentStructure}. Limited by the GPU's memory, the maximum number of layers within the configurable structure could not exceed \SI{84}{}.

\begin{figure}[H]
	\centering
	\includegraphics[width=0.95\columnwidth]{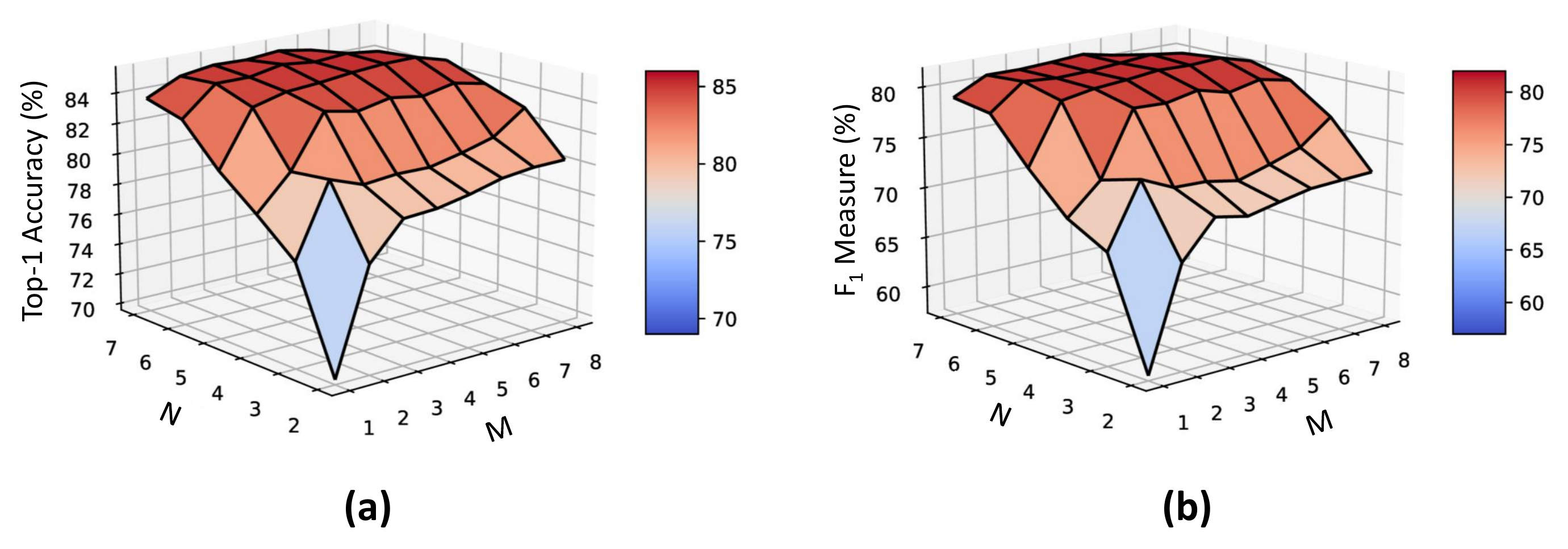}
	\caption{Classification performance of the ResNet-based AF detector with various pairs of $M$ and $N$: (\textbf{a}) Mean saturated validation top-1 accuracies. (\textbf{b}) Mean saturated validation $F_{1}$ measures.}
	\label{fig:AFDetectorPerformanceDifferentStructure}
\end{figure}

As illustrated in Figure~\ref{fig:AFDetectorPerformanceDifferentStructure}, it is apparent that the mean values of the saturated top-1 accuracies or $F_{1}$ measures generally increased along with the growth of $M$ and $N$, which is expected because a classification task usually benefits from a network with greater depth. The largest mean value of the saturated validation top-1 accuracies was achieved when $M=4$ and $N=6$ (i.e., \SI{48}{} convolutional layers), while the largest mean value of the saturated validation $F_{1}$ measures was achieved when $M=5$ and $N=5$ (i.e., \SI{50}{} convolutional layers). No noticeable improvement was observed from the models with greater network depths. Further considering that the $F_{1}$ measure was used for scoring in the 2017 PhysioNet CinC challenge, the configurable structure with $M=5$ and $N=5$ was empirically selected. The final structure of the proposed ResNet-based AF detector with \SI{52}{} layers is depicted in Figure~\ref{fig:StructuralDiagramAFDetector}.

\renewcommand\thespecialtable{C\arabic{specialtable}}
\setcounter{specialtable}{0}
\subsection{Results from the Five 10-Fold Cross-Validation Experiments}
\label{SubApped:AFDetectorPerformanceFiveExperiments}

The proposed AF detector's classification performances from the five 10-fold cross-validation experiments are presented in this section.

\end{paracol}
\renewcommand\thespecialtable{C\arabic{specialtable}}
\begin{specialtable}[H]
\caption{Classification performance of the ResNet-based AF detector in the first 10-fold cross-validation experiment.}
\label{Tab:AFDetectorFirst10FoldResult}
\small
\begin{tabular}{ccccccccccccc}
\toprule
& \textbf{Fold 1} & \textbf{Fold 2} & \textbf{Fold 3} & \textbf{Fold 4} & \textbf{Fold 5} & \textbf{Fold 6} & \textbf{Fold 7} & \textbf{Fold 8} & \textbf{Fold 9} & \textbf{Fold 10} & \textbf{Mean} & \textbf{SD} \\
\midrule
\textbf{Training $A_{1}$ (\%)} & 92.19 & 95.31 & 93.75 & 90.63 & 95.32 & 93.75 & 93.75 & 93.75 & 98.44 & 93.75 & 94.06 & 1.95 \\
\textbf{Training $F_{1}$ (\%)} & 91.81 & 96.46 & 94.45 & 89.03 & 94.20 & 91.81 & 92.23 & 91.52 & 92.68 & 91.60 & 92.58 & 1.93 \\
\midrule
\textbf{Validation $A_{1}$ (\%)} & 86.18 & 85.27 & 85.64 & 86.91 & 87.09 & 88.55 & 85.27 & 85.27 & 83.09 & 86.91 & 86.02 & 1.40 \\
\textbf{Validation $F_{1}$ (\%)} & 83.13 & 80.68 & 81.56 & 84.09 & 84.50 & 86.24 & 82.19 & 80.73 & 78.24 & 84.50 & 82.59 & 2.25 \\
\midrule
\textbf{Testing $A_{1}$ (\%)} & 86.16 & 85.53 & 86.58 & 85.53 & 87.00 & 86.16 & 85.32 & 85.74 & 86.58 & 86.58 & 86.12 & 0.54 \\
\textbf{Testing $F_{1}$ (\%)} & 83.03 & 82.49 & 83.64 & 83.29 & 84.04 & 82.08 & 81.08 & 82.50 & 81.91 & 82.99 & 82.71 & 0.83 \\
\bottomrule
\end{tabular}
\\
\footnotesize
SD is the abbreviations of standard deviation.
\end{specialtable}
\begin{paracol}{2}
\switchcolumn

\begin{specialtable}[H]
\caption{Classification Performance of the ResNet-based AF detector on the testing dataset using different voting strategies in the first 10-fold cross-validation experiment.}
\label{Tab:AFDetectorFirst10FoldStrategies}
\small
\begin{tabular}{ccccccc}
\toprule
& \multicolumn{6}{c}{\textbf{Classification Performance on Testing Dataset (\%)}} \\
\cmidrule[0.6pt]{2-7}
& \textbf{$F_{1}(N)$} & \textbf{$F_{1}(A)$} & \textbf{$F_{1}(O)$} & \textbf{$F_{1}(\sim)$} & \textbf{$A_{1}$} & \textbf{$F_{1}$} \\
\midrule
\textbf{Average}		& -- & -- & -- & -- & 86.12 & 82.71 \\
\midrule
\textbf{Strategy 1} 	& 91.64 & 80.00 & 75.95 & 76.19 & 87.00 & 82.86 \\
\midrule
\textbf{Strategy 2} 	& 91.93 & \textbf{81.93} & 75.52 & \textbf{78.26} & 86.58 & 83.12 \\
\midrule
\textbf{Strategy 3} 	& \textbf{92.43} & 80.95 & \textbf{76.67} & 72.73 & \textbf{87.00} & \textbf{83.35} \\
\bottomrule
\end{tabular}
\end{specialtable}

\begin{specialtable}[H]
\caption{Average classification performances of the ResNet-based AF detector from five 10-fold cross-validation experiments.}
\label{Tab:AFDetectorFive10FoldResult}
\small
\begin{tabular}{cccccccc}
\toprule
& \textbf{EXP 1} & \textbf{EXP 2} & \textbf{EXP 3} & \textbf{EXP 4} & \textbf{EXP 5} & \textbf{Mean} & \textbf{SD} \\
\midrule
\textbf{Training $\overline{A_{1}}$ (\%)} & 94.06 & 92.03 & 92.97 & 92.66 & 91.25 & 92.59 & 0.94 \\
\textbf{Training $\overline{F_{1}}$ (\%)} & 92.58 & 91.81 & 92.31 & 91.17 & 89.86 & 91.55 & 0.97 \\
\midrule
\textbf{Validation $\overline{A_{1}}$ (\%)} & 86.02 & 85.84 & 86.25 & 86.18 & 85.47 & 85.95 & 0.28 \\
\textbf{Validation $\overline{F_{1}}$ (\%)} & 82.59 & 82.83 & 82.98 & 82.65 & 82.49 & 82.71 & 0.18 \\
\midrule
\textbf{Testing $\overline{A_{1}}$ (\%)} & 86.12 & 87.15 & 86.81 & 86.84 & 88.09 & 87.00 & 0.64 \\
\textbf{Testing $\overline{F_{1}}$ (\%)} & 82.71 & 84.82 & 84.25 & 83.12 & 83.57 & 83.69 & 0.76 \\
\bottomrule
\end{tabular}
\\
\footnotesize
EXP and SD are the abbreviations of experiment and standard deviation, respectively.
\end{specialtable}

\end{paracol}
\renewcommand\thespecialtable{C\arabic{specialtable}}
\begin{specialtable}[H]
\caption{Classification Performance of the ResNet-based AF detector on the testing dataset using different voting strategies in the five 10-fold cross-validation experiments.}
\label{Tab:AFDetectorFive10FoldStrategies}
\small
\begin{tabular}{ccccccccccccccc}
\toprule
& \multicolumn{14}{c}{\textbf{Classification Performance on Testing Dataset (\%)}} \\
\cmidrule[0.6pt]{2-15}
& \multicolumn{2}{c}{\textbf{EXP 1}} & \multicolumn{2}{c}{\textbf{EXP 2}} & \multicolumn{2}{c}{\textbf{EXP 3}} & \multicolumn{2}{c}{\textbf{EXP 4}} & \multicolumn{2}{c}{\textbf{EXP 5}} & \multicolumn{2}{c}{\textbf{Mean}} & \multicolumn{2}{c}{\textbf{SD}} \\
\cmidrule[0.6pt]{2-15}
& \textbf{$A_{1}$} & \textbf{$F_{1}$} & \textbf{$A_{1}$} & \textbf{$F_{1}$} & \textbf{$A_{1}$} & \textbf{$F_{1}$} & \textbf{$A_{1}$} & \textbf{$F_{1}$} & \textbf{$A_{1}$} & \textbf{$F_{1}$} & \textbf{$A_{1}$} & \textbf{$F_{1}$} & \textbf{$A_{1}$} & \textbf{$F_{1}$} \\
\midrule
\textbf{Average}		& 86.12 & 82.71 & 87.15 & 84.82 & 86.81 & 84.25 & 86.84 & 83.12 & 88.09 & 83.57 & 87.00 & 83.69 & 0.63 & 0.76 \\
\midrule
\textbf{Strategy 1} 	& 87.00 & 82.86 & 88.05 & 86.41 & 87.63 & 85.39 & 86.79 & 83.61 & 89.10 & 84.43 & 87.71 & 84.54 & 0.83 & 1.23 \\
\midrule
\textbf{Strategy 2} 	& 86.58 & 83.12 & 88.05 & 86.63 & 88.05 & 85.29 & \textbf{87.63} & \textbf{85.18} & 89.31 & 84.76 & 87.92 & 85.00 & 0.88 & 1.13 \\
\midrule
\textbf{Strategy 3} 	& \textbf{87.00} & \textbf{83.35} & \textbf{88.89} & \textbf{87.31} & \textbf{88.05} & \textbf{86.01} & 87.00 & 83.90 & \textbf{89.31} & \textbf{84.92} & \textbf{88.05} & \textbf{85.10} & 0.95 & 1.43 \\
\bottomrule
\end{tabular}
\\
\footnotesize
EXP and SD are the abbreviations of experiment and standard deviation, respectively.
\end{specialtable}
\begin{paracol}{2}
\switchcolumn

\end{paracol}
\reftitle{References}


\externalbibliography{yes}
\bibliography{./Citations/citation.bib}

\end{document}